# Усовершенствованный метод прямого статистического моделирования для решения современных задач динамики разреженных газов

Роман. В. Мальцев

rmaltsev@gmail.com

**Аннотация.** В настоящей работе, прежде всего, предлагаются усовершенствования метода ПСМ в виде схем и подходов, позволяющих повысить его производительность и уменьшить требовательность к ресурсам ЭВМ для широкого класса задач. Наиболее важным усовершенствованием является схема временных множителей, позволяющая использовать разный временной шаг для различных сортов частиц, что позволяет снизить трудоёмкость и/или ресурсоёмкость расчётов для стационарных задач с сильно отличающимися сечениями столкновений компонент смеси между собой. Другие усовершенствования включают в себя обоснованный критерий подобия для оценки требуемого числа частиц для задач различной размерности, новые схемы решения осесимметричных задач, способы ограничения флуктуаций числа частиц и отбраковки повторных столкновений. Также предложены некоторые рекомендации по технической оптимизации алгоритма для современных ЭВМ.

Данная работа основана на материалах кандидатской диссертации по специальности 01.02.05, в настоящий момент подготавливаемой к защите.

В данную работу включены только те материалы диссертации, которые относятся непосредственно к доработке метода ПСМ.

# Оглавление

# Введение

Метод прямого статистического моделирования (далее ПСМ) является одним из важнейших численных методов в газовой динамике. Метод предназначен для расчета течений разреженного газа и может трактоваться как численное решение уравнения Больцмана. Основоположник метода – Грэм Бёрд (1).

Метод ПСМ основан на представлении газа множеством дискретных частиц (каждая их которых представляет собой большое количество реальных молекул), для которых задан стохастический процесс их столкновения друг с другом. Эволюция множества частиц описывается как равномерное прямолинейное движение, прерываемое в случайные моменты времени мгновенными актами парных столкновений, поэтому используются, как правило, модели столкновения с полным конечным сечением. После достижения стационарного режима течения, макропараметры течения вычисляются осреднением параметров частиц в течение достаточно длительного времени.

Метод имеет три основных параметра дискретизации: временной шаг (для ускорения счёта, фазы перемещения и столкновения разделены между собой и чередуются), размер ячейки (столкновительные партнеры выбираются в пределах одной ячейки), число частиц. Классические рекомендации по выбору дискретизации следующие: размер ячейки должен быть меньше длины свободного пробега, временной шаг – меньше времени между столкновениями, а желательно – не больше времени пребывания частицы в ячейке, в каждой ячейке среднее число частиц должно быть велико (по разным рекомендациям, от 3 до 30).

В целом, трудоёмкость метода ПСМ непосредственно связана со степенью разреженности газа, которая определяется числом Кнудсена (отношение длины свободного пробега к характерному размеру системы). Трудоёмкость быстро растёт при уменьшении числа Кнудсена, т.е. с повышением плотности газа. Ситуация осложняется тем, что установление стационарного режима в более плотном газе происходит медленнее, в то время как временной шаг при этом необходимо уменьшать. Как следствие, метод ПСМ применяется, в первую очередь, тогда, когда предположение о предельно малом локальном отклонении газа от равновесия не работает и, как следствие, не применимы уравнения Навье-Стокса, и требуется решение уравнений Больцмана.

Последние 20 лет, в связи с развитием вычислительной техники, ускоряющимися темпами растёт и применение метода ПСМ для решения задач динамики разреженных газов. В свою очередь, ускоряющимися темпами растёт и сложность решаемых задач. В XX веке метод ПСМ использовался, прежде всего, для исследования простых идеализированных проблем. Последнее время метод ПСМ всё чаще применяют как для расчета газодинамики технологических процессов внутри вакуумных систем, так и для моделирования газодинамических условий эксперимента при научных исследованиях. Сюда же можно отнести некоторые исследования процессов получения органических полупроводников и синтеза наносистем, в которых газовая динамика играет немаловажную роль.

Вот некоторые из особенностей современных задач:

- **Сильно отличающиеся массы компонент.** Так, если молекула водорода имеет массу 2 а. е., то масса, например, молекулы пентацена $C_{22}H_{14}$ – важного вещества-прекурсора в исследованиях по получению органических полупроводников – составляет 278 а. е, т. е. более чем в 100 раз тяжелее. Масса же наночастицы серебра диаметром 2 нм составляет уже более 26000 а. е., что на порядки превышает массу молекул газообразных веществ, состоящих лишь из нескольких атомов. Тепловая скорость таких наночастиц более чем в 100 раз меньше таковой для молекул водорода.

- **Сильно отличающиеся размеры молекул.** Газодинамическое сечение столкновения молекул водорода между собой – порядка 0.3 нм$^2$. Сечение же столкновения той же молекулы водорода с большой органической молекулой или наночастицей может составлять десятки нм$^2$ и более. При этом, как правило, почти столь же сильно отличаются частоты столкновений и длины свободного пробега компонент.

- **Сильно отличающиеся концентрации компонент.** Часто мольная концентрация разных компонент смеси отличается на несколько порядков. Например, при сверхзвуковом ускорении тяжелых молекул или наночастиц легким несущим газом требуется весьма сильно разбавлять смесь несущим газом, чтобы средняя масса молекулы смеси была достаточно малой по сравнению с массой тяжелой компоненты. Например, добавление всего 0.1 % примеси пентацена к чистому гелию повышает среднюю массу молекулы смеси почти на 7 %.

- **Сильные перепады плотности.** Последние характерны для сверхзвуковых течений вообще. Однако, проблема часто усугубляется большими перепадами давления в моделируемой системе. Так, в реакторе-испарителе вещества-прекурсора давление может составлять сотни тысяч паскаль, фоновое давление в вакуумной камере – единиц паскаль, а за скиммером молекулярно-пучковой системы, где обычно и происходит осаждение вещества на подложку – глубокий вакуум.

- **Повышенные требования к точности.** Получение полей макропараметров с точностью 10 % – 20 % часто считается достаточным для большинства практических приложений метода ПСМ. Однако, для современных задач такая точность не всегда удовлетворительна. В тех же случаях, когда указанной точности достаточно, обычно желательна предсказуемость величины ошибки и понимание её природы.

- **Повышенные требования к быстродействию кода.** Значительная доля современных газодинамических задач требует использования не менее 1 миллиона ячеек сетки и нескольких миллионов расчетных частиц. При характерных числах Кнудсена порядка 0.001, установление стационарного решения требует расчета огромного количества временных шагов. Как следствие, быстродействие используемого DSMC кода имеет очень важное значение. Классический DSMC-код, основанный, например, на примерах программ Бёрда (1), крайне неэффективен при таких условиях, так как недостаточно учитывает особенности устройства процессоров и памяти современных компьютеров.

В настоящей работе представлен разработанный комплекс новых подходов и схем для метода ПСМ, позволяющий во многих случаях снизить рост трудоёмкости вычислений и ресурсных требований к ЭВМ, связанный с перечисленными выше особенностями.

Первый раздел настоящей работы является кратким изложением основ метода ПСМ, некоторых уже существующих подходов к повышению его эффективности, известных исследований о влиянии дискретизации на его точность.

Небольшой по объёму второй раздел излагает способ повысить сходимость величины мажорантной частоты, а также содержит сравнение среднего числа столкновительных партнёров в схемах мажорантной частоты и no time counter.

В третьем разделе предложены рекомендации по реализации метода ПСМ, учитывающие особенности современных ЭВМ.

Четвертый раздел направлен на изучение влияния дискретизации на точность метода ПСМ при использовании схемы мажорантной частоты. В нём также выводится и тестируется критерий подобия, позволяющий эффективно оценивать требуемое число расчетных частиц.

В пятом разделе предлагается новая схема («временные множители»), позволяющая использовать разный временной шаг для различных молекул в пределах одной ячейки, например, рассчитывать разные компоненты смеси с разным временным шагом. Определены условия, когда применение новой схемы даёт выигрыш в ресурсоёмкости и/или производительности метода ПСМ.

Шестой и седьмой разделы посвящены изучению схем решения осесимметричных задач, для которых характерно многократное различие объёмов ячеек на разном расстоянии от оси. Описывается альтернативный подход, заключающийся в размножении частиц в фазе столкновения, а не в фазе перемещения. Затем, на его базе, а также не основе новой схемы временных множителей, разрабатывается новый подход к решению осесимметричных задач.

В восьмом разделе предложена новая схема размножения-уничтожения частиц, позволяющая многократно уменьшить размах флуктуаций плотности, связанных со случайным блужданием числа частиц при стохастическом процессе их рождения-уничтожения.

В девятом разделе предлагается простой и эффективный способ детектирования повторных столкновений, позволяющий вплоть до двух раз снизить артефакты, вызванные повторными столкновениями при использовании схем размножения-уничтожения частиц.

# 1 Основные сведения о методе ПСМ

## 1.1 Модели столкновений, применяемые в методе ПСМ

Недостаток плавно убывающих потенциалов – бесконечность полного сечения столкновения, причём при больших прицельных расстояниях изменение скорости частиц при столкновении стремится к нулю. В методе ПСМ используют модели столкновения с конечным полным сечением, так как в этом случае столкновения можно рассматривать как дискретную последовательность событий. Простое обрезание потенциала обычно не является эффективным решением проблемы полного сечения, поэтому используются специальные модели столкновения, которым, вообще говоря, не всегда можно поставить в соответствие какой-либо физический потенциал. Параметры моделей настраиваются таким образом, чтобы макроскопические коэффициенты переноса для этих моделей, вычисленные методами кинетической теории (2), соответствовали свойствам моделируемых газов. Ещё одним критерием согласованности является совпадение как можно большего числа *эффективных сечений*, которые, для случая центрально-симметричного потенциала, определены следующим образом:

$$\sigma^{(l)}(E_\infty) = \int_0^\infty \left(1 - cos^l \chi(\rho, E_\infty)\right) 2\pi\rho d\rho.$$

Здесь $\chi$ – угол отклонения, $\rho$ – прицельный параметр, $E_\infty$ – кинетическая энергия относительного движения, $l$ – номер эффективного сечения.

Как правило, использование «искусственных» моделей оказывается достаточно для приемлемого вычисления макропараметров даже сильно неравновесных течений, но вычисленная функция распределения в зонах сильной неравновесности может существенно зависеть от выбора модели (3; 4).

В монографии (1) Бёрд описывает 3 модели столкновений молекул с конечным полным сечением: твердые сферы (Hard Spheres, HS), переменные твердые сферы (Variable Hard Spheres, VHS), переменные мягкие сферы (Variable Soft Spheres, VSS). Последняя была предложена авторами (5).

В работе (3) предложена модель переменных сфер (Variable Spheres, VS). Кроме того, в этой работе произведено численное сравнение результатов, рассчитанных при помощи численной модели, построенной для потенциала Леннарда-Джонса, с результатами, рассчитанными по модели VS, параметры которой задаются таким образом, чтобы первые два эффективных сечения (диффузионное и вязкостное) совпадали. Показано, что в ударной волне макропараметры течения практически совпадают для двух моделей, в то время, как локальная функция распределения может заметно отличаться.

*Модель твердых сфер* – наиболее простая модель, единственный параметр – полное сечение столкновения. Рассеяние частиц изотропно по углам (в системе центра масс). Модель обеспечивает пропорциональность коэффициентов переноса квадратному корню температуры. Число Шмидта (отношение коэффициентов кинематической вязкости и самодиффузии) для газа твердых сфер составляет 5/6.

*Модель VHS* отличается от модели твердых сфер тем, что вводится параметр $\omega$, при этом полное сечение столкновения пропорционально степени $1-2\omega$ относительной скорости сталкивающихся частиц. Это позволяет учесть зависимость вязкости или диффузии от температуры. Коэффициенты переноса такого газа зависят от температуры по степенному закону с показателем $\omega$. Число Шмидта также зависит от $\omega$ и, для $0.5 < \omega < 1$, лежит в диапазоне 5/6 ÷ 1. Модель твердых сфер можно рассматривать как частный случай модели VHS при $\omega$ = 1/2.

*Модель VSS* отличается от предыдущих степенным законом распределения для косинуса угла отклонения, который более не изотропен, и описывается дополнительным параметром $\alpha$. Модель VHS является частным случаем модели VSS с показателем $\alpha$ = 1. Число Шмидта при этом также зависит от $\alpha$.

Формулы, связанные со свойствами модели VSS, приведены в **Табл. 1.1**.

**Табл. 1.1.** Кинетические формулы и коэффициенты переноса для модели VSS.

| | |
|---|---|
| Зависимость полного сечения столкновения $\sigma$ от относительной скорости $c_r$. | $\sigma(c_r) = \frac{\sigma_{ref}}{\Gamma\left(\frac{5}{2}-\omega\right)}\left(\frac{\mu c_r^2}{2kT_{ref}}\right)^{\frac{1}{2}-\omega}$ |
| Плотность вероятности $\frac{df}{dcos\Theta}$ для угла отклонения $\Theta$. | $\frac{df}{dcos\Theta} = \frac{\alpha}{2}\left\|cos\frac{\Theta}{2}\right\|^{2\alpha-2}$ |
| Средняя частота столкновений $\upsilon$ | $\upsilon = \frac{n^2}{2}\overline{\sigma c_r} = \frac{n^2}{2}\sigma_{ref}\sqrt{\frac{8kT_{ref}}{\pi\mu}}\left(\frac{T}{T_{ref}}\right)^{1-\omega}$ |
| Средняя длина свободного пробега $\lambda$ | $\lambda = \frac{1}{\sqrt{2}n\sigma_{ref}}\left(\frac{T}{T_{ref}}\right)^{\omega-\frac{1}{2}}$ |
| Зависимость коэффициента самодиффузии $\mathfrak{D}$ (первое приближение) от температуры $T$. | $\mathfrak{D}(T) = \frac{(\alpha+1)}{\left(\frac{5}{2}-\omega\right)}\left(\frac{T}{T_{ref}}\right)^{\omega}\cdot\frac{3}{8}\frac{\sqrt{\pi mkT_{ref}}}{\sigma_{ref}}\frac{1}{nm}$ |
| Зависимость коэффициента вязкости $\eta$ (первое приближение) от температуры $T$. | $\eta(T) = \frac{(\alpha+1)(\alpha+2)}{\alpha\left(\frac{5}{2}-\omega\right)\left(\frac{7}{2}-\omega\right)}\left(\frac{T}{T_{ref}}\right)^{\omega}\cdot\frac{5}{16}\frac{\sqrt{\pi mkT_{ref}}}{\sigma_{ref}}$ |

Обозначения:

$m$ – масса частицы

$\mu$ – приведенная масса пары ($m/2$ для тождественных частиц)

$n$ – молекулярная плотность газа

$\Gamma(\mathrm{x})$ – гамма-функция от x

$k$ – постоянная Больцмана

$T_{ref}$ – температура, которой соответствует характеристическое сечение $\sigma_{ref}$

$\gamma$ – показатель адиабаты газа

*Модель VS* является упрощением модели VSS. В модели VS частицы при столкновении всегда отклоняются на один и тот же угол $\chi_{ref}$. Так же как и модель VSS, модель VS позволяет сбалансировать между собой вязкостные и диффузионные процессы. При этом, полное сечение, а значит, и частота столкновений для модели VS оказываются минимальными, в т.ч. меньше таковых для VSS. Кроме того, вычисление столкновения требует меньше операций, а также на одно случайное число меньше. В этом отношении модель VS наиболее эффективна в вычислительном плане.

Формулы, связанные со свойствами модели VS, приведены в **Табл. 1.2**.

**Табл. 1.2.** Кинетические формулы и коэффициенты переноса для модели VS.

| | |
|---|---|
| Угол отклонения $\Theta$ | $\Theta = \chi_{ref}$ |
| Зависимость коэффициента самодиффузии $\mathfrak{D}$ (первое приближение) от температуры $T$. | $\mathfrak{D}(T) = \frac{2}{\left(1-\cos\chi_{ref}\right)\left(\frac{5}{2}-\omega\right)}\left(\frac{T}{T_{ref}}\right)^{\omega} \cdot \frac{3}{8}\frac{\sqrt{\pi m k T_{ref}}}{\sigma_{ref}}\frac{1}{nm}$ |
| Зависимость коэффициента вязкости $\eta$ (первое приближение) от температуры $T$. | $\eta(T) = \frac{4}{\left(1-\cos^2\chi_{ref}\right)\left(\frac{5}{2}-\omega\right)\left(\frac{7}{2}-\omega\right)}\left(\frac{T}{T_{ref}}\right)^{\omega} \cdot \frac{5}{16}\frac{\sqrt{\pi m k T_{ref}}}{\sigma_{ref}}$ |

## 1.2 Схемы выбора столкновительных пар в методе ПСМ

В методе ПСМ акт столкновения пары частиц является единичным, мгновенным событием. В результате парного столкновения параметры обоих частиц изменяются в соответствии с моделью столкновений. В модели столкновений прицельные параметры столкновения выбираются по вероятностному закону, игнорируя взаимное расположение частиц в пространстве.

Пространство делится на ячейки небольшого размера. Обе частицы пары выбираются из одной ячейки. Используется допущение, что в пределах ячейки функция распределения параметров частиц однородна по пространству. Взаимное пространственное положение частиц в ячейке игнорируется. Задается временной шаг $\delta t$.

Задача алгоритма столкновения – случайно выбрать подряд некоторое количество пар моделируемых частиц, соответствующих столкновениям за интервал времени $\delta t$. Наиболее известны три подхода: со счетчиком времени (time counter), без счетчика времени (no time counter), а также схема мажорантной частоты (6) – наиболее современная из трех перечисленных, именно она подразумевается в данной работе.

Схема мажорантной частоты (MFS – majorant frequency scheme) определяет для каждой пары частиц (i,j) независимо вероятность столкновения в единицу времени (среднюю частоту) $\nu_{ij}$:

$$\nu_{ij} = \frac{\sigma_{ij}(\mathrm{c}_{ij}) \cdot \mathrm{c}_{ij}}{V/F}.$$

Здесь $\mathrm{c}_{ij}$ – относительная скорость, $\sigma_{ij}(\mathrm{c}_{ij})$ – сечение столкновения (зависящее от скорости), $V$ – объем ячейки, $F$ – количество молекул, соответствующих одной модельной частице. Считается, что вероятность однородна по времени. Тогда при наличии единственной пары частиц, вероятность того, что ни одного столкновений за время $\delta t$ не произойдет, равна $e^{-\nu_{ij}\delta t}$. Количество $P$ возможных пар составляет $P = \frac{N(N-1)}{2}$ для частиц одного сорта в количестве $N$, и $P = N_1 \cdot N_2$ при наличии частиц двух сортов в соответствующих количествах.

Идея схемы мажорантной частоты заключается в том, что задается некоторая оценка сверху для частоты одного столкновения $\nu_{max}$, тогда среднее ожидаемое число событий за время $\delta t$ при их частоте $\nu_{max}$ на каждую пару, составит $\overline{K_{max}} = P \cdot \nu_{max} \cdot \delta t$. Эта величина служит основой для вычисления случайного количества тестов $K_{max}$ согласно вероятностного закона распределения Пуассона: $w_K = \frac{\bar{K}^K}{K!} e^{-\bar{K}}$. Теперь при каждом тесте достаточно равновероятно выбрать пару (i,j) и с вероятностью $\nu_{ij}/\nu_{max}$ принять это столкновение и обработать его, а иначе отклонить и проигнорировать.

Используется следующая реализация выборки из распределения Пуассона. Каждая итерация начинается с вычисления случайного интервала между двумя тестами согласно плотности вероятности:

$$df(\tau) = P \cdot \nu_{max} \cdot e^{-P \cdot \nu_{max} \cdot \tau} d\tau.$$

Затем счетчик времени уменьшается на величину $\tau$, и, если он всё еще составляет положительную величину, производится один тест на столкновение, после чего запускается следующая итерация.

Для схемы мажорантной частоты необходима оценка $\nu_{max}$. Ее можно было бы вычислить, перебрав все возможные пары и взяв наибольшее значение. Однако, это неэффективно с вычислительной точки зрения, так как число столкновений в ячейке за временной шаг должно быть малым по сравнению с числом пар, поэтому перебор всех пар сильно замедляет вычисления. На практике удобно задать в каждой ячейке некоторое изначальное разумное значение $\nu_{max}$, и корректировать его каждый раз, когда оно превышено. В этом случае, частота событий превышения предыдущего значения $\nu_{max}$ весьма низка. Как правило, средняя величина $\nu / \nu_{max}$ при расчете быстро снижается к значению от 10 % до 50 % и в дальнейшем практически не меняется со временем.

Некоторые соображения для улучшения сходимости $\nu_{max}$ приведены в разделе 2.

Схема мажорантной частоты пригодна для моделирования нестационарных течений.

Схема со счетчиком времени предполагает одинаковую вероятность столкновений для всех пар, причём счетчик времени уменьшается на величину, обратно пропорциональную вероятности столкновения данной пары в единицу времени. В настоящее время признана устаревшей и не используется.

Схема без счетчика времени (NTC – no time counter) была создана для избавления от недостатков предыдущей схемы и до сих пор используется во многих исследованиях. Она похожа на схему мажорантной частоты, но количество столкновений $K_{ab}$ между частицами видов $(a, b)$ вычисляется по другой формуле:

$$K_{ab} = \frac{N_a \overline{N_b} \nu_{max} \delta t}{2} + K'_{ab}.$$

Если $a \neq b$, то учет пары производится дважды: $(a, b)$ и $(b, a)$. $K'_{ab}$ – разница между рассчитанным и фактическим числом столкновений при предыдущем шаге. Так как фактическое количество столкновений всегда целое, то $K_{ab}$ округляется до целого значения, а разница переносится на следующий шаг. Если частиц в ячейке нет, то $K_{ab}$ полностью переносится на следующий шаг. В простой модификации схемы NTC – без подъячеек – то же самое происходит и в том случае, когда в ячейке находится только 1 частица.

Широко применяется схема с подъячейками, когда ячейка состоит из нескольких подъячеек. При этом среднее число частиц $\overline{N_b}$ и счетчик дробных частей $K'_{ab}$ являются общими для всей составной ячейки. После того, как выбрана первая частица для столкновения, поиск её партнера производится сперва в той же самой подъячейке, что и первая частица. Если же исходная частица была в подъячейке единственной, то поиск продолжается в соседних подъячейках, в порядке их удаленности от исходной.

Схема с подъячейками позволяет использовать размер ячейки больше длины свободного пробега, если нет резких градиентов плотности. Но размер подъячеек всё же не должен быть больше длины свободного пробега.

Схема NTC требует знания среднего числа частиц в ячейке (т.е. плотности), которое приходится вычислять осреднением за всё время счета. По этой же причине он непригоден для расчета нестационарных течений. С вычислительной точки зрения, хранение наряду с $v_{max}$ величин $\overline{N_b}$ и $\acute{K_{ab}}$ для каждой ячейки требует лишних затрат памяти (что, правда, частично решается использованием логики подъячеек). Для $v_{max}$, напротив, можно использовать одну величину на несколько ячеек без ущерба корректности счета.

## 1.3 Рекомендации к выбору дискретизации

Свобода выбора числа модельных частиц, размера ячейки, временного шага при использовании метода ПСМ ограничена условиями достоверности метода. Часто рекомендуют исходить из следующих соотношений:

- Размер ячейки $\delta x$ меньше локальной длины свободного пробега $\lambda$: $\delta x \ll \lambda$.
- Расстояние $v_{max}\delta t$, проходимое каждой частицей за временной шаг $\delta t$, меньше размера ячейки: $v_{max}\delta t < \delta x$ , также временной шаг много меньше времени между столкновениями (что автоматически выполняется при $\delta x \ll \lambda$ и $v_{max}\delta t < \delta x$).
- Среднее число частиц $\overline{N_\text{C}}$ в каждой ячейке велико: $\overline{N_\text{C}} \gg 1$.

В литературе влияние дискретности параметров исследовалось как теоретически, так и численно.

### 1.3.1 Теоретические предсказания

Согласно теории (7), основанной на изучении пространственных корреляций, дискретность пространства и связанная с ним конечная дистанция между сталкивающимися частицами должны приводить к искажению коэффициентов переноса (вязкости $\eta$ и теплопроводности $\kappa$), что в простейшем случае однокомпонентного газа твердых сфер можно описать формулами:

$$\frac{\kappa}{\kappa_{|\delta x \to 0}} = 1 + \frac{32}{225\pi} \cdot \left(\frac{\delta x}{\lambda}\right)^2,$$

$$\frac{\eta}{\eta_{|\delta x \to 0}} = 1 + \frac{16}{45\pi} \cdot \left(\frac{\delta x}{\lambda}\right)^2.$$

Здесь $\kappa$ – коэффициент теплопроводности, $\eta$ – коэффициент вязкости, $h$ – размер ячейки, $\lambda$ – локальная длина свободного пробега. Теория предсказывает второй порядок точности по шагу сетки.

Прямого искажения процессов диффузии из-за дискретности пространства не происходит.

Существует также теория (8), описывающая влияние дискретности времени:

$$\frac{\kappa}{\kappa_{|\delta t \to 0}} = 1 + \frac{64}{675\pi} \cdot \frac{2kT}{m}\left(\frac{\delta t}{\lambda}\right)^2,$$

$$\frac{\eta}{\eta_{|\delta t \to 0}} = 1 + \frac{32}{150\pi} \cdot \frac{2kT}{m}\left(\frac{\delta t}{\lambda}\right)^2,$$

$$\frac{D}{D_{|\delta t \to 0}} = 1 + \frac{4}{27\pi} \cdot \frac{2kT}{m}\left(\frac{\delta t}{\lambda}\right)^2.$$

Здесь $D$ – коэффициент самодиффузии, $\frac{2kT}{m}$ – значение наиболее вероятной тепловой скорости, $\delta t$ – временной шаг. Теория предсказывает второй порядок точности по временному шагу.

Что касается зависимости ошибки от числа частиц в ячейке – на уровне гипотезы полагается, что ошибка обратно пропорциональна числу частиц в ячейке: $\sim \frac{1}{N_\text{C}}$.

Автором настоящей работы проведено специальное исследование для обобщения влияния числа частиц в ячейке на случаи большей размерности при использовании схемы мажорантной частоты и предложен критерий $\overline{N_C}\mathrm{Kn}_C \gg 1$, где $\overline{N_C}$ – число частиц в ячейке, $\mathrm{Kn}_C$ – количество ячеек вдоль длины свободного пробега. Эти результаты представлены в разделе 4.

### 1.3.2 Численные исследования

Численное исследование погрешностей произведено в (9). Исследование проведено для схемы NTC и только для случая с одним пространственным измерением. Анализировался вычисленный тепловой поток между двух пластин. При этом градиент температур был слабый (перепад $\approx 1.45$ раза), число Кнудсена же составляло $\approx 0.025$. Вычисления проводились при различных числе частиц, временном шаге, шаге сетки. Затем вычислялся эффективный коэффициент теплопроводности. После регрессии, авторы получили следующее соотношение:

$$\frac{\bar{\kappa}_{DSMC}}{\kappa_{|h\to0,\Delta t\to0,\overline{N_C}\to\infty}} = 1.0001 + 0.0287\Delta\tilde{t}^2 + 0.0405\Delta\tilde{x}^2 - 0.0009\Delta\tilde{x}^4 - 0.016\Delta\tilde{t}^2\Delta\tilde{x}^2 + 0.0081\Delta\tilde{t}^4\Delta\tilde{x}^2$$
$$+\frac{-0.083 + 1.16\Delta\tilde{x} - 0.220\Delta\tilde{x}^2 + 1.56\Delta\tilde{t}^2 - 2.55\Delta\tilde{t}^2\Delta\tilde{x} + 1.14\Delta\tilde{t}^2\Delta\tilde{x}^2}{\overline{N_C}}$$
$$+\frac{-0.92\Delta\tilde{t}^3 + 1.91\Delta\tilde{t}^3\Delta\tilde{x} - 0.94\Delta\tilde{t}^3\Delta\tilde{x}^2}{\overline{N_C}} + \frac{0.095\Delta\tilde{t}^2}{\overline{N_C}^2} \quad (3).$$

Здесь $\frac{\bar{\kappa}_{DSMC}}{\kappa_{|h\to0,\Delta t\to0,\overline{N_C}\to\infty}}$ – отношение эффективного и реального коэффициентов теплопроводности, $\Delta\tilde{t} = \sqrt{\frac{2kT}{m}}\frac{\delta t}{\lambda}$, $\Delta\tilde{x} = \frac{\delta x}{\lambda}$, $\overline{N_C}$ – среднее число частиц в ячейке.

Авторы (9) сравнили также и непосредственно полученную ими зависимость коэффициента теплопроводности с теорией, получив, соответственно, различие в 5 % для временного шага и 9 % для шага сетки.

Что касается числа частиц, то наибольший вклад вносят члены $\sim\frac{1}{\overline{N_C}}$ и $\sim\frac{\Delta\tilde{x}}{\overline{N_C}}$.

В работе (10) производился расчет ударной волны, также на одномерной сетке, но уже с использованием схемы мажорантной частоты. Кроме того, подсчитывалась доля повторных столкновений. Результаты были получены следующие. Во-первых, какой-либо существенной зависимости ошибки от числа частиц в ячейке не обнаружилось. Во-вторых, было обнаружено уменьшение ошибки при уменьшении доли повторных столкновений. В-третьих, показано, что величина $N_\lambda$ хорошо характеризует как долю повторных столкновений, так и ошибку поля макропараметров, связанную с конечностью числа частиц. $N_\lambda$ – число частиц в элементарном объёме с линейным размером порядка локальной длины свободного пробега.

Оба приведенных выше результата по изучению влияния числа частиц были получены для одномерного случая, в котором $\frac{1}{N_\lambda} = \frac{\Delta\tilde{x}}{\overline{N_C}}$. Однако, для иного числа пространственных измерений это равенство не выполняется, поэтому два критерия на большее количество измерений обобщаются по разному.

## 1.4 Дополнительные приёмы для улучшения дискретизации

В методе ПСМ точность вычисления макропараметров находится в прямой зависимости от плотности расчетных частиц. При расчетах часто возникают ситуации, когда частицы сосредоточены совсем не там, где макропараметры особо интересны. Кроме того, могут играть роль и ограничения на минимальное количество расчетных частиц. Примеры проблемных ситуаций:

- смесь газов, в которой одна или несколько интересующих компонент имеют очень низкую концентрацию по сравнению с остальными компонентами, из-за чего статистическая погрешность их макропараметров велика,

- течения, в которых плотность меняется на порядки,

- осесимметричные течения, в которых основная масса частиц сосредоточена на периферии, в том время как в приосевой области частиц слишком мало.

Возникает потребность искусственно перераспределить распределение плотности расчетных частиц. Это достигается применением весовых множителей. Идея в том, чтобы в интересующих областях было сконцентрировано большое число частиц малого веса, в то время как в менее интересных областях располагаются частицы большого веса. Для этого необходимо задать закон, по которому частицы разных весов будут взаимодействовать друг с другом, и закон, по которому частицы меняют свой вес. Как правило, при этом неизбежно возникает необходимость так или иначе размножать частицы большого веса и уничтожать частицы малого.

### 1.4.1 Схема Бёрда компонентных весовых множителей

Количество физических молекул на одну расчетную частицу часто называют весом или весовым множителем частицы. Различающиеся компонентные весовые множители полезны для ситуации, когда концентрация интересующей компоненты очень мала.

Схема Бёрда столкновения частиц разного веса никак не затрагивает перемещение частиц. При столкновении двух частиц 1 и 2 весами $F_1$ и $F_2$ соответственно (считаем $F_1 < F_2$), вероятность столкновения умножается на $F_2$ – максимальный из двух весов. После столкновения, по-прежнему, остается две частицы с теми же весами. Параметры частицы 1 всегда изменяются в соответствии с законом столкновения. Параметры частицы 2 изменяются с вероятностью $F_1/F_2$, в противном случае остаются теми же, что были до столкновения. Таким образом, частица 2 должна совершить в среднем $F_2/F_1$ столкновений с частицами типа 1, чтобы ее параметры изменились. Отсюда же произрастает недостаток, характерный для всех схем с размножением частиц: так как исходные частицы не всегда удаляются из ячейки при столкновении, возможен вариант, когда одна и та же частица 2 может столкнуться с одной и той же исходной частицей 1 несколько раз подряд, что нефизично. В этом случае, повторные столкновения могут нарушать статистику обмена частиц энергией и импульсом, что приводит к искажению функции распределения частиц даже в равновесном случае и к неправильным значениям вычисленных макропараметров.

Ещё одна проблема весовых схем с размножением частиц – неконсервативность. В описанной выше схеме количество частиц всё ещё сохраняется. Однако, энергия и импульс сохраняются лишь в среднем и потому подвержены случайным блужданиям, что может создать проблему при расчете замкнутых течений.

### 1.4.2 Пространственно-зависимые весовые множители

Использование разного веса частиц в разных областях пространства позволяет как решить проблему нехватки частиц в ячейках малого объёма, так и просто улучшить сбор статистики в таких ячейках. Подход наиболее полезен при решении осесимметричных задач, в которых, при том же размере ячеек, их объём пропорционален расстоянию до оси симметрии. В этом случае пространственно-зависимые веса называют радиальными весами, и частицам сопоставляется вес, связанный с расстоянием от оси.

После того, как частица удалилась от оси, её вес $F$ становится меньше $\acute{F}$ – веса, соответствующего её положению. В этом случае, частица уничтожается с вероятностью $1 - \frac{F}{\acute{F}}$, а иначе приобретает вес $\acute{F}$. Если же частица движется к оси, то её необходимо размножить – превратить в несколько частиц. Параметры этих размножившихся частиц будут совпадать, что является недостатком схемы, приводящим к тем же артефактам, что и схема Бёрда. Дополнительно, появляются флуктуации числа частиц в расчетной области, в некоторых случаях довольно значительные (11).

Существует две вариации подхода в задании весов. В первом варианте, пространственный вес постоянен в пределах ячейки, но между ячейками меняется скачками. Тогда все частицы в ячейке имеют один и тот же вес, что удобно при обработке столкновений. Во втором варианте, вес частиц привязывается к их координатам. Этот вариант считается более эффективным (1). Однако, сталкивающиеся частицы при этом имеют различный вес. При столкновении частиц разного веса может применяться схема, аналогичная схеме Бёрда компонентных весовых множителей. Чаще всего, вес частиц вовсе игнорируется, что может быть допустимо, если в пределах ячейки веса частиц отличаются слабо (1).

Как правило, радиальные весовые множители вводятся лишь начиная с некоторого минимального радиуса.

*Задержка копий частиц* на случайный интервал времени – один из способов избавиться от взаимосвязи параметров размножившихся частиц. Обычно это достигается применением дополнительного массива конечного размера, в который каждая копия частицы помещается в случайную ячейку, в то время как частица, занимавшая эту ячейку ранее, возвращается в поток. Такой подход «смазывает» функцию распределения по времени, что ограничивает его использование для расчета нестационарных течений. Для устранения взаимозависимости, достаточно задержать частицы на время порядка интервала между столкновениями (12). Что, впрочем, для течений с большими градиентами плотности может быть непросто.

В разделе 6 настоящей работы представлено несколько новых способов применения радиальных весовых множителей. В разделе 8 предложен метод ограничения размаха флуктуаций числа частиц.

### 1.4.3 Математические исследования схем весовых множителей

Существует множество статей, исследующих свойства схем весовых множителей на основе строгого математического подхода. Кратко рабочую математическую модель (13) можно описать следующим образом:

- Каждой частице приписывается статистический вес.
- Каждому столкновению также приписывается свой статистический вес как функция от параметров исходных и/или послестолкновительных частиц. Вес столкновения является свободным параметром. Он положителен, но не может превышать весов исходных частиц. Частота столкновения пары обратно пропорциональна весу столкновения, так, чтобы их произведение соответствовало физической частоте столкновений.
- В процессе столкновения, генерируется две новых частицы с весом, равным весу столкновения. Вес обоих исходных частиц уменьшается, также на величину веса столкновения, но они остаются в потоке. Таким образом, каждое столкновение рождает 2 новых частицы, кроме предельных случаев, когда веса исходных частиц уменьшаются до нуля. Как следствие, число частиц в расчетной области растёт в процессе счёта.
- Заданный таким образом процесс столкновения всё ещё обладает свойствами консервативности.
- Для ограничения числа частиц в расчётной области (редукция), необходимо задать процедуру уничтожения избыточных частиц. Именно процедура уничтожения частиц привносит неконсервативность.

В работе (14) доказывается сходимость некоторого класса стохастических схем редукции, в частности двух классических подходов. Первый из них состоит в том, что каждая частица с весом $W_i$ ниже некоторого заданного веса $W_0$, с вероятностью ${W_i}/{W_0}$ выживает и приобретает вес $W_0$, в противном случае уничтожается. Второй подход состоит в том, что сперва по некоторому принципу выбирается подмножество частиц, подлежащее редукции, а затем из этого подмножества выделяется одна случайная частица, которая выживает и приобретает суммарный вес всего подмножества. Вероятность частицы быть избранной пропорциональна её весу.

Для достижения консервативности, предлагается перераспределять параметры уничтожаемых частиц между выживающими частицами. Так, в (15) предлагается поделить множество частиц на кластеры частиц, имеющих близкие скорости, и проводить редукцию в каждом кластере отдельно. При этом может быть достигнута консервативность в пределах каждого кластера, а вносимая ошибка будет ограничена.

### 1.4.4 Методы устранения неконсервативности

Для компонентных весовых множителей, наиболее известен подход (16). Неконсервативность импульса устраняется искусственно – импульс частицы большого веса осредняется: значение до столкновения берётся с весом $1 - F_1/F_2$, значение после столкновения берётся с весом $F_1/F_2$. Такое осреднение импульса неизбежно приводит к убытку кинетической энергии осредняемой частицы. Убыток пропорционален произведению $\frac{F_1}{F_2}\left(1 - \frac{F_1}{F_2}\right)$, и, при $F_1 \ll F_2$, довольно мал. Для устранения энергетической неконсервативности, этот убыток накапливается в каждой ячейке. Затем, при столкновении частиц большого веса между собой, накопленная величина искусственно подмешивается к кинетической энергии относительного движения.

Схема Бойда может быть расширена и на случай химических реакций в потоке и/или на поверхности (17).

Следует помнить, что устранение неконсервативности не гарантирует исчезновение численных артефактов, таких как искажение функции распределения.

Способ на основе кластеризации (13) сложен и в исходной формулировке непрактичен, так как кластеризация частиц – затратная операция.

Есть и более простой подход (18). Идея его в том, чтобы распределить все частицы внутри 6 групп. Суммарные величины 5 моментов для каждой из 6 групп образуют матрицу $\mathrm{M}$ размером $5 \times 6$ элементов. Фаза редукции запускается тогда, когда частиц скопилось слишком много. Во время фазы редукции, веса групп корректируются таким образом, чтобы суммарные значения 5 моментов (плотность, импульсы, энергия) для совокупности групп не изменились, но вес одной из групп обращался в ноль. Все частицы группы с нулевым весом (т.е. около 1/6 частиц) уничтожаются. Используется следующий алгоритм:

1. генерируется случайный 6-мерный единичный вектор $\vec{g}$.
2. Сгенерированный вектор преобразуется по формуле: $\vec{g} \coloneqq \vec{g} - \mathrm{M}^{\mathrm{T}}(\mathrm{M}\mathrm{M}^{\mathrm{T}})^{-1}\mathrm{M}\vec{g}$.
3. Определяются величины $\alpha_1 = \min_{g_k<0}\left(-\frac{1}{g_k}\right)$ и $\alpha_2 = \min_{g_k>0}\left(\frac{1}{g_k}\right)$.
4. С вероятностью $\frac{\alpha_1}{\alpha_1+\alpha_2}$, используется вектор $h_m = 1 - \alpha_2 g_m$. Иначе (с вероятностью $\frac{\alpha_2}{\alpha_1+\alpha_2}$), используется вектор $h_m = 1 + \alpha_1 g_m$.
5. Вес m-й группы умножается на $h_m$. Одна из групп при этом обретёт нулевой вес.

Авторы, однако, не дают рекомендаций, как именно следует распределять частицы по группам.

### 1.4.5 Пространственно-зависимый временной шаг

Как видно из приведенных рекомендаций, в плотных областях течения необходимо использовать маленькие ячейки и временной шаг, так как длина свободного пробега и интервал между столкновениями там малы. Кроме того, плотность расчетных частиц в плотной области должна быть достаточно большой.

В то же время, для сверхзвуковых течений характерны перепады плотности на несколько порядков. В разреженных областях использование таких же мелких ячеек и короткого временного шага избыточно.

Шаг сетки можно многократно увеличить в таких областях, так как логика ПСМ допускает использование переменного шага сетки. Увеличение ячеек позволяет уменьшить необходимую для счета плотность расчетных частиц (при размерности пространства два и более). Однако, при фиксированном числе реальных молекул на одну расчетную частицу, напрямую избежать избытка частиц в разреженных областях нельзя. Один из способов обойти проблему с плотностью расчетных частиц – использовать пространственно-зависимые веса. Но, имея присущие им недостатки, проблему слишком короткого временного шага весовые множители не решают.

В работе (19) предложено использовать пространственно-зависимый временной шаг. При этом в разных областях расчетной области временной шаг отличается – в плотных областях используются мелкие ячейки и короткий временной шаг, в то время как в разреженных областях – крупные ячейки и увеличенный временной шаг. При пересечении границы двух областей с разным временным шагом (очевидно, в фазе перемещения), частица сохраняет долю истраченного временного шага, соответственно, остаток времени на перемещение масштабируется согласно отношению временных шагов в двух областях. Метод прост в реализации. Очевидно, для нестационарных расчетов он неприменим.

Побочным эффектом пространственно-зависимого временного шага является изменение веса частиц. Другими словами, при сохраняющемся количестве расчетных частиц, в областях с разным временным шагом одна и та же частица соответствует разному числу частиц реального газа – прямо пропорционально временному шагу. Однако, количество пересечения частицами границ ячеек и столкновения с поверхностями за счетный временной шаг не изменяется – что нужно учитывать при накоплении статистики по пересечениям.

При одной и той же физической плотности, в областях с малым временным шагом частицы проходят меньшее расстояние за временной шаг, в результате чего и скапливаются там. Частицам нужно больше временных шагов, чтобы покинуть ячейку, в то время как частота пересечения границ ячейки частицами не зависит от временного шага. Напротив, в областях с большим временным шагом частиц мало, но перемещаются они быстро (и быстро покидают ячейку).

Указанный побочный эффект оказывается полезен, и использование пространственно-зависимого временного шага способствует решению сразу двух проблем: и малого временного шага, и избыточно большого числа частиц. Наилучший баланс этих двух эффектов складывается при двух пространственных измерениях.

Однако, имеется ещё один побочный эффект. Сохранение числа расчетных частиц теперь не означает сохранение числа физических молекул. При расчёте замкнутых течений с заданным числом частиц это означает, что, в процессе установления стационарного режима, меняется физическое число молекул в системе, по мере того как частицы перераспределяются по областям с разным временным шагом. То же самое касается сохранения импульса и энергии. Это создает определенные неудобства при применении пространственно-зависимого временного шага для расчета замкнутых систем.

В разделе 5 настоящей работы представлено обобщение метода для смесей газов, позволяющее использовать разный временной шаг для разных компонент смеси. Кроме того, в разделе 6, идея переменного шага по времени используется для построения новых схем решения осесимметричных задач.

# 2 Улучшение сходимости мажорантной частоты

В наиболее распространённой реализации метода, сперва в каждой ячейке задаётся некоторое разумное значение мажорантной частоты $\nu_{max}$, затем, каждый раз при превышении частотой столкновения пары $\nu_{ij}$ величины $\nu_{max}$, происходит коррекция мажорантной частоты: $\nu_{max} \coloneqq \nu_{ij}$. При этом, вычисление $\nu_{ij}$ выполняется уже после выбора первой столкновительной пары, который, в свою очередь, происходит только в том случае, если решение о необходимости хотя бы одного (первого) столкновительного теста обусловлено логикой алгоритма при использовании прежнего (заниженного) значения $\nu_{max}$. Практические вычисления показали, что, в случае малого среднего числа частиц в ячейке и изначально сильно заниженном значении $\nu_{max}$, сходимость $\nu_{max}$ достаточно медленна, что привносит существенную ошибку. Такая проблема возникала при изучении поведения метода ПСМ при очень малом среднем числе частиц в ячейке. Автором настоящей работы предложено изменение логики алгоритма, уменьшающее этот эффект. Для этого, если в ячейке имеется хотя бы одна пара частиц, выбор тестовой пары и коррекция $\nu_{max}$ осуществляется до того, как принимается решение о необходимости одного столкновительного теста. В этом случае сходимость $\nu_{max}$ при малом среднем числе частиц в ячейке существенно ускоряется.

## 2.1 Сравнение среднего числа столкновительных партнеров в схеме мажорантной частоты и классической схеме без счетчика времени

Отличие MFS и классической схемы NTC[1] заключается в способе вычисления числа тестовых столкновений в некоторой ячейке на некотором временном шаге. В схеме MFS оно пропорционально мгновенному числу возможных пар частиц в этой ячейке: $\sim\frac{N(N-1)}{2}$, где $N$ – мгновенное число частиц в ячейке. В схеме NTC число тестовых столкновений пропорционально мгновенному числу частиц в ячейке: $\sim N\cdot\frac{\bar{N}}{2}$, где $\bar{N}$ – среднее число частиц в ячейке и не зависит от мгновенного числа частиц. В схеме MFS при $N=1$ столкновения не производятся. По логике NTC, столкновения при $N=1$ возможны, но, так как нет ни одного партнера для столкновений, их приходится переносить на следующие временные шаги. В результате, при том же самом среднем числе частиц в ячейке, в схеме NTC, по сравнению со схемой MFS, при малом мгновенном числе частиц в ячейке столкновения происходят чаще, а при большом – реже (в среднем же, частота столкновений в обеих схемах полагается $\sim\frac{\bar{N}^2}{2}$). Как следствие, количество доступных столкновительных партнеров для схемы NTC должно быть в среднем (по столкновениям) несколько меньше, что должно приводить к более высокой вероятности повторных столкновений.

В общей постановке аналитически сравнить схемы NTC и MFS трудно. Однако, сделаем попытку произвести сравнение среднего числа столкновительных партнеров в упрощенной постановке. Для этого предположим, что каждый временной шаг число частиц в каждой подъячейке статистически независимо и подчиняется распределению Пуассона. Также, для упрощения вычислений, считаем, что $\frac{\bar{N}\nu_{max}\delta t}{2}=1$.

Вычислим теперь среднее число доступных партнеров при столкновении по схеме NTC. Пусть имеется S подъячеек в ячейке, и, в среднем, в каждой подъячейке находится $\bar{N}$ частиц. При каждом временном шаге возможны 4 варианта:

1) Частиц в подъячейке нет. Вероятность: $e^{-\bar{N}}$.

2) В подъячейке находится $N\geq 2$ частиц. Вероятность: $p_N=\frac{\bar{N}^N}{N!}e^{-\bar{N}}$. Среднее число партнеров: $L_N=N-1$.

3) В подъячейке находится 1 частица, и в других подъячейках частиц нет. Вероятность: $\bar{N}e^{-\bar{N}}\cdot e^{-\bar{N}(S-1)}$.

4) В подъячейке находится 1 частица, и есть возможность найти партнера в других подъячейках. Вероятность: $p_1=\bar{N}e^{-\bar{N}}\left(1-e^{-\bar{N}(S-1)}\right)$. Среднее число частиц в непустой подъячейке, оно же – среднее число партнеров: $L_1=\frac{\bar{N}}{1-e^{-\bar{N}}}$.

[1] Речь идёт о схеме, описанной в (1). Впоследствии Бёрд также порекомендовал вычислять число тестовых столкновений пропорционально числу пар.

Сперва вычислим среднее число тестовых столкновений, накопленных в течение последовательности временных шагов, когда партнеры отсутствовали. Вероятность того, что непосредственно перед данным случайно выбранным временным шагом было совместно $k$ временных шагов с $N = 0$ (случай 1) и $m$ временных шагов с $N = 1$ и без партнеров (случай 3), составит:

$$w_{km} = \left[\left(e^{-\bar{N}} + \bar{N}e^{-\bar{N}S}\right)^{k+m}\left(1 - e^{-\bar{N}} - \bar{N}e^{-\bar{N}S}\right)\right] \cdot \left[C_{k+m}^{m}\left(\frac{\bar{N}e^{-\bar{N}S}}{e^{-\bar{N}} + \bar{N}e^{-\bar{N}S}}\right)^{m}\left(\frac{e^{-\bar{N}}}{e^{-\bar{N}} + \bar{N}e^{-\bar{N}S}}\right)^{k}\right].$$

Здесь $C_{k+m}^{m}$ – число сочетаний $m$ из $k+m$. Осредняя по всем $k$ и $m$, получаем: $\bar{M} = \sum_{k,m} m \cdot w_{km} = \frac{\bar{N}}{e^{\bar{N}S} - \bar{N}e^{\bar{N}(S-1)} - 1}$. Таким образом, если в подъячейке находится $N \geq 1$ частиц, и есть партнеры для столкновения, то ожидаемое число столкновений составит: $K_N = (\bar{M} + N)$. Среднее число столкновений, соответственно: $K = \sum_N K_N \cdot p_N = \bar{N}$. Таким образом, среднее число партнеров:

$$\bar{L}_{NTC} = \sum_N L_N \frac{K_N}{\bar{N}} = \bar{N} + \frac{e^{-\bar{N}} + \bar{N}e^{\bar{N}(S-1)} - 1}{e^{\bar{N}S} - e^{\bar{N}(S-1)} - \bar{N}}.$$

Выделим два предельных случая: ячейка, не разделенная на подъячейки ($S = 1$), и подъячейка большой ячейки ($S \to \infty$):

$$\bar{L}_{NTC-I|S=1} = \bar{N} + \frac{e^{-\bar{N}} + \bar{N} - 1}{e^{\bar{N}} - \bar{N} - 1}$$

$$\bar{L}_{NTC|S\to\infty.} = \frac{\bar{N}}{1 - e^{-\bar{N}}}$$

Оговорим отдельно случай $S = 1$ и $\delta t \to 0$. Особенность здесь в том, что в фазе перемещения положение частиц изменяется мало, поэтому число частиц в ячейке $N$ на смежных временных шагах нельзя считать статистически независимым, так как оно меняется с шагом 1. В этом случае все накопленные столкновения при $N < 2$, скорее всего, будут реализованы при $N = 2$, откуда:

$$\bar{L}_{NTC-II|S=1} = \bar{N} + e^{-\bar{N}}.$$

Вычислить среднее число партнеров при столкновении для MFS проще:

$$K_N = \frac{1}{\bar{N}}\frac{N(N-1)}{2} \cdot \frac{\bar{N}^N}{N!}e^{-\bar{N}}; \quad \bar{L}_{MFS} = \sum_{N\geq 2}(N-1)\frac{K_N}{\bar{N}} = \bar{N} + 1.$$

На **Рис. 2.1-1-А** изображены вычисленные зависимости среднего по столкновениям числа доступных партнеров от среднего числа частиц в ячейке. Можно убедиться, что в схеме MFS в каждом столкновении доступно в среднем больше партнеров, чем в схеме NTC. В пределе $\bar{N} \gg 1$ достигается разница на одного партнера. Также можно видеть, что в модификациях схемы NTC с использованием подъячеек и без, среднее число партнеров отличается незначительно.

На **Рис. 2.1-1-Б** показано отношение среднего числа партнеров для обеих схем. Без использования подъячеек, максимальное отношение достигается при $\bar{N} \approx 1.15$ и малом временном шаге, при этом в схеме MFS доступное число партнеров выше на $\approx 47$ %. Если же подъячейки используются, то максимальное отношение смещается к $\bar{N} \approx 1.8$, а разница числа партнеров незначительно снижается – до $\approx 30$ %.

На этом же рисунке нанесены точки фактических значений, полученных тестовыми DSMC расчетами. Все точки довольно хорошо ложатся на теоретические кривые.

Фактические точки получены следующим образом. Моделировался газ твёрдых сфер, покоящийся между двумя изотермическими диффузно-рассеивающими пластинами. При таких условиях газ находится в равновесии, его плотность постоянна и известна. Была реализована схема NTC, в которой средняя плотность не вычислялась, а задавалась априори. В расчетной области содержалось 1000 частиц. Число Кнудсена и шаг сетки менялись таким образом, чтобы на ячейку приходилось в среднем нужное число частиц, а ячейка была не более половины длины свободного пробега. Временной шаг брался либо порядка 1/10 от времени, за которое частица в среднем пролетает расстояние, равное шагу сетки, либо сравнимое с временем между столкновениями.

Проведенный анализ показывает, что при использовании схемы NTC следует ожидать повышенной частоты повторных столкновений и, следовательно, меньшей достоверности вычислений при одинаковом числе частиц в ячейке. При этом, использование довольно сложной логики подъячеек дает заметное преимущества только при очень малом временном шаге.

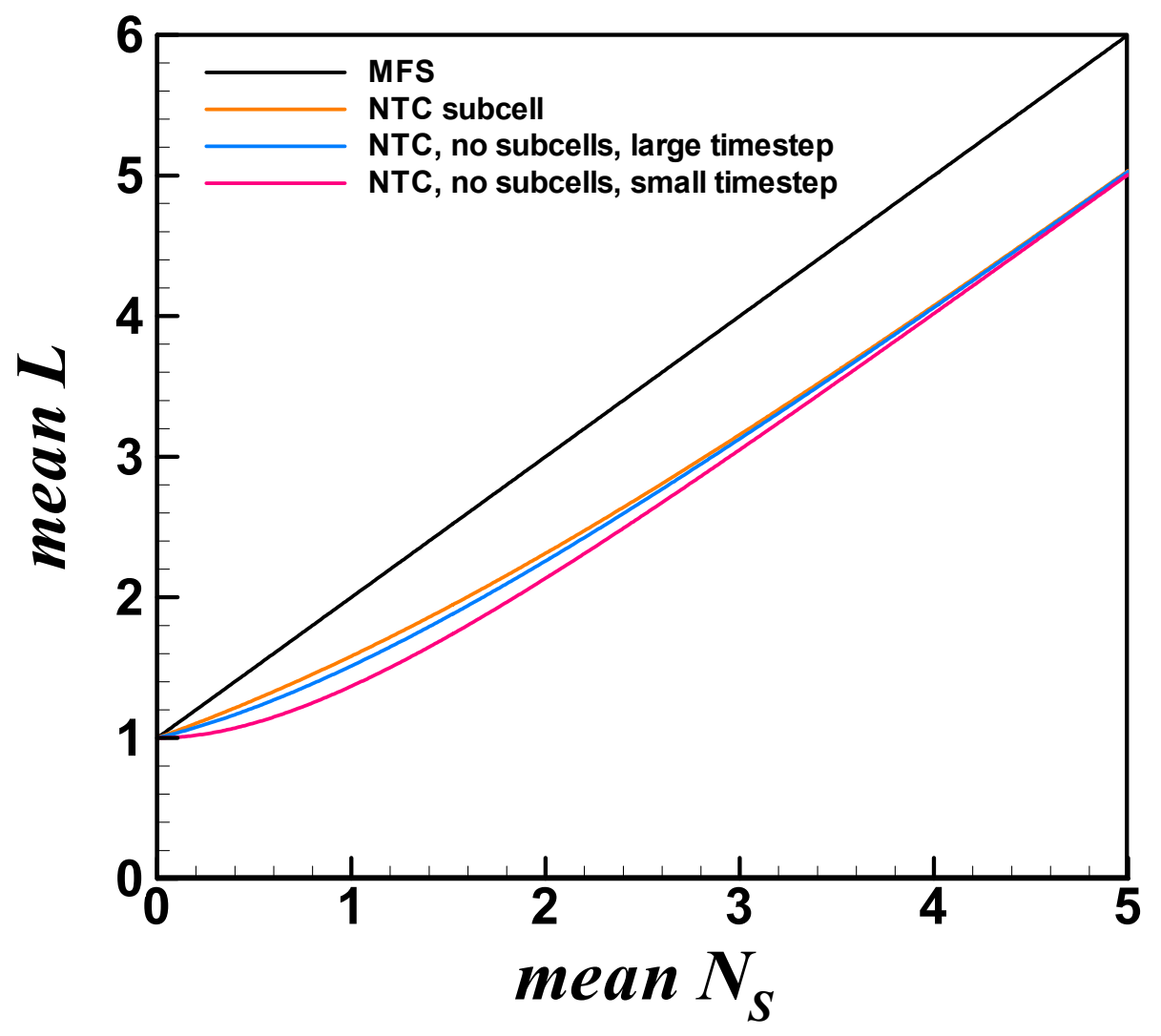

**Рис. 2.1-1-А.** Аналитически полученная зависимость среднего числа партнеров при столкновении от среднего числа частиц в ячейке для схем NTC и MFS.

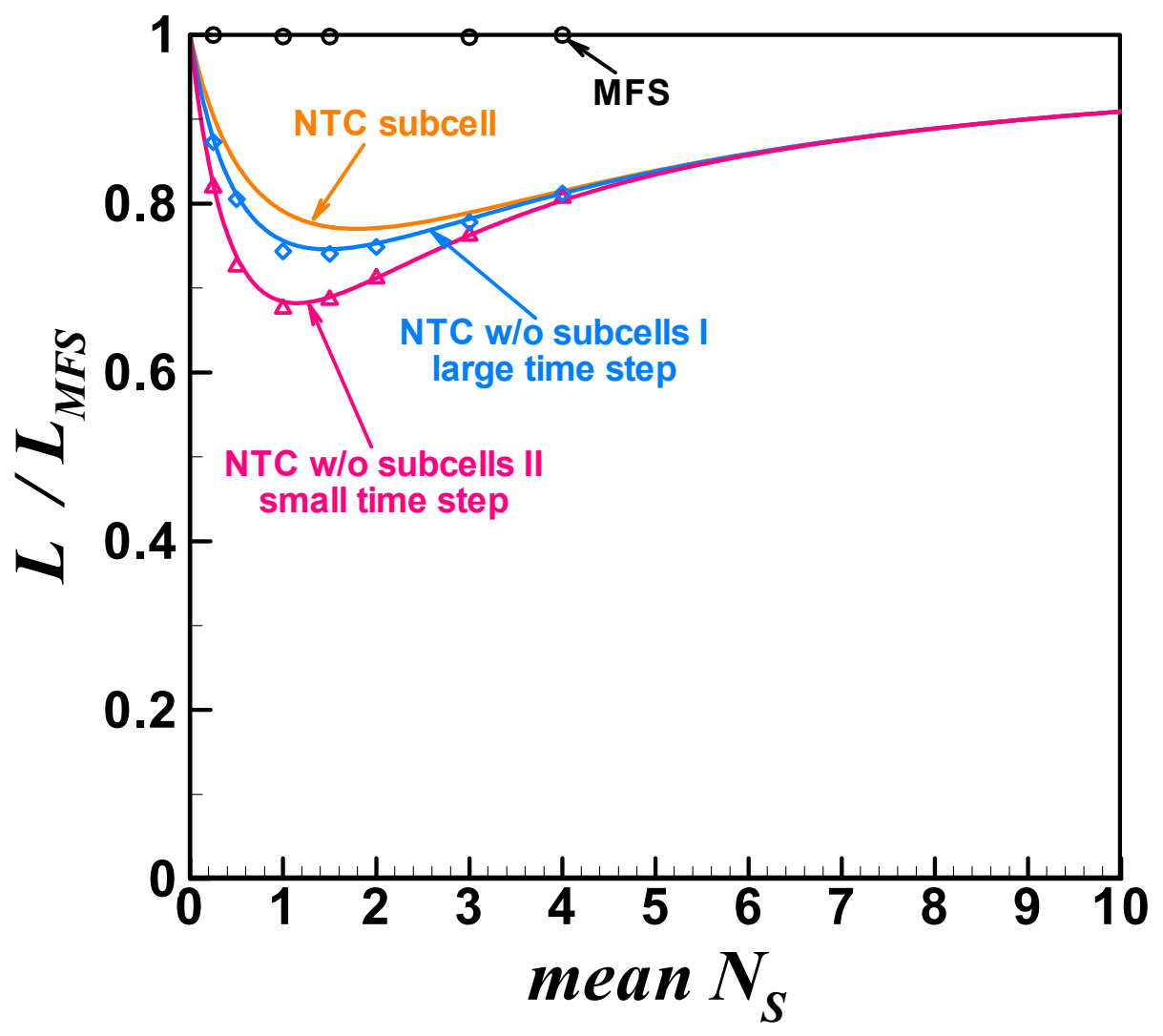

**Рис. 2.1-1-Б.** Отношение среднего числа столкновительных партнеров в схемах MFS и NTC в зависимости от среднего числа частиц в ячейке.

Если же учесть, что логика схемы MFS проще в реализации, чем логика NTC, то обоснованный выбор статистической схемы обработки столкновений однозначно падает на схему MFS.

# 3 Техническая оптимизация алгоритма

Моделирование современных задач динамики разреженных газов нередко вынуждает проводить весьма трудоёмкие расчеты. Те из задач, которые ещё целесообразно моделировать на современных персональных компьютерах, часто требуют использования порядка миллиона ячеек сетки и десятка миллионов частиц. При этом размер требуемой памяти может превышать 2 гигабайта, причём основная доля приходится на хранение параметров расчетных частиц. Фактически, каждая частица должна быть обработана в течение каждого временного шага. При таких требованиях к памяти, существенную роль в скорости исполнения алгоритма играет организация данных в памяти и порядок доступа к ним.

В приложении представлена более полная информация о принципах работы процессоров и памяти в современных компьютерах.

## 3.1 Оптимизация структур данных

Скорость доступ к памяти в современных компьютерах значительно зависит от локальности данных. При этом выгодно работать одновременно лишь с небольшим объёмом данных, расположенных в памяти достаточно плотно.

Заранее следует пояснить, что существует два подхода к расположению данных в виде массива объектов-записей в памяти:

1) Элементы одной записи хранятся в смежных ячейках памяти. Записи хранятся в массиве друг за другом. Такое расположение данных характерно для объектно-ориентированных языков вроде C/C++.

2) Каждый элемент записи хранится в отдельном массиве, массивы элементов следуют друг за другом. Такое расположение данных характерно для языка FORTRAN.

Intel в руководстве для программистов рекомендует придерживаться варианта 2), если важно быстродействие. Кроме того, реализация варианта 1) несколько осложняется требованиями к выравниванию данных. Однако, на самом деле выбор следует делать в зависимости от того, насколько совместно используются разные элементы записей.

Размер строки кэша – 64 байта. Одно число двойной точности с плавающей запятой (тип double), например, занимает 8 байт памяти. Если несколько элементов типа double используются совместно, их имеет смысл хранить рядом, так как при помещении строки данных в кэш при запросе одного числа, смежные числа также автоматически оказываются в кэше.

Так, в методе ПСМ, в фазе перемещения используются координаты и скорости частиц, но практически не используется информация, например, о внутренней энергии, в то время, как при столкновении частиц используются скорости и внутренняя энергия, но не координаты частиц. Однако, в обоих случаях 3 компоненты скорости каждой частицы используются совместно, поэтому их разумно хранить в смежных ячейках памяти.

Есть ещё несколько аргументов против универсальности варианта 2). Во-первых, при больших размерах массива, обращение к нескольким элементам одной записи будет задействовано по 1 записи TLB на каждый элемент, так как разные элементы окажутся в разных страницах виртуальной памяти. При смежном хранении элементов будет задействована лишь 1 запись TLB. Во-вторых, размер массива может случайно оказаться кратен числу индексов в кэш-памяти. В этом случае, число элементов записи, одновременно находящихся в кэш-памяти, ограничено её ассоциативностью. Впрочем, ассоциативность кэш-памяти современных процессоров достаточно велика. В-третьих, при загрузке записи из ОЗУ вследствие промаха кэш-памяти, все элементы записи с большой вероятностью окажутся в разных страницах ОЗУ.

Особенность ПСМ в том, что частицы активно перемешиваются в пространстве. Постоянно сохранять упорядоченность массива частиц по ячейкам сетки довольно дорого с вычислительной точки зрения. При отсутствии упорядоченности, информация о каждой частице одной ячейки неизбежно окажется в случайной области памяти. И наоборот, частицы, занимающие смежные области памяти, соответствуют случайным ячейкам сетки. Как следствие, число проходов алгоритма ПСМ резко отрицательно сказывается на быстродействии, и необходимо предпринимать меры к их объединению с целью сокращения доли случайных обращений к памяти.

## 3.2 Базовые подходы к оптимизации кода

Несмотря на существенные отличия в деталях архитектуры, основной принцип увеличения производительности современных процессоров так или иначе связан с разбитием программы на независимые друг от друга потоки хорошо распараллеливаемых инструкций. Не поддающиеся распараллеливанию программы – с непрерывными и неразрешимыми зависимостями в последовательности инструкций – работают в разы или даже десятки раз медленнее, так как процессор каждый раз ожидает результат предыдущей инструкции. Это вынуждает избегать строго последовательных алгоритмов, предпочитая им хорошо распараллеливающиеся алгоритмы, даже если они требуют большего числа операций. Как частный случай, приходится избегать условных переходов.

Основные подходы к оптимизации кода программы следующие:

1) Уменьшение требуемого числа арифметических операций. Подход в основном заключается в упрощении формул и устранении повторного вычисления констант. Так, оригинальный код Бёрда содержит немало арифметических операций, которых можно было бы избежать. Справедливости ради надо сказать, что код Бёрда играет больше академическую, чем прикладную роль, поэтому наглядности и понятности кода придавалось большее значение, чем его скорости исполнения.

2) Особое внимание следует уделить процедурам выборки случайных чисел из заданного распределения методом Монте-Карло – так называемый подход acception-rejection – в котором значительное влияние оказывает среднее число попыток при генерации числа.

3) Минимизация числа условных переходов. Они оказывают неблагоприятный эффект на быстродействие программы, если результат ветвления каждый раз непредсказуем.

4) Использование оптимизирующего компилятора. Например, современные процессоры обладают сложными специализированными наборами векторных (так называемых single-instruction-multiple-data) инструкций для параллельных арифметических операций, а также командами условного исполнения, позволяющими избегать условных переходов. Далеко не все компиляторы эффективно их используют. Кроме того, их автоматическому использованию компилятором способствует правильная организация кода программы, позволяющая компилятору распознать взаимную независимость распараллеливаемых операций.

## 3.3 Оптимизация сортировки частиц по ячейкам

Наименее благоприятной операцией метода ПСМ в отношении локальности доступа к памяти является сортировка частиц по ячейкам. Сортировка частиц, к тому же, является одной из наиболее затратных операций. Неблагоприятно также и то, что частицы, находящиеся в одной ячейке, располагаются в случайных областях памяти, что важно при сборе статистики и обработке столкновений. При неизбежности случайного доступа целесообразно сокращать число проходов, его использующих.

Упростим структуру алгоритма Бёрда из примеров программ на случай 1 сорта молекул и 1 подъячейки на ячейку, причём пусть столкновительные ячейки совпадают с ячейками сбора статистики. Тогда шаблон доступа к памяти будет выглядеть следующим образом:

- Для каждой молекулы хранится номер ячейки, в которой молекула находится. Кроме того, имеется массив упорядоченных номеров молекул – по одному элементу на молекулу – в котором для каждой ячейки находящиеся в ней молекулы перечислены по порядку.
- Для каждой ячейки хранятся: а) счетчик числа молекул в данной ячейке, б) стартовый индекс в массиве упорядоченных номеров, начиная с которого перечисляются молекулы именно этой ячейки. Вообще говоря, достаточно хранить только индекс, вычисляя количество молекул как разницу индексов между следующей и текущей ячейкой.

1) Обрабатывается перемещение частиц по порядку их следования в массиве частиц. В конце фазы перемещения определяется и сохраняется номер ячейки, куда переместилась частица.

2) По порядку зануляются счетчики частиц во всех ячейках.

3) Частицы перебираются по порядку следования в массиве частиц. Для каждой частицы считывается номер ячейки. Счетчик в соответствующей ячейке увеличивается на единицу. Как следствие, ячейки на этом шаге перебираются в случайном порядке.

4) Ячейки перебираются по порядку, для каждой ячейки определяется стартовый индекс путём суммирования счетчиков всех предыдущих ячеек. К сохраняемому индексу прибавляется количество частиц в, т.е., фактически, это индекс не первой частицы в упорядоченном массиве, а последней частицы + 1.

5) Частицы перебираются по порядку. Для каждой частицы считывается номер ячейки. Для этой ячейки индекс в упорядоченном массиве уменьшается на 1 и по получившемуся индексу элемента массива записывается номер частицы. Как следствие, ячейки и элементы упорядоченного массива перебираются в случайном порядке.

6) Обработка столкновений. Ячейки перебираются по порядку. Считывается индекс первой частицы в упорядоченном массиве и количество частиц в ячейке. Сталкивающиеся частицы, как и элементы упорядоченного массива перебираются в случайном порядке. Однако, число частиц в ячейке небольшое. Как следствие, почти наверняка параметры частиц будут успешно «оседать» в кэш-памяти.

7) Сбор статистики. Ячейки перебираются по порядку. Для каждой ячейки перебираются находящиеся в ней частицы и их параметры суммируются. Частицы перебираются в случайном порядке.

Итак, за одну итерацию алгоритма, производится 3 последовательных и 2 случайных перебора частиц, а также 4 последовательных и 2 случайных перебора ячеек.

*Альтернативный алгоритм.* Предлагается следующий альтернативный алгоритм, оптимизирующий шаблон доступа к памяти:

- Для каждой ячейки хранится индекс *mfirst* первой молекулы в массиве частиц, который находится в этой ячейке. При использовании нескольких сортов частиц может быть удобно использовать больше одного списка, каждый из которых имеет свою величину *mfirst.*

- Для каждой молекулы хранится индекс следующей молекулы *mnext*, находящейся в той же самой ячейке (и в том же списке). Как следствие, молекулы одного сорта из одной ячейки объединены в односвязный список. Конец списка помечается особым числом, например, минус единицей.

- Для каждого списка имеется упорядоченный массив номеров частиц – однако, общий для всех ячеек. Размер массива равен максимально возможному числу частиц одного сорта в одной ячейке.

0) Для каждой ячейки индексу первой молекулы *mfirst* присваивается значение конца списка. Этот шаг производится только перед самой первой итерацией алгоритма.

1) Обрабатывается перемещение частиц по порядку их следования в массиве частиц. В конце фазы перемещения определяется номер ячейки, куда переместилась частица. Частица помещается в начало списка этой ячейки, т.е.: а) считывается *mfirst* – старый индекс первой молекулы в ячейке, б) *mfirst* записывается в поле *mnext* данной частицы, в) индекс текущей частицы записывается в *mfirst* ячейки. Ячейки при этом перебираются в случайном порядке. Если для сбора статистики используется независимая сетка, то именно на этом шаге удобно вычислить номер ячейки второй сетки.

2) Обрабатываются за один проход как столкновения, так и сбор статистики. Ячейки перебираются по порядку. Производятся следующие действия:

   a. Список частиц только этой ячейки «разворачивается» в массив. Это производится перебором всех частиц, начиная с *mfirst*, затем последовательно считывая *mnext*, пока не встретится значение конца списка. Если статистика собирается дважды (до и после фазы столкновений), то одновременно суммируются параметры частиц. Частицы при этом перебираются в случайном порядке, однако, с большой вероятностью их параметры при этом «оседают» в кэш-памяти. Также на этом шаге подсчитывается количество частиц в ячейке.

Использование независимой сетки для сбора статистики не является препятствием для объединения выборки с разворачиванием списка – находящиеся в одной и той же столкновительной ячейке частицы с большой вероятностью будут распределены в пределах всего нескольких ячеек сбора статистики, также умещающихся в кэш-памяти.

b. Опустошение списка – величине *mfirst* присваивается значение конца списка.

c. Обрабатываются столкновения в ячейке. Развернутый массив позволяет быстро найти индексы частиц в глобальном массиве по их локальному номеру. Частицы перебираются в случайном порядке, однако, велика вероятность, что параметры частиц уже имеются в кэш-памяти.

d. Сбор статистики. Вновь перебираются частицы в ячейке и суммируются их параметры. На этот раз индексы частиц считываются из развернутого массива. По-прежнему, параметры частиц могут всё ещё храниться в кэш-памяти.

Можно подсчитать, что, в усовершенствованном алгоритме, как частицы, так и ячейки, перебираются лишь 1 раз по порядку и 1 раз в случайном порядке. Все остальное время работа с данными производится в основном в пределах кэш-памяти. На практике, при переходе на усовершенствованный алгоритм достигается практически двукратное ускорение счета. Кроме того, на многоядерных процессорах наблюдается заметно меньшее взаимное замедление потоков за счёт уменьшения частоты обращений к ОЗУ.

В **Табл. 3-1** приведены результаты теста производительности. Результаты получены на двух двухъядерных компьютерах. В одних тестах было задействовано только одно ядро процессора, в то время как второе простаивало. В других тестах, одновременно запускалось две копии программы – по одной на ядро. Во всех случаях рассчитывалось обтекание плоской пластины (4).

Как видно из таблицы, при использовании очередности MSCS (сбор статистики до и после столкновений) и работе обоих ядер, достигается ускорение около 85 %. Для очередности MCS (сбор статистики только после столкновений) и при работе одного ядра, ускорение составляет 33 %. Неожиданным оказалось то, что ускоренный алгоритм при стандартной очередности MCS работает не только несколько медленнее, чем при очередности MSC, но даже медленнее очередности MSCS – двукратный сбор статистики при ускоренном алгоритме обходится бесплатно. По-видимому, такой эффект связан с тем, что чтение параметров частиц на шаге сбора статистики более эффективно заполняет кэш-память, чем выбор случайных пар.

Далее условия расчета. Характерное число Кнудсена составляло 0.04, рассчитанное по ширине пластины $H$ (взятой также за единицу длины), плотности набегающего потока, сечению столкновения He–He при температуре торможения. Временной шаг 0.0025 (за единицу принят интервал $H \cdot \sqrt{\frac{\mu_{He}}{RT_0}}$). Состав смеси: 90 % He + 10 % Xe. Параметры набегающего потока соответствуют бесконечному числу Маха. Размеры расчетной области: 2 единицы перед пластиной, 4 единицы за пластиной, 5 единиц от зеркальной плоскости симметрии. Шаг сетки 0.005 единиц.

**Табл. 3-1.** Время выполнения одного временного шага в миллисекундах.

| Алгоритм | Очередность шагов | Intel Core2Duo E6600, 2.4 GHz | | Intel Core2Duo E8400, 3.0 GHz | |
|---|---|---|---|---|---|
| | | Одно ядро | Два ядра | Одно ядро | Два ядра |
| Ускоренный | MSC | 2686 | 3240 | 2009 | 2438 |
| Классический | | 3787 | 4931 | 2853 | 3752 |
| Ускоренный | MCS | 2872 | 3389 | 2195 | 2574 |
| Классический | | 3807 | 4955 | 2899 | 3755 |
| Ускоренный | MSCS | 2784 | 3315 | 2089 | 2474 |
| Классический | | 4735 | 6139 | 3548 | 4647 |

## 3.4 Дополнительные оптимизации

*Генератор случайных чисел* – первый очевидный кандидат на оптимизацию. Бёрд использует ГСЧ на основе целочисленной последовательности типа Фибоначчи: $x_n = (x_{n-55} - x_{n-24})\ mod\ 10^9$. Целые числа затем масштабируются в число с плавающей точкой из диапазона 0..1. Модификация состоит в отбраковке результатов (число генерируется заново), не превышающих 0.00000001, либо достигающих 0.99999999.

В настоящей работе использовался ГСЧ на основе той же самой последовательности Фибоначчи, только при этом использовался полный диапазон 32-битных целых чисел. Это, по сути, один из широко известных и весьма распространённых генераторов. Масштабирование производилось прибавлением 0.5 и последующим делением суммы на $2^{32}$.

При каждом запуске программы массив считывается из файла, и в конце снова сохраняется в файл.

В реализации, удобно выделить отдельную функцию, генерирующую сразу 55 чисел последовательности в массив, а функцию генерации свести к извлечению следующего числа из массива и вызову генерирующей функции, если массив опустошен. В этом случае, процедуру генерации удобно сделать inline (когда вместо команды вызова подпрограммы компилятор использует копию кода).

Такая реализация ГСЧ обладает неплохим быстродействием и легко адаптируется под другие последовательности.

*Реализация модели Боргнакке-Ларсена*, предложенная Бёрдом, далека от оптимальной, прежде всего для молекул с большим количеством степеней свободы. Так, для выборки из бета-распределения при $\xi_{tr} = 3$ и $\xi_{int} = 40$ алгоритм производит в среднем 10 попыток, а при $\xi_{int} < 2$ вовсе неприменим. Эту трудность можно обойти сведением задачи к использованию гамма-распределения. Имея две независимые случайные величины $y_1 \sim \Gamma\left(\frac{\xi_{tr}}{2}, 1\right)$ и $y_2 \sim \Gamma\left(\frac{\xi_{int}}{2}, 1\right)$, можно получить $x = \frac{y_1}{y_1+y_2} \sim \mathrm{B}(\frac{\xi_{tr}}{2}, \frac{\xi_{int}}{2})$ (20).

Ещё одним преимуществом использования гамма-распределения является простое обобщение для случая перераспределения энергии при наличии двух и более внутренних степеней свободы.

Алгоритм выборки из гамма-распределения в реализации Бёрда при $\xi_{int} < 2$ также не применим вовсе, а при $\xi_{int} > 7$ дает ошибку более 1%, которая быстро нарастает с ростом $\xi_{int}$. Причина в ограничении величины энергии значением $10kT$.

Был разработан собственный алгоритм выборки из гамма-распределения, основанный на принципе принятия-отклонения. Охватывающая функция распределения строится из двух частей: левая часть представляет собой степенную функцию, правая – затухающую экспоненту. Коэффициенты двух кривых и точка раздела между ними вычисляются по эмпирическим формулам.

Алгоритм выборки из гамма-распределения включает в себя около десятка констант, рассчитываемых для заданного числа степеней свободы. Так как выборка производится в основном при одних и тех же постоянных количествах степеней свободы, оправдано разделить алгоритм на две части. Первая часть заполняет структуру с набором констант. Вторая часть получает эту структуру в качестве аргумента.

Реализация алгоритма генерирует около 8 млн. чисел в секунду на одном ядре процессора Core2Duo E6600 2.4 GHz. Алгоритм был также протестирован численно. Код алгоритма и методика тестирования описаны в приложении.

# 4 Исследование ошибок дискретизации в методе ПСМ, основанном на схеме мажорантной частоты

## 4.1 Задача Фурье. Измерение теплового потока на стенку

Все описанные в этом параграфе численные исследования относятся к моделированию задачи Фурье: теплопередача между двумя бесконечными параллельными неподвижными пластинами разной температуры. При этом использовалась простейшая модель газа – твёрдые сферы. Отражение от пластин задавалось как диффузное, с полной аккомодацией. Отношение температур стенок во всех расчетах составляло 4 раза: $T_2/T_1 = 4$. Расчеты проводились для двух чисел Кнудсена $\mathrm{Kn_o}$ (определяемых по средней плотности и расстоянию между пластинами): 0.1 и 0.025. Значение $\mathrm{Kn_o} = 0.1$ задаёт весьма разреженный поток, позволяет использовать небольшое число расчетных частиц, а также позволяет сравниться с результатами (21). Значение $\mathrm{Kn_o} = 0.025$ описывает уже довольно плотный газ, в котором неплохо работает уравнение Фурье, если учитывать тепловой скачок на стенках.

Наблюдаемыми величинами являлись, прежде всего, тепловые потоки на холодную и горячую пластины. Именно посредством изучения отклонения фактического результата от предельного целевого значения и определялись вклады дискретности по пространству, времени, числу частиц в погрешность теплового потока.

Если говорить о тепловом потоке в газе, то его можно условно разделить на 2 слагаемых. Одно из слагаемых (кинетический член) соответствует физическому переносу тепла при тепловом движении молекул. Второе слагаемое (столкновительный член) описывает перенос тепла посредством столкновений молекул. Физически, в разреженном газе столкновения полагаются точечными, и второе слагаемое обращается в ноль. Однако, в методе ПСМ сталкивающиеся молекулы находятся на расстоянии друг от друга, и столкновительный член также даёт свой вклад. Его можно считать артефактом метода. Значение кинетического члена, даже при заметном вкладе нефизичного столкновительного переноса тепла, может быть довольно близко к физическому тепловому потоку (9). Это, впрочем, не даёт гарантии того, что столкновительный перенос тепла не исказит поле макропараметров.

Другой способ – определять поток энергии через некоторую поверхность, суммируя энергию частиц, пересекающих её (со знаком плюс или минус, в зависимости о того, с какой стороны частица пересекает поверхность). Если газ покоится относительно этой поверхности, то поток энергии совпадает с нормальной компонентой теплового потока. В задаче Фурье, газ как раз покоится, что позволяет измерять таким способом тепловой поток через неподвижную поверхность. Удобнее всего расположить поверхность на границе между ячейками – перенос тепла столкновениями через такую поверхность отсутствует. Измерение энергетического баланса на стенках является частным случаем этого подхода.

*Определение теплового потока и его дисперсии.* Для упрощения выкладок, будем говорить о балансе энергии $\mathcal{E}$ и его дисперсии. Тепловой поток: $q = \mathcal{E}/\tau$. Здесь $\tau$ – временной промежуток наблюдения. Поделим $\tau$ на $K$ равных интервалов длительностью $\delta\tau = \tau/K$, достаточно большое, чтобы за время $\delta\tau$ со стенкой столкнулось либо 0, либо 1 частица, а вероятность прибытия хотя бы 2 частиц была пренебрежимо малой. Баланс энергии на временном интервале $k$ составляет $\delta\mathcal{E}_k$, а полный баланс $\mathcal{E} = \sum_{k=1}^{K} \delta\mathcal{E}_k$. Теперь посчитаем средние значения $\delta\mathcal{E}_k$ и $(\delta\mathcal{E}_k)^2$ по этим интервалам. На каждом временном интервале $k$, либо $\delta\mathcal{E}_k = (\delta\mathcal{E}_k)^2 = 0$, если не прибыло ни одной частицы, либо $\delta\mathcal{E}_k = \delta E_i$ и $(\delta\mathcal{E}_k)^2 = (\delta E_i)^2$, если прибыла частица $i$, внесшая в энергетический баланс вклад $\delta E_i$. Таким образом, $\langle\delta\mathcal{E}\rangle = \frac{\sum_{i=1}^{N}\delta E_i}{K}$, $\langle(\delta\mathcal{E})^2\rangle = \frac{\sum_{i=1}^{N}(\delta E_i)^2}{K}$, где $N$ – полное количество частиц. Соответственно, математическое ожидание и дисперсию можно оценить следующим образом: $\mathcal{M}[\delta\mathcal{E}] \approx \langle\delta\mathcal{E}\rangle = \frac{\sum_{i=1}^{N}\delta E_i}{K}$, $\mathfrak{D}[\delta\mathcal{E}] \approx \langle(\delta\mathcal{E})^2\rangle - \langle\delta\mathcal{E}\rangle^2 = \frac{\sum_{i=1}^{N}(\delta E_i)^2}{K} - \frac{\left(\sum_{i=1}^{N}\delta E_i\right)^2}{K^2}$. Тогда $\mathcal{M}[\mathcal{E}] = K\cdot\mathcal{M}[\delta\mathcal{E}] \approx \sum_{i=1}^{N}\delta E_i$, $\mathfrak{D}[\mathcal{E}] = K\cdot\widetilde{D}\cdot\mathfrak{D}[\delta\mathcal{E}] \approx \widetilde{D}\cdot\left[\sum_{i=1}^{N}(\delta E_i)^2 - \frac{\left(\sum_{i=1}^{N}\delta E_i\right)^2}{K}\right]$. В пределе, при $K\to\infty$, $\mathfrak{D}[\mathcal{E}] \approx \widetilde{D}\cdot\sum_{i=1}^{N}(\delta E_i)^2$. Здесь $\widetilde{D}$ – поправочный коэффициент дисперсии. В простейшем случае, когда все $\delta\mathcal{E}_k$ статистически независимы, $\widetilde{D} = 1$.

Все приведенные в настоящем параграфе численные значения среднеквадратических отклонений и доверительных интервалов посчитаны именно исходя из предположения $\widetilde{D} = 1$. Таким образом, для определения теплового потока и его среднеквадратического отклонения использовались формулы: $q = \frac{\sum_i \delta E_i}{\tau}$, $\sigma[q] = \frac{\sqrt{\sum_i(\delta E_i)^2}}{\tau}$. Все приведенные далее доверительные интервалы составляют $3\sigma$.

В конце раздела, тем не менее, представлены некоторые оценки коэффициента $\widetilde{D}$. Читатель может использовать их для самостоятельного уточнения приведенных доверительных интервалов.

Для того, чтобы несколько увеличить точность теплового потока, были использованы обе определяемые величины теплового потока – для холодной и горячей стенки. Параметры частиц, столкнувшихся с левой и правой пластиной, также полагались статистически независимыми. Это позволяет использовать коэффициенты взвешенного среднего для объединения двух независимых выборок, и определить дисперсию объединенной выборки, используя стандартные формулы:

$$q = \frac{\frac{q_1}{\sigma^2[q_1]} - \frac{q_2}{\sigma^2[q_2]}}{\frac{1}{\sigma^2[q_1]} + \frac{1}{\sigma^2[q_2]}},\ \sigma^2[q] = \frac{1}{\frac{1}{\sigma^2[q_1]} + \frac{1}{\sigma^2[q_2]}}.$$

Тепловые потоки $q_1$ и $q_2$ имеют противоположный знак, поэтому в формуле и присутствует вычитание.

## 4.2 Число частиц в ячейке. Критерий $\overline{N_C}\cdot Kn_C$

В литературе можно встретить разные рекомендации, касающиеся необходимого количества расчетных частиц. Обычно рекомендуют от 3 до 30 частиц в среднем в каждой ячейке. Упомянем также вскользь тот факт, что, технически, строгое выполнение этого требования является непростой задачей. Кроме того, эти рекомендации были отработаны на схеме NTC, в то время как в настоящей работе используется более перспективная схема MFS. При решении современных задач динамики разреженного газа, наивный подход к выбору числа частиц не годится – требуется предсказуемость ошибки. Общепринятый эвристический подход – сравнение результатов нескольких расчетов, полученных с заметно отличающимся числом частиц – не всегда удовлетворителен, так как кратно увеличивает стоимость вычислений. Хорошим решением этого вопроса была бы некоторая асимптотическая оценка величины ошибки от числа частиц.

Так как схема NTC требует оценку среднего числа частиц в ячейке $\overline{N_C}$, то зависимость статистического разброса этой оценки, действительно, зависит от $\overline{N_C}$. В результате, разброс оценки может являться дополнительным источником ошибок. Возможно, именно этим объясняется природа члена $\sim\frac{1}{\overline{N_C}}$ в полученной численно зависимости ошибки в коэффициенте теплопроводности (9). Ещё одним объяснением может быть меньшее количество среднего числа доступных партнеров для столкновения в схеме NTC по сравнению со схемой MFS. Исследование схемы MFS не выявляет подобного члена, о чём будет сказано ниже.

Следующий член зависимости (9), включающий число частиц, имеет порядок $\sim\frac{1}{\overline{N_C}\mathrm{Kn}_C}$, где $\mathrm{Kn}_C = \frac{h}{\lambda}$ есть количество ячеек (линейным размером $h$), умещающихся на длине свободного пробега $\lambda$. Следует напомнить, что речь идёт о численном результате, полученным исключительно на одномерных сетках, и при использовании схемы NTC.

В работе (10) предложено пользоваться критерием $N_\lambda > 1$: число частиц в объёме с линейным размером порядка $\lambda$ должно быть большим.

Сопоставим эти два критерия. Можно легко подсчитать, что объём с линейным размером $\lambda$ содержит $\sim(Kn_C)^d$ ячеек, где $d$ – пространственная размерность расчетной области. Отсюда следует, что $N_\lambda \sim N_C (Kn_C)^d$. Если $d = 1$, то $N_\lambda = N_C Kn_C$.

Итак, критерий $\overline{N_C}\mathrm{Kn}_C$ и критерий $N_\lambda$ по-разному обобщаются на пространства большей размерности, но совпадают в одномерном случае.

### 4.2.1 Аналитическая оценка вероятности повторных столкновений.

Попробуем оценить вид зависимости вероятности повторных столкновений из физических соображений. Отталкиваться будем от неравенства $\nu_{ij}\Delta t_P \ll 1$ – требование пренебрежимо малой доли повторных столкновений. Здесь $\Delta t_P$ – продолжительность времени, которое пара частиц находится в одной ячейке. $\nu_{ij} = \frac{\sigma_{ij}(\mathrm{c}_{ij})\cdot \mathrm{c}_{ij}}{V/F}$ – частота (вероятность в единицу времени) столкновения пары. Здесь $\sigma_{ij}$ - сечение столкновения (зависит от относительной скорости), $\mathrm{c}_{ij}$ – относительная скорость двух частиц, $V$ – объём ячейки, $F$ – число физических молекул, представляемое 1 расчетной частицей. Простыми словами, неравенство $\nu_{ij}\Delta t_P \ll 1$ означает, что в течение времени, за которое содержимое ячейки обновляется за счет хаотического теплового движения, лишь небольшая часть возможных пар должна столкнуться. Фактически, это является дискретным аналогом одного из допущений уравнения Больцмана: $n \cdot d^3 \ll 1$.

Итак, пусть пара молекул (i, j) только что столкнулась. Вероятность в единицу времени того, что одна из двух молекул успеет претерпеть ещё одно столкновение за время $\Delta t_P$, составляет:

$$\nu_{i\vee j} = \sum_{\substack{k\neq i,\\ k\neq j}} \nu_{ik} + \sum_{\substack{k\neq i,\\ k\neq j}} \nu_{kj} + \nu_{ij} \sim (2N-3)\bar{\nu}.$$

Здесь $\bar{\nu}$ – средняя частота столкновения пары. Для максвелловских молекул частота для всех пар одинакова $\nu_{ij} = \bar{\nu}$, поэтому $\nu_{i\vee j} = (2N-3)\bar{\nu}$. Если такое столкновение произойдёт, то вероятность, что оно окажется повторным, будет составлять:

$$p_{rpt-cond} = \frac{\nu_{ij}}{\nu_{i\vee j}} \sim \frac{1}{2N-3}.$$

Однако, молекулы пары могут удалиться друг от друга раньше, чем какая-либо из них испытает столкновение. Вероятность, что повторное столкновение всё же состоится, составит:

$$p_{rpt} = \frac{\nu_{ij}}{\nu_{i\vee j}} \cdot \left[1 - \exp\left(-\nu_{i\vee j}\Delta t_P\right)\right] \leq \nu_{ij}\Delta t_P.$$

Как видно, величина $\nu_{ij}\Delta t_P$ является вполне работоспособной верхней оценкой вероятности повторных столкновений. При этом число частиц в ячейке не оказывает какого-либо заметного влияния на вероятность повторных столкновений, если $\nu_{i\vee j}\Delta t_P \ll 1$.

Оценим $\Delta t_P \sim h/\overline{c_r}$, т.е. как отношение размера ячейки и средней относительной скорости. Теперь можно заключить, что $\nu_{i\vee j}\Delta t_P \ll 1$ выполняется, если $h \ll \lambda$. Т.е., если размер ячейки мал по сравнению с длиной свободного пробега, то вероятность повторных столкновений практически не зависит от числа частиц в ячейке.

Запишем оценку для частоты столкновения пары: $\nu_{ij} \sim \bar{\nu} \sim \frac{\bar{\sigma}\cdot\overline{\mathrm{c}_r}}{V/F}$.

Теперь неравенство $\nu_{ij}\Delta t_P \ll 1$ можно переписать в виде:

$$\nu_{ij}\Delta t_P = \frac{\sigma_{ij}(c_{ij})\cdot c_{ij}}{\bar{\sigma}\cdot\overline{c_r}}\cdot\frac{\bar{\sigma}\cdot\overline{c_r}}{V/F}\cdot\frac{h}{\overline{c_r}}\cdot\frac{\overline{c_r}\cdot\Delta t_P}{h} = \frac{\bar{\sigma}\cdot h}{V/F}\cdot\frac{\sigma_{ij}(c_{ij})\cdot c_{ij}}{\bar{\sigma}\cdot\overline{c_r}}\cdot\frac{\overline{c_r}\cdot\Delta t_P}{h} \sim \frac{\bar{\sigma}\cdot h}{V/F} \ll 1.$$

Для того, чтобы привести неравенство к несколько более удобным величинам, таким как средняя длина свободного пробега и среднее число частиц в ячейке, сделаем следующие преобразования:

$$\frac{\bar{\sigma}\cdot h}{V/F} = \frac{\sqrt{2}n\cdot\bar{\sigma}\cdot h}{\sqrt{2}n\cdot V/F} = \frac{h/\lambda}{\sqrt{2}\cdot nV/F} = \frac{1}{\sqrt{2}}\cdot\frac{h/\lambda}{\overline{N_C}} = \frac{1}{\sqrt{2}}\cdot\frac{1}{\overline{N_C}\mathrm{Kn}_C} \ll 1.$$

Таким образом, аналитическая оценка говорит в пользу критерия $\overline{N_C}\cdot\mathrm{Kn}_C = \frac{V}{\sqrt{2}\cdot F\bar{\sigma}\cdot h} \gg 1$: произведение среднего числа частиц в ячейке и числа Кнудсена ячейки должно быть большим.

### 4.2.2 Свойства полученного критерия.

Легко показать, что критерий $\overline{N_C}\cdot\mathrm{Kn}_C$ инвариантен относительно локальной плотности. Из этого следует, что использование популярного подхода с адаптивной сеткой, размер ячеек которой в разных областях расчетной области различен и зависит от локальной плотности, не всегда оправдано, и широкий класс задач может быть решен при использовании равномерной сетки со столкновительными ячейками постоянного объема, даже если плотность течения сильно меняется.

В одномерном случае критерий $\overline{N_C}\cdot\mathrm{Kn}_C$ инвариантен также относительно размера ячейки. В результате, среднее число частиц в ячейке не имеет значения, важно лишь число частиц на длине свободного пробега, т.е., фактически, общее число частиц. Как уже было сказано, в одномерном случае критерий $\overline{N_C}\cdot\mathrm{Kn}_C$ совпадает с критерием $\overline{N_\lambda}$.

Следует сделать также важное замечание о зависимости среднего сечения столкновения от локальной температуры. В отличие от локальной плотности, сечение столкновения влияет на значение критерия $\overline{N_c}\cdot\mathrm{Kn}_c$. Пусть, например, решается задача, в которой газ, представляющий собой VSS-аналог максвелловских молекул ($\omega = 1$, т.е. частота столкновений не зависит от относительной скорости молекул), разгоняется от состояния практически покоя в камере торможения до числа Маха $\mathrm{M}\sim 7$. Тогда, при сохранении энтальпии газа, следует ожидать уменьшение локальной температуры в 16 раз. Среднее сечение столкновения при этом должно увеличиться в 4 раза, соответственно, значение критерия $\overline{N_C}\cdot\mathrm{Kn}_C$ снижается в 4 раза. Плотность же при этом упадет в 70 раз. Классический подход, базирующийся на среднем числе частиц в ячейке, диктует необходимость увеличения размера столкновительной ячейки в 70 раз в одномерном, в 8.5 раз в двумерном плоском и в 4 раза в трехмерном случаях. Применение же критерия $\overline{N_C}\cdot\mathrm{Kn}_C$ дает более мягкие рекомендации: в трехмерном случае достаточно увеличить шаг сетки в 2 раза, в двумерном – в 4 раза. Такое смягчение условий предоставляет дополнительное «пространство для манёвра» и ослабляет требования к алгоритмам построения адаптивных сеток.

В одномерном же случае, увеличение шага сетки ничего не даст, необходимо увеличивать число модельных частиц.

Аналогично сечению, на значение критерия влияет параметр $F_N$ – весовой множитель модельных частиц, что важно при вычислениях с переменным весом, например, при использовании радиальных весовых множителей. Особое «коварство» весовых множителей заключается в том, что, при столкновении частиц разных весов, наличие повторных столкновений может приводить к развитию ярко выраженных нефизических счетных эффектов. Их происхождение связано с тем, что при столкновении параметры одной из модельных частиц меняются не всегда, а лишь с некоторой вероятностью. Как следствие, нарушаются статистика обмена частиц импульсом и энергией. В частности, возможно двойное столкновение с некоторой частицей, которая при этом будет иметь одни и те же начальные параметры и передаст свой импульс дважды, что можно трактовать, как изменение её эффективной массы. Всё это приводит к искажению вида функции распределения скоростей, наиболее заметным следствием чего является разница температур компонент с различными массами – завышение температуры тяжелой компоненты и занижение легкой, даже в равновесии.

### 4.2.3 Численное испытание критерия.

Произведено тестирование работоспособности критерия на примере задачи Фурье о теплопередаче между двух пластин. Были взяты параметры задачи (21): $T_2/T_1 = 4$, $\mathrm{Kn}_0 = 0.1$. Тестирование проводилось как на одномерной, так и на двумерной сетке. В последнем случае использовалась квадратная расчетная область, ограниченная двумя зеркальными границами, перпендикулярными параллельным между собой диффузно рассеивающим стенкам. Выбор числа Кнудсена обусловлен возможностью произвести двумерный расчет с 0.1 частицами в ячейке и разумной точностью, что подтвердило бы отсутствие необходимости наличия в ячейках большого числа частиц.

**Табл. 4-1.** Ошибка теплового потока при расчете на разных сетках и с различным числом частиц. Красным цветом выделены результаты расчетов по модели VS.

| Ошибка значения теплового потока, · 100 % | Критерий $\overline{N_c} \cdot \mathrm{Kn}_c$ | $\overline{N_\lambda}$ | $\overline{N_c}$ | $\mathrm{Kn}_c$ | Частиц всего | Ячеек сетки |
|---|---|---|---|---|---|---|
| эталон | 10000 | 10000 | 100 | 100 | 100000 | 1000 x 1 |
| *эталон VS: 0.17 ± 0.18* | *10000* | *10000* | *100* | *100* | *100000* | *1000 x 1* |
| 0.41 ± 0.25 | 100 | 1000 | 10 | 10 | 100000 | 100 x 100 |
| 0.58 ± 0.29 | 100 | 100 | 4 | 25 | 1000 | 250 x 1 |
| *2.94 ± 0.48* | *13.3 / 10*[2] | *133 / 100*[2] | *1* | *13.3 / 10*[2] | *10000* | *100 x 100* |
| 3.16 ± 0.62 | 10 | 1000 | 0.1 | 100 | 100000 | 1000 x 1000 |
| 3.17 ± 0.35 | 10 | 100 | 1 | 10 | 10000 | 100 x 100 |
| *5.09 ± 0.58* | *13.3 / 10*[2] | *13.3 / 10*[2] | *5* | *2.66 / 2*[2] | *100* | *20 x 1* |
| 5.36 ± 0.47 | 10 | 10 | 5 | 2 | 100 | 20 x 1 |
| 5.57 ± 1.05 | 10 | 10 | 0.1 | 100 | 100 | 1000 x 1 |
| *11.62 ± 0.80* | *5.33 / 4*[2] | *5.33 / 4*[2] | *2* | *2.66 / 2*[2] | *40* | *20 x 1* |
| 12.13 ± 0.87 | 4 | 4 | 2 | 2 | 40 | 20 x 1 |
| 27.71 ± 3.35 | 1 | 100 | 0.01 | 100 | 10000 | 1000 x 1000 |
| 28.32 ± 0.71 | 1 | 10 | 0.1 | 10 | 1000 | 100 x 100 |
| 40.56 ± 1.40 | 1 | 1 | 0.5 | 2 | 10 | 20 x 1 |

Результаты тестирования систематизированы в **Табл. 4-1**. Дополнительно, в таблице приведены результаты вычисления с использованием модели VS (отмечены красным цветом). Ошибка вычисляется как относительное отклонение от эталона: $\varepsilon = \left|\frac{q}{q_{\text{этал.}}} - 1\right|$. В качестве доверительного интервала используется разброс $3\sigma$. Эталонный расчет дал результат теплового потока, превышающий на 0.3 % данные (21). Однако, следует отметить, что результаты (21) получены на более слабой вычислительной технике.

| Величина | Полученный результат | Результат (21) |
|---|---|---|
| Отношение теплового потока к значению при свободно-молекулярном режиме | 0.295886 ± 0.22 % | 0.295 |
| Отношение давления на стенки к значению при свободно-молекулярном режиме | 1.12311 ± 0.080 % | 1.124 |

[2] Первое число – истинное, соответствующее модели VS. Второе число показывает значение величины, вычисленное для первоначальной модели твердых сфер, замененной VS-аналогом.

Обнаружена однозначная корреляция критерия $\overline{N_C} \cdot \mathrm{Kn}_C$ с точностью величины теплового потока, в отличие от общепринятых критериев, таких как среднее число частиц в ячейке. Так, удалось достигнуть хорошей точности теплового потока при среднем числе частиц в ячейке 0.1. Значение ошибки, как и ожидалось, оказалось практически обратно пропорциональным значению критерия.

Следует пояснить, что вычисления по модели VS формально относятся к числу Кнудсена 0.133, так как полное сечение столкновения для VS-аналога модели твердых сфер на 25 % ниже сечения столкновения твердых сфер. Однако, значение ошибки при этом уменьшается лишь на 4 — 8 %, а не на 25 %, как можно было бы предполагать из значения критерия. Это свидетельствует о том, что при одном и том же значении критерия $\overline{N_C} \cdot \mathrm{Kn}_C$ (и, соответственно, при схожей вероятности повторных столкновений), величина ошибки теплового потока зависит в том числе и от модели столкновений. Результат же «эталонного» расчета VS отличается от результата расчета HS на 0.17 %, что укладывается в $3\sigma$. Следует отметить, что, даже с учётом второго приближения, коэффициенты вязкости и теплопроводности модели HS и её VS-аналога отличаются менее чем на 0.003 %.

Вычисленные точки Табл. **4-1** (относящиеся к модели твердых сфер), в зависимости от $\overline{N_C} \cdot \mathrm{Kn}_C$, изображены на Рис. 4.2-1, отдельно для одномерного и двумерного случая. Там же показан результат регрессии (многочлен второй степени от величины, обратной $\overline{N_C} \cdot \mathrm{Kn}_C$ ). Можно видеть, что при использовании двумерной сетки ошибка оказалась примерно на 40 % меньше, чем при одномерной. Очевидно, это связано с тем, что, с ростом размерности, пропорционально растет отношение площади границы ячейки к ее объему, и паре частиц становится проще покинуть ячейку. Так как вектор относительной скорости всегда трехмерный, то величина $\frac{c_r \Delta t_P}{h}$ будет иметь повышенные значения для тех пар, вектор относительной скорости которых почти перпендикулярен одно- или двухмерному пространству расчетной области.

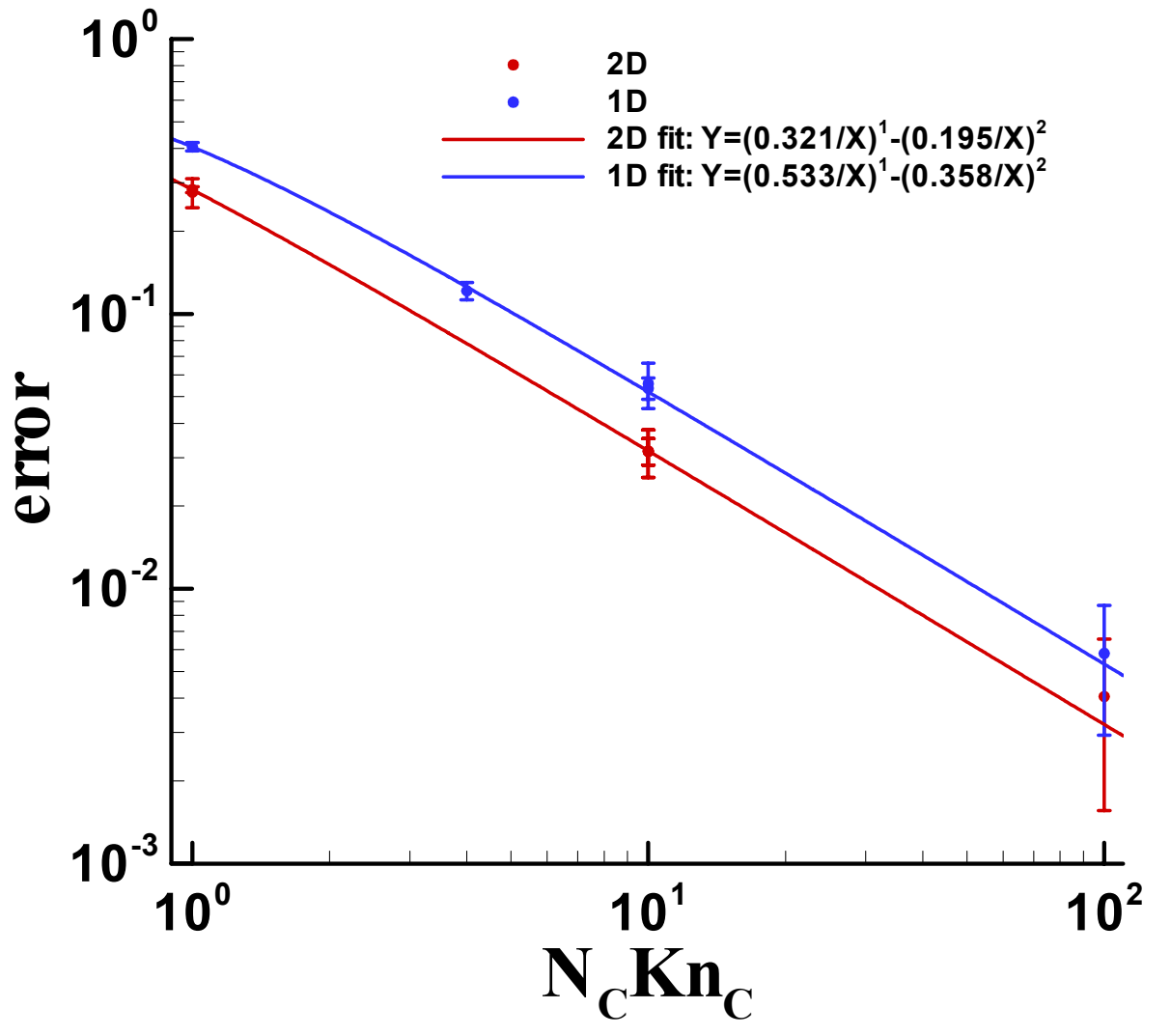

**Рис. 4.2-1.** Зависимость ошибки теплового потока от критерия $\overline{N_C} \cdot \mathrm{Kn}_C$ в одномерном и двумерном случае. Фактические точки и регрессия.

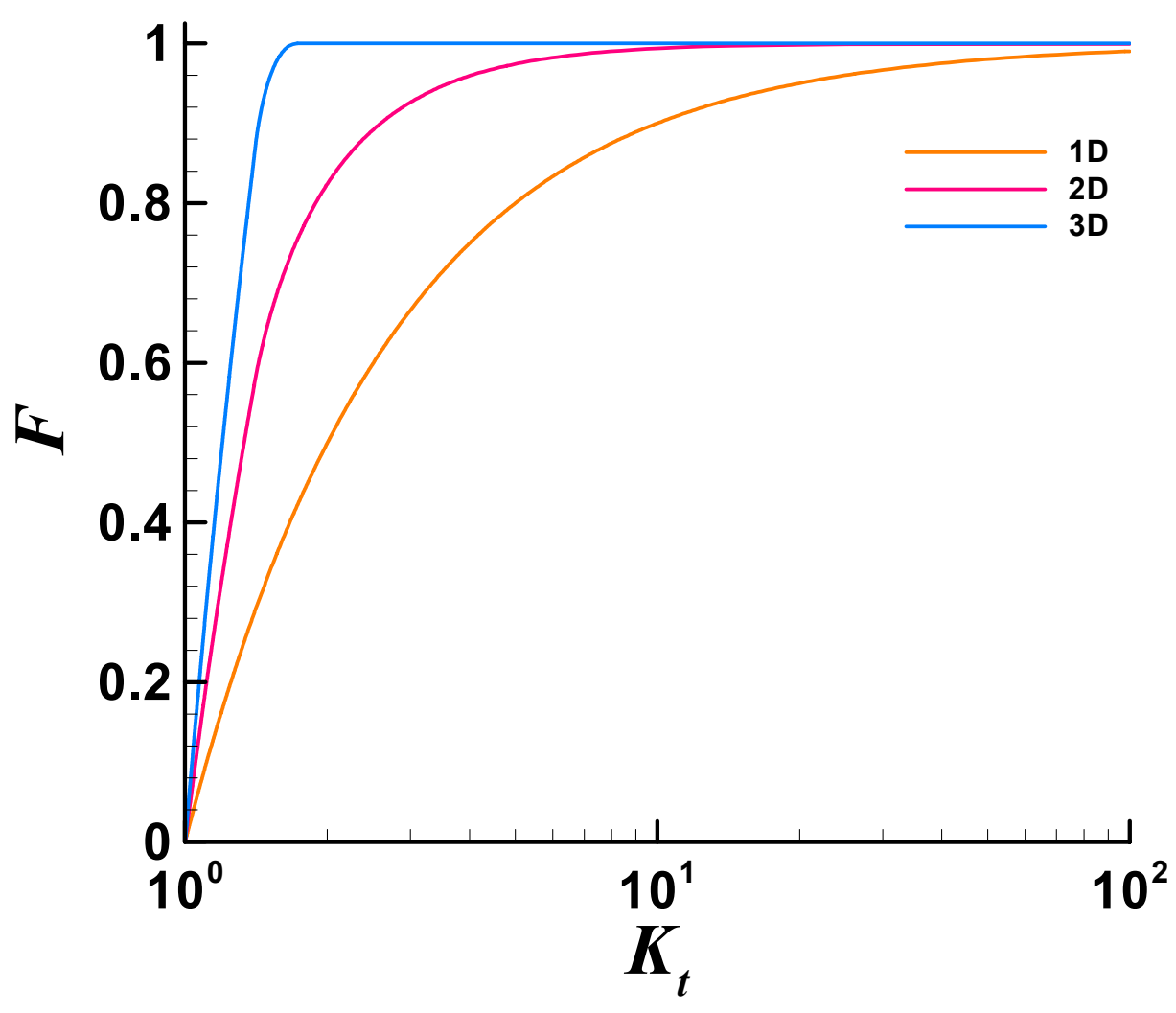

**Рис. 4.2-2.** Аналитически полученные функции распределения (интегралы плотности вероятности) величины $K_t = \min_i \frac{c_r}{\vec{c_r} \cdot \vec{e_i}}$, для разных размерностей.

Для иллюстрации рассмотрим одномерный случай. Пусть две частицы, двигаясь навстречу друг другу, попали в одну ячейку и столкнулись. Других частиц в ячейке при этом нет. Может оказаться, что, после столкновения, эти две частицы приобрели скорости, практически параллельные границам между ячейками, из-за чего могут находиться в одной ячейке довольно долго. Всё это время две частицы склонны к совершению повторного столкновения. Если к этому времени другие частицы в ячейку так и не прибудут, то повторное столкновение становится неизбежным.

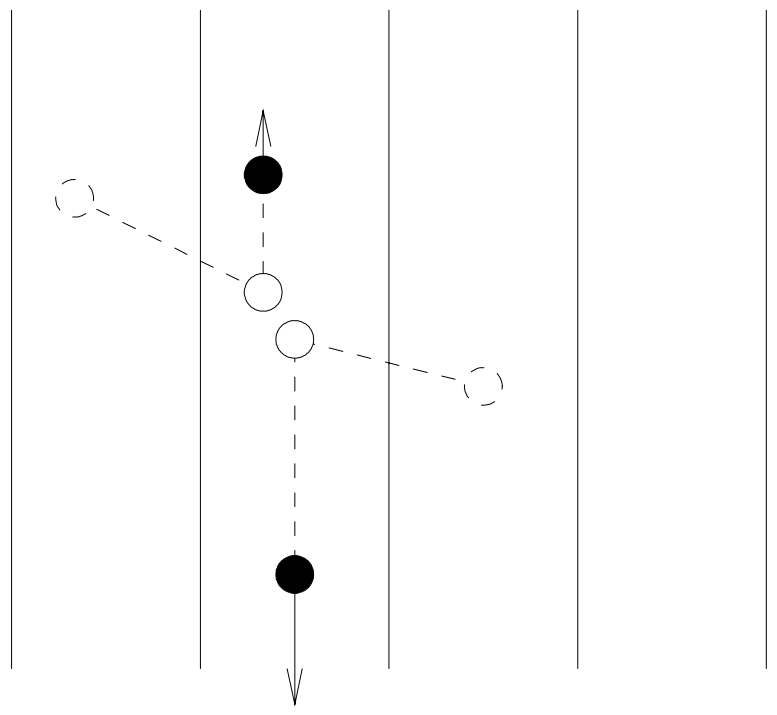

Предполагая распределение направления относительной скорости изотропным, интересно изучить функцию распределения (Рис. 4.2-2) величины $K_t = \min_i \frac{c_r}{\vec{c_r}\cdot\vec{e_i}}$ – минимального по всем пространственным измерениям $i$ относительного времени удаления пары на расстояние $h$, полагая поведение величины $K_t$ качественно похожим на поведение $\frac{c_r \Delta t_P}{h}$. Так, в одномерном случае, $K_t > 2$ с вероятностью 50 %, и $K_t > 10$ с вероятностью 10 %. В двумерном, соответственно, 17.6 % и 0.6 %. В трехмерном случае, очевидно, $K_t \leq \sqrt{3} \approx 1.73$.

## 4.3 Размер ячейки

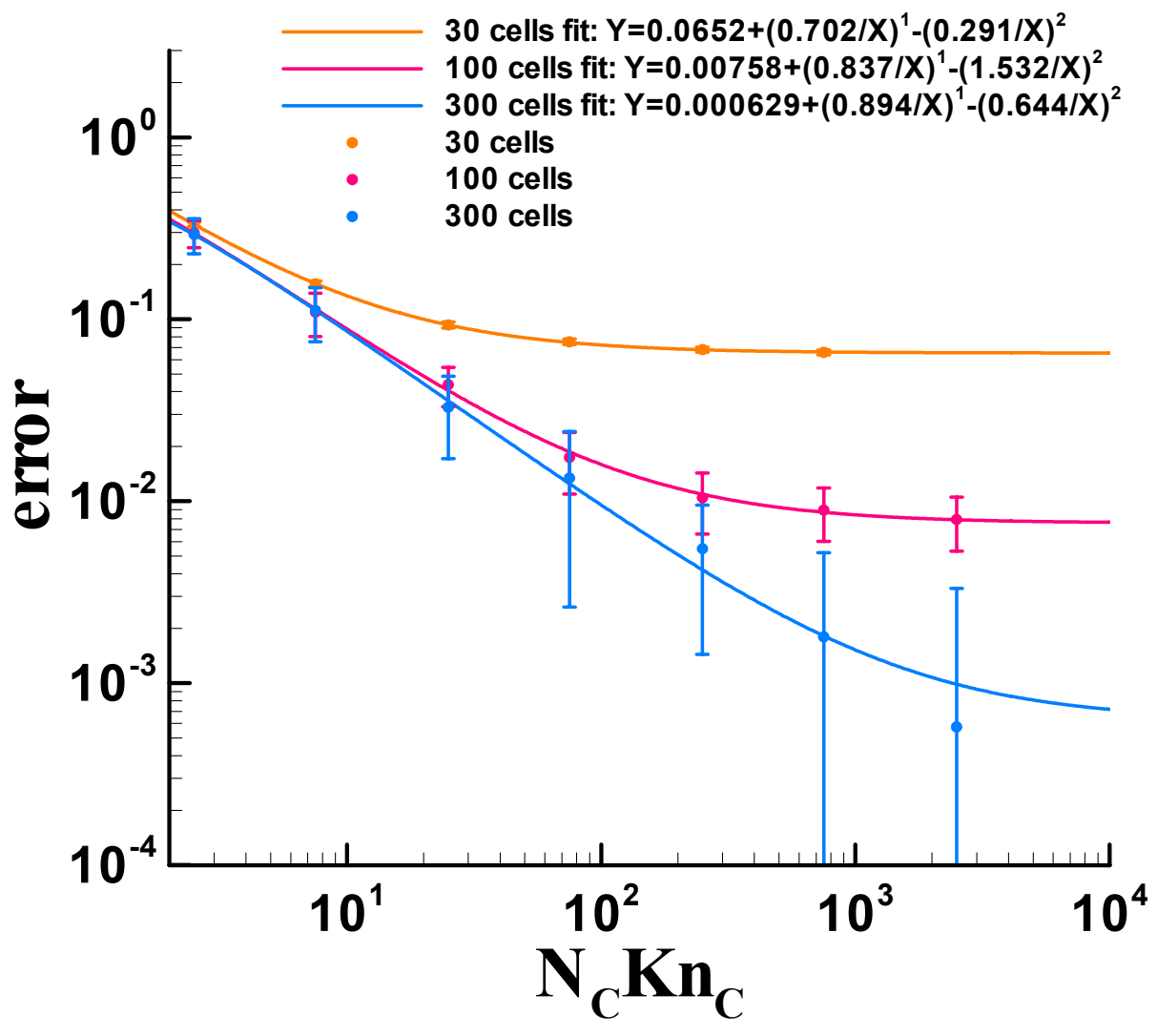

**Рис. 4.3-1.** Зависимость ошибки теплового потока от критерия $\overline{N_C} \cdot \mathrm{Kn}_C$ при разном шаге сетке, $\mathrm{Kn}_0$ = 0.025. Фактические точки и регрессия.

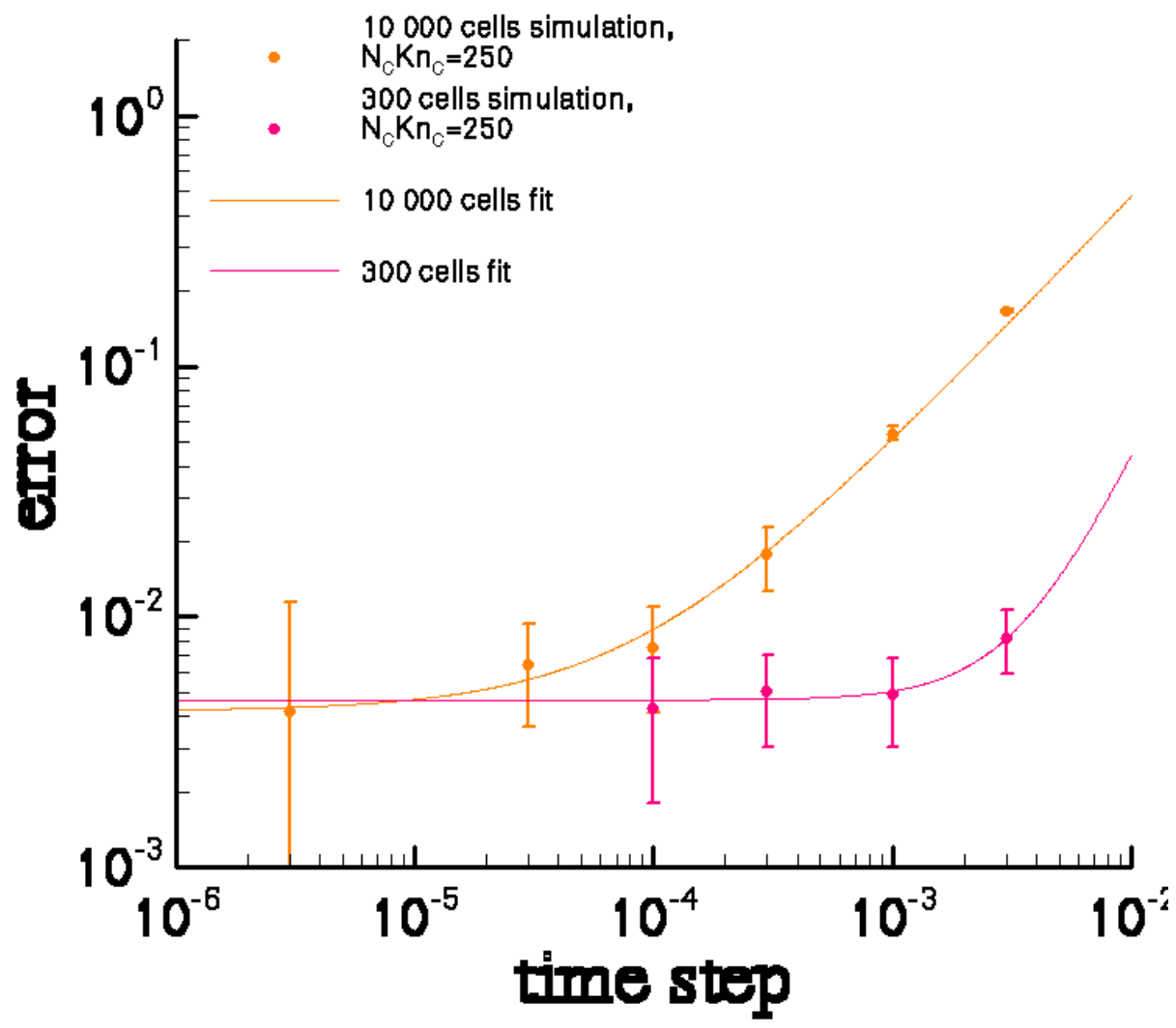

**Рис. 4.3-2.** Зависимость ошибки теплового потока от временного шага при двух размерах сетки: 300 и 10 000 ячеек.

Исследование зависимости погрешности от шага сетки для метода мажорантной частоты показало результаты, согласующиеся с работами (9) и (7). На **Рис. 4.3-1** показана зависимость ошибки теплового потока от $\overline{N_C} \cdot \mathrm{Kn}_C$ при разном шаге сетки. Как видно, зависимости почти совпадают при малых значениях $\overline{N_C} \cdot \mathrm{Kn}_C$, но затем постепенно выходят на «полочки», когда определяющий вклад в погрешность привносит именно конечность шага сетки.

На **Рис. 4.3-4** показаны результаты регрессии как по данным **Рис. 4.3-1**, так и по данным **Рис. 4.3-2**. Полученные двумя способами точки совпадают в пределах статистической погрешности и согласуются с теоретическим предсказанием на основе зависимости (7).

## 4.4 Временной шаг

Эмпирически принято считать, что временной шаг должен быть достаточно мал, чтобы большинство частиц не успевали пролетать более одной ячейки. Временной шаг также должен быть меньше времени свободного пробега.

Исследование погрешности метода мажорантной частоты в зависимости от временного шага также показало согласие с работами (9) и (8). На **Рис. 4.3-3** показана зависимость погрешность от временного шага при разном шаге сетки. Там же приведена теоретическая кривая на основе зависимости (8). Как и ранее для шага сетки, зависимости очень схожи до тех пор, пока определяющий вклад в погрешность не будет оказывать конечность шага сетки.

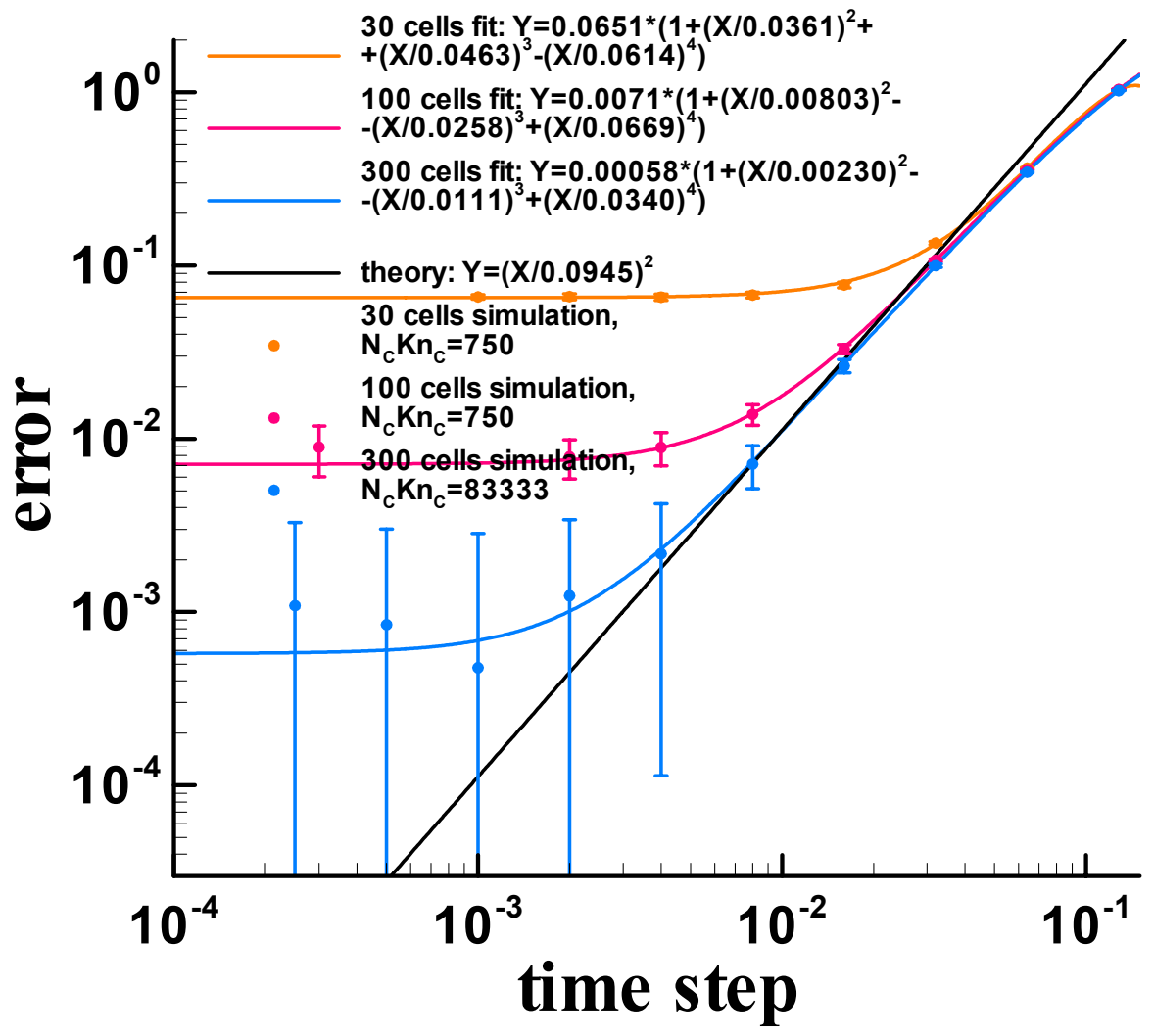

**Рис. 4.3-3.** Зависимость ошибки теплового потока от временного шага при различном шаге сетки.

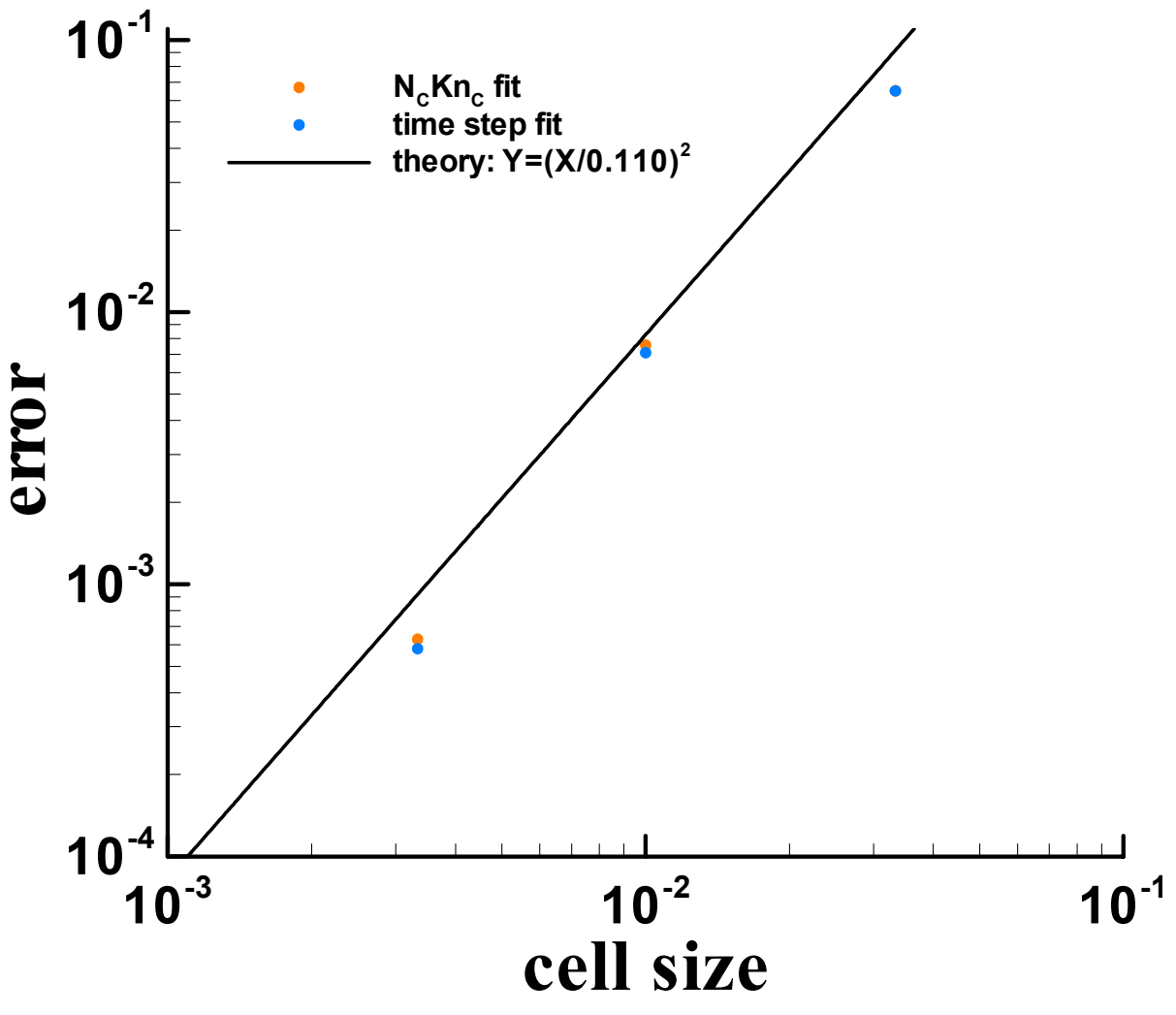

**Рис. 4.3-4.** Зависимость ошибки теплового потока от шага сетки.

Дополнительно, была исследована принципиальность требования того, чтобы за временной шаг частицы проходили менее 1 ячейки сетки. На **Рис. 4.3-2** показана зависимость погрешности от временного шага при двух шагах сетки. В обоих случаях значение критерия $\overline{N_C} \cdot \mathrm{Kn}_C$ было выбрано таким, чтобы конечность числа частиц привносила определяющий вклад в погрешность. Как можно видеть, при достаточно больших временных шагах, когда частица проходит сразу несколько ячеек, появляется дополнительный линейный член в зависимости погрешности от временного шага, который может превосходить квадратичную часть. Появление линейного члена объясняется ростом эффективного значения критерия $\overline{N_C} \cdot \mathrm{Kn}_C$. При очень малом временном шаге, характерное время пребывания пары в одной ячейке $\delta t_C = \frac{h}{c_r}$. Но это время не может быть меньше одного временного шага. Поэтому, если временной шаг велик $\delta t > \delta t_C$, то вероятность повторных столкновений определяется уже не критерием $\overline{N_C} \cdot \frac{\lambda}{h}$, а скорее величиной $\overline{N_C} \cdot \frac{\lambda}{h} \cdot \frac{h}{c_r \delta t}$.

Таким образом, можно сделать вывод, что требование о том, чтобы частицы проходили не более 1 ячейки за временной шаг, не является строго обязательным. Однако, нарушение этого требования может повысить вероятность повторных столкновений линейно с ростом среднего числа ячеек, проходимых частицами за временной шаг.

## 4.5 Определение теплового потока приближенным решением уравнения теплопроводности

Выше приводились «теоретические» кривые зависимости теплового потока от временного шага и шага сетки. Построить эти кривые можно при помощи приближенного решения уравнения теплопроводности, с учётом температурных скачков на стенках и малой поправкой на коэффициент теплопроводности.

Для начала пренебрежем температурными скачками у стенок. Уравнение Фурье для газа твёрдых сфер, а также уравнение постоянства давления с учётом уравнения состояния идеального газа:

$$q = \kappa_0 \sqrt{\frac{T}{T_0}} \frac{dT}{dx} = \kappa_0 T_0 \frac{2}{3} \frac{d}{dx} \left(\frac{T}{T_0}\right)^{\frac{3}{2}} = \frac{\kappa_0 T_0}{L_0} \cdot \frac{2}{3} \frac{d}{d\tilde{x}} \tilde{T}^{\frac{3}{2}} = const,$$

$$P = nkT = const$$

Здесь и далее безразмерные величины отмечены волной и отнесены к $T_0$ (единичная температура), $L_0$ (расстояние между пластинами), $n_0$ (единичная плотность, которая установится при отсутствии градиента температур). Профили температуры и плотности, а также давление $P$ и плотность потока тепла на единицу площади $q$:

$$\tilde{T}(\tilde{x}) = \left[\tilde{T}_1^{\frac{3}{2}}(1-\tilde{x}) + \tilde{T}_2^{\frac{3}{2}}\tilde{x}\right]^{\frac{2}{3}},$$

$$\frac{1}{\tilde{n}} \frac{d\tilde{T}}{d\tilde{x}} = \frac{\tilde{q}}{\tilde{P}} \sqrt{\tilde{T}(\tilde{x})},$$

$$q = \frac{\kappa_0 T_0}{L_0} \cdot \frac{2}{3} \cdot \left[\tilde{T}_2^{\frac{3}{2}} - \tilde{T}_1^{\frac{3}{2}}\right],$$

$$P = n_0 k T_0 \cdot \frac{1}{3} \cdot \left[\tilde{T}_1 + \tilde{T}_2 + \sqrt{\tilde{T}_1 \tilde{T}_2}\right].$$

Таким образом, для условий рассмотренного выше расчета: $\tilde{T}_1 = 0.5, \tilde{T}_2 = 2, \mathrm{Kn}_0 = 0.025$.

В дальнейшем, будем приводить численные значения к единицам, использовавшимся в расчете ПСМ. Единицы температуры, плотности и давления равны, соответственно, $T_0$, $n_0$ и $n_0 k T_0$. За единицу длины принято расстояние между пластинами $L_0$. За единицу поверхностной плотности теплового потока принята величина $m n_0 \left(\frac{kT_0}{m}\right)^{3/2}$.

В численных единицах, использовавшихся при расчете ПСМ (с учётом второго приближения для коэффициента теплопроводности):

$$\tilde{\kappa_0} = \frac{\kappa_0 T_0}{m n_0 L_0} \left(\frac{kT_0}{m}\right)^{-3/2} = \sqrt{2\pi} \frac{75}{64} \mathrm{Kn}_0 [1 + 0.024819] \approx 0.0752590,$$

а тепловой поток, полученный при решении уравнения Фурье, составляет 0.124171 единиц, что на 11.6 % больше величины, полученной методом ПСМ, аналитическое значение давления составило 7/6 в численных единицах, на 0.85 % больше результата ПСМ.

Попробуем учесть температурные скачки у стенок. Согласно формуле из (22), температурный скачок у стенки при полной аккомодации составляет:

$$\Delta T \approx 1.173 \cdot \frac{1}{2nk} \sqrt{\frac{\pi m}{2kT}} \cdot \kappa \frac{dT}{dx}, \text{ или:}$$

$$\Delta \tilde{T} \approx 1.202 \cdot \frac{75\pi}{128} \cdot \mathrm{Kn}_0 \cdot \frac{1}{\tilde{n}} \frac{d\tilde{T}}{d\tilde{x}} \approx 0.0553 \cdot \frac{1}{\tilde{n}} \frac{d\tilde{T}}{d\tilde{x}} = 0.0553 \cdot 2\sqrt{\tilde{T}} \left(\sqrt{\tilde{T}_2} - \sqrt{\tilde{T}_1}\right).$$

С учётом температурных скачков, условия на границе должны теперь быть заданы следующими: $\widetilde{T_1} \approx 0.5523, \widetilde{T_2} \approx 1.9029$, что получается решением соответствующей нелинейной системы алгебраических уравнений. Теперь, с корректированными граничными условиями, тепловой поток составляет 0.111108 единиц (отличие от результата ПСМ теперь 0.13 %, причём другого знака). Давление же составляет 1.16012, превышая результат ПСМ всего на 0.29 %. Учёт теплового скачка существенно уменьшил разницу результатов.

Дальнейший шаг – снова решить уравнение Фурье, но с учётом малой поправки на коэффициент теплопроводности. Воспользуемся методом Гильберта разложения по малому параметру. В качестве малого множителя используем переменную $\alpha$. Тогда, при исследовании дискретности по пространству и по времени соответственно, будет использоваться следующее представление коэффициента теплопроводности:

$$\kappa = \kappa_0 \tilde{T}^{\frac{1}{2}}[1 + \alpha \tilde{P}^2 \tilde{T}^{-2}], \qquad \alpha = \frac{32}{225\pi} \cdot \left(\frac{\tilde{h}}{\mathrm{Kn}_0}\right)^2 \approx 72.433 \cdot \tilde{h}^2,$$
$$\kappa = \kappa_0 \tilde{T}^{\frac{1}{2}}[1 + \alpha \tilde{P}^2 \tilde{T}^{-1}], \qquad \alpha = \frac{128}{675\pi} \cdot \left(\frac{\widetilde{\delta t}}{\mathrm{Kn}_0}\right)^2 \approx 96.578 \cdot \widetilde{\delta t}^2.$$

Влияние отклонения в коэффициенте теплопроводности на величину скачков температуры у границ требует специального исследования, поэтому его учитывать не будем. Температурный профиль теперь представлен в виде:

$$\tilde{T} = \left[\widetilde{T_{1(0)}}^{\frac{3}{2}}(1 - \tilde{x}) + \widetilde{T_{2(0)}}^{\frac{3}{2}}\tilde{x} + \alpha F(x)\right]^{\frac{2}{3}},$$

Тепловой поток:

$$\tilde{q} = \widetilde{q_{(0)}}[1 + \alpha A].$$

Представив коэффициент теплопроводности в виде $\kappa = \kappa_0 \tilde{T}^{\frac{1}{2}}[1 + \alpha\beta\tilde{T}^{\gamma}]$, для первого приближения получаем:

$$F(x) = B + \frac{3}{2} A \widetilde{q_{(0)}} \tilde{x} - \beta \cdot \frac{3}{3+2\gamma}\left[\widetilde{T_{1(0)}}^{\frac{3}{2}}(1 - \tilde{x}) + \widetilde{T_{2(0)}}^{\frac{3}{2}}\tilde{x}\right]^{1+\frac{2}{3}\gamma}.$$

Здесь $A, B$– коэффициенты, подбираемые так, чтобы удовлетворить граничным условиям. Для $A$ имеем:

$$A = \beta \cdot \frac{3}{3 + 2\gamma} \cdot \frac{T_2^{\frac{3}{2}+\gamma} - T_1^{\frac{3}{2}+\gamma}}{T_2^{\frac{3}{2}} - T_1^{\frac{3}{2}}}$$

Таким образом, в первом случае, для шага по пространству: $\beta = \tilde{P}_{(0)}^{\ 2}$, $\gamma = -2$, $A = 1.1316$, $\tilde{q} = \widetilde{q_{(0)}}[1 + 81.968 \cdot \tilde{h}^2]$. Аналогично, во втором случае, для шага по времени: $\beta = \tilde{P}_{(0)}^{\ 2}$, $\gamma = -1$, $A = 1.1601$, $\tilde{q} = \widetilde{q_{(0)}}[1 + 112.04 \cdot \widetilde{\delta t}^2]$.

Именно по этим формулам получены теоретические кривые, изображенные на **Рис. 4.3-3** и **Рис. 4.3-4**.

Применим эту же методику к условиям работы (9) с более слабыми градиентами и сравним с её численными результатами. Температуры на стенках: 0.81695 и 1.18305 (перепад $\approx 1.45$ раза), число Кнудсена 0.02417. $\frac{\kappa_0 T_0}{L_0}$ = 0.072780. Скачки температур вычисляем с коэффициентом 0.0535, получая следующие значения температур у стенки: 0.8332 и 1.1639. Имеем давление и тепловой поток: $\tilde{q} = 0.02402$ (1527.1 Вт/м$^2$) $\tilde{P} = 0.9940$ (265.04 Па). Численный результат авторов – 1512.0 Вт/м$^2$ и 264.96 Па – показывают хорошее согласие. Проделаем все остальные действия, получаем: $A = 0.9940$ для шага по времени и $A = 1.00938$ для шага по пространству. Оба коэффициента, как и следовало ожидать, слабо отличаются от единицы, поэтому для этих условий поправочный множитель к тепловому потоку будет практически совпадать с таковым для коэффициента теплопроводности.

## 4.6 Автокорреляционные свойства пристеночных макропараметров. Поправочный коэффициент дисперсии

В начале раздела, использовалось несколько предположений о статистической независимости параметров и времени прибытия сталкивающихся со стенками частиц. Что, конечно же, было лишь допущением. Однако, уже после того, как основная часть работы, представленной в настоящем разделе, была выполнена, были предприняты попытки определить действительное значение поправочного коэффициента дисперсии $\widetilde{D}$.

Вычислить его можно по следующей формуле:

$$\widetilde{D}(\tau) = \widetilde{D}(\delta\tau)\left[1 + 2\sum_{k'=1}^{\frac{\tau}{\delta\tau}-1} \operatorname{corr}\left[\delta\mathcal{E}_{k_0}, \delta\mathcal{E}_{k_0+k'}\right] \cdot \left(1 - \frac{\delta\tau}{\tau}\cdot k'\right)\right]$$
$$= \frac{\delta\tau}{\tau}\widetilde{D}(\delta\tau)\left[1 + \sum_{m=2}^{\tau/\delta\tau}\left[1 + 2\sum_{k'=1}^{m-1} \operatorname{corr}\left[\delta\mathcal{E}_{k_0}, \delta\mathcal{E}_{k_0+k'}\right]\right]\right]$$

Как видно, коэффициент зависит от длины выборки $\tau/\delta\tau$. Здесь $\operatorname{corr}\left[\delta\mathcal{E}_{k_0}, \delta\mathcal{E}_{k_0+k'}\right]$ означает автокорреляцию теплового потока (скорости изменения энергетического баланса) $\delta\mathcal{E}_{k_0}$, где $k_0$ – свободный параметр; $\tau$ – время наблюдения, $\delta\tau$ – время, приходящееся на один отсчёт.

Автокорреляционная функция для моделируемого процесса неизвестна. Значит, её придётся получать численно. Однако, для хорошей оценки автокорреляционной функции по конечной выборке, нужна выборка существенно длиннее $\tau$, для которого будет оцениваться $\widetilde{D}$. Следует обратить внимание на то, что $\widetilde{D}$ является, фактически, вторым интегралом автокорреляционной функции. Это значит, что нижняя часть спектра шумовой компоненты автокорреляционной функции после двукратного интегрирования способна усилиться до довольно большой амплитуды.

«Потренируемся» сперва на псевдослучайных числах. Что, к тому же, будет лишним поводом проверить качество генератора. Вычисление корреляционной функции будем проводить при помощи алгоритма быстрого преобразования Фурье, по следующей схеме:

1. Создать массив комплексных чисел необходимой длины.
2. Загрузить данные в действительную часть массива (в данном случае сгенерировать случайные числа), мнимую часть занулить.
3. Посчитать среднее значение элемента и вычесть его из каждого элемента массива.
4. Посчитать сумму квадратов модулей элементов массива и нормировать данные так, чтобы сумма квадратов модулей стала равна единице.
5. Увеличить массив вдвое, дописав нужное количество нулевых значений.
6. Увеличить массив до размера, равного ближайшей степени двойки. Дописав, аналогично, нулевые значения.
7. Выполнить над массивом стандартный алгоритм быстрого преобразования Фурье.

8. Для каждой гармоники, вычислить квадрат модуля комплексного числа и присвоить как новое значение гармоники.

9. Выполнить обратное преобразование Фурье.

10. Поделить k-й элемент массива на величину $1-\frac{|k|}{K}$, где $k=-(K-1)\ldots+(K-1)$, $K$ – первоначальная длина выборки.

11. Вычисляется двойная сумма полученной автокорреляционной функции и строится зависимость $\widetilde{D}(K')$.

На **Рис. 4.6-1** показаны полученные численно зависимости $\widetilde{D}(K')$, посчитанные для нескольких выборок случайных чисел длиной $K=1048576$. Как видно, только при $K'\lesssim 1024=\sqrt{K}$ флуктуации $\widetilde{D}(K')$ лежат в более-менее приемлемом диапазоне. При $K'\to K$, $\widetilde{D}(K')\to 0$, что есть особенность использованного метода построения автокорреляционной функции, связанная с занулением среднего значения исходной выборки. Таким образом, для уверенной оценки $\widetilde{D}(K')$, необходимо иметь выборку длиной $K>(K')^2$. В данном случае, изучая зависимость $\widetilde{D}(K')$ при $K'\lesssim 1024$, можно заключить, что последовательности, вырабатываемые использованным генератором случайных чисел, не обладают выраженными автокорреляционными свойствами.

На **Рис. 4.6-2** представлены зависимости $\widetilde{D}_N(\tau)/\widetilde{D}_N(\delta\tau)$, определенные для частоты соударений молекул с холодной стенкой, для режима $Kn_0$=0.025. Использованы четыре выборки. Во всех случаях, при расчете использовался шаг сетки 0.005 и временной шаг 0.001. Первая выборка (**красная** кривая) получена при 1000 частицах в расчетной области, $\delta\tau=0.001$. Это соответствует в среднем 0.62 соударениям за время $\delta\tau$. Вторая выборка (**голубая** кривая) получена при 100 000 частицах и том же $\delta\tau$. Третья выборка (**сиреневая** кривая) получена также при 100 000 частицах, но при $\delta\tau=\frac{0.001}{60}$. Длина этих трёх выборок – $2^{24}$ значений. Четвертая выборка (**коричневая** кривая) получена при 10 000 частицах, но $\delta\tau=0.01$, а длина выборки составляет чуть более 11 млн. элементов. Именно из-за большого $\delta\tau$, и, следовательно, повышенного $\widetilde{D}_N(\delta\tau)$, зависимость для этой выборки лежит на графике несколько ниже. Однако, использование столь большого $\delta\tau$ позволило продвинуться в сторону бо́льших $\tau$. Дело в том, что созданная для расчета $\widetilde{D}_N(\tau)$ программа была способна работать только с выборками не длиннее $2^{24}$. Первые две выборки (для $\delta\tau=0.001$) дали зависимости, совпадающие при малых $\tau$, несмотря на 100-кратную разницу в числе расчетных частиц. Это свидетельствует о том, что число частиц мало влияет на свойства $\widetilde{D}_N(\tau)$.

О поведении зависимости $\widetilde{D}_N(\tau)$ можно заключить следующее. Вплоть $\tau=0.01$ (характерная частота столкновений в газе, вблизи стенки), $\widetilde{D}_N$ близок к единице и растёт медленно. При этом, основной вклад в дисперсию вносит «белый шум», связанный с дискретностью и случайностью соударений. Затем, рост ускоряется. $\widetilde{D}_N$ достигает плато при $\tau\approx 0.3$, после чего начинает снижаться. Минимум достигается при $\tau\approx 1.4$, что хорошо совпадает со временем, за которое звуковая волна проходит путь от холодной пластины к горячей и обратно. После минимума, рост $\widetilde{D}_N$ возобновляется. По мере приближения к $\tau\approx 40$, влияние интерференции шума ослабевает, и зависимость $\widetilde{D}_N(\tau)$ демонстрирует тенденцию к насыщению. Величина насыщения $\widetilde{D}_N$ лежит в диапазоне 1.75 – 2.

В целом, анализ автокорреляционных свойств частоты столкновения частиц с пластиной показал отсутствие каких либо корреляций на временах, характерных для кинетических процессов. С другой стороны, обнаруживаются заметные долгоживущие автокорреляции, связанные с макроскопическими релаксационными свойствами моделируемой системы.

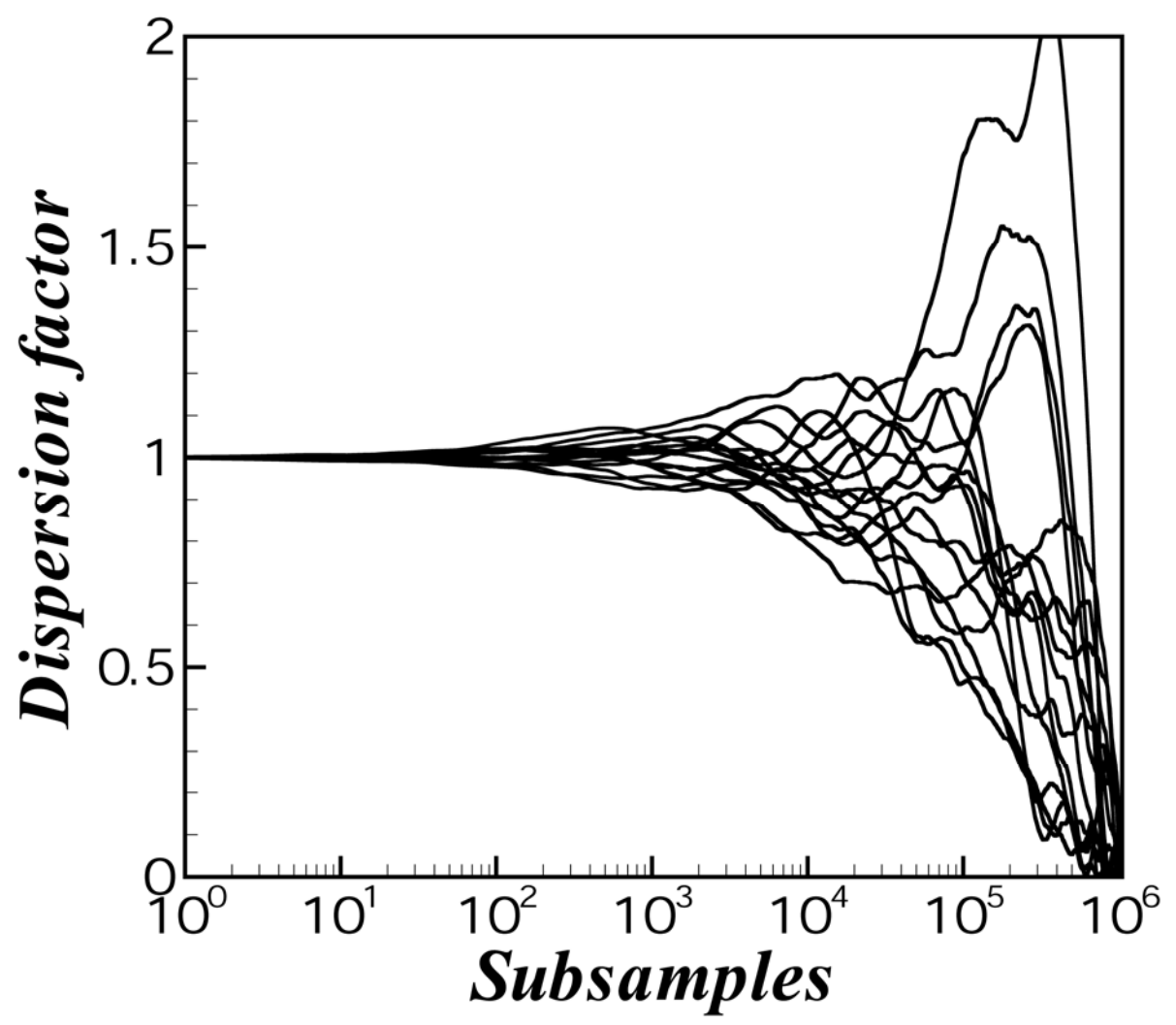

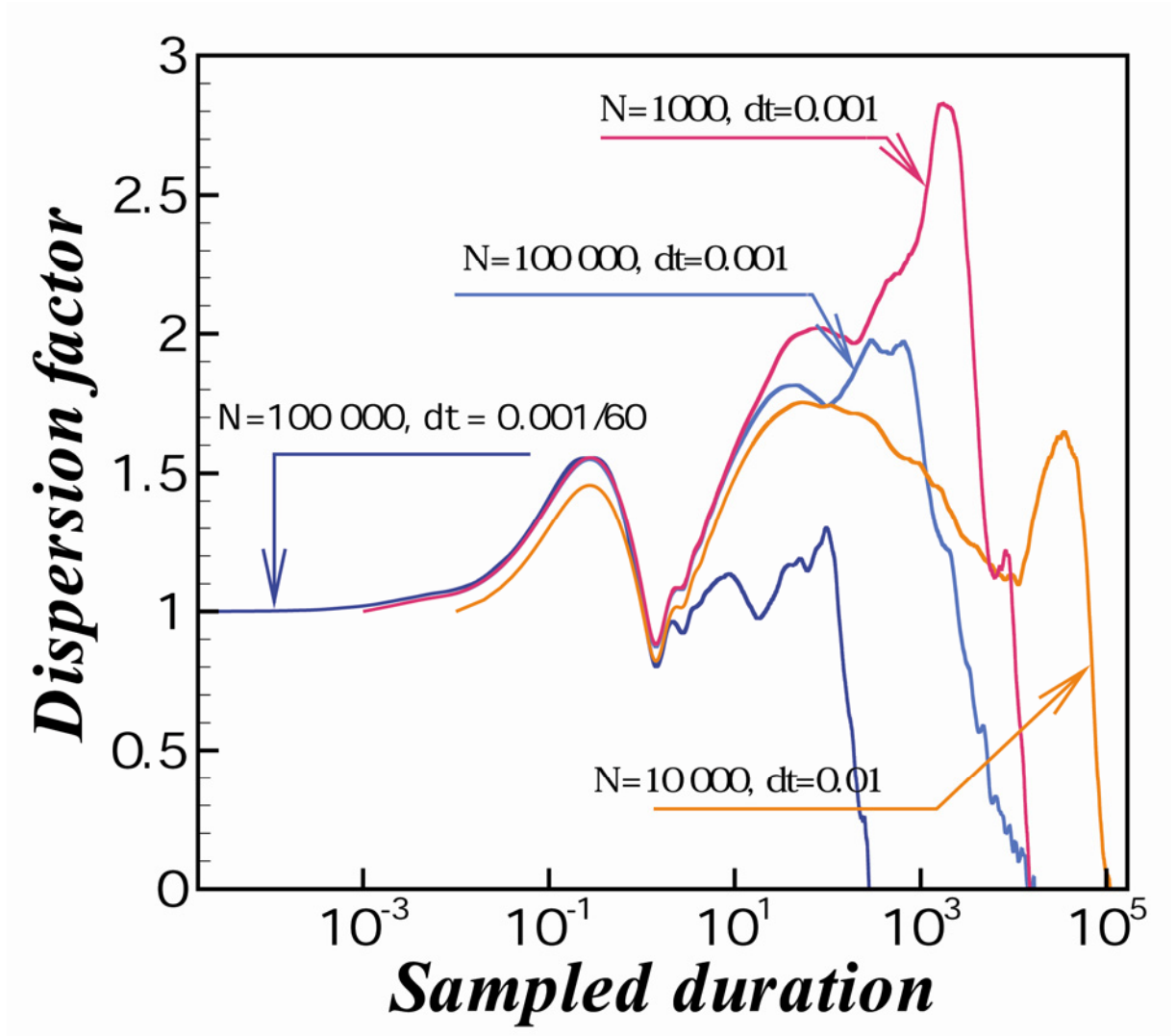

**Рис. 4.6-1.** Результаты вычисления поправочного коэффициента дисперсии для псевдослучайного белого шума.

**Рис. 4.6-2.** Зависимость поправочного коэффициента дисперсии от длительности осреднения для частоты столкновений частиц с холодной пластиной.

На **Рис. 4.6-3** представлены кривые зависимости $\widetilde{D}_N(\tau)/\widetilde{D}_N(\delta\tau)$ для давления на холодную пластину. В целом, картина качественно повторяет таковую для частоты соударений, но коэффициент принимает несколько меньшие значения. Насыщение происходит при $\widetilde{D}_N \approx 1.7$.

На **Рис. 4.6-4** представлены кривые зависимости $\widetilde{D}_N(\tau)/\widetilde{D}_N(\delta\tau)$ для теплового потока на холодную пластину. На этот раз, наблюдается качественно иная картина. С самого начала величина $\widetilde{D}_N(\tau)$ монотонно снижается. При $\tau \approx 2.8$ (время релаксации первой моды уравнения теплопроводности) величина $\widetilde{D}_N$ уже практически достигла минимума. Насыщение происходит при $\widetilde{D}_N \approx 0.4$.

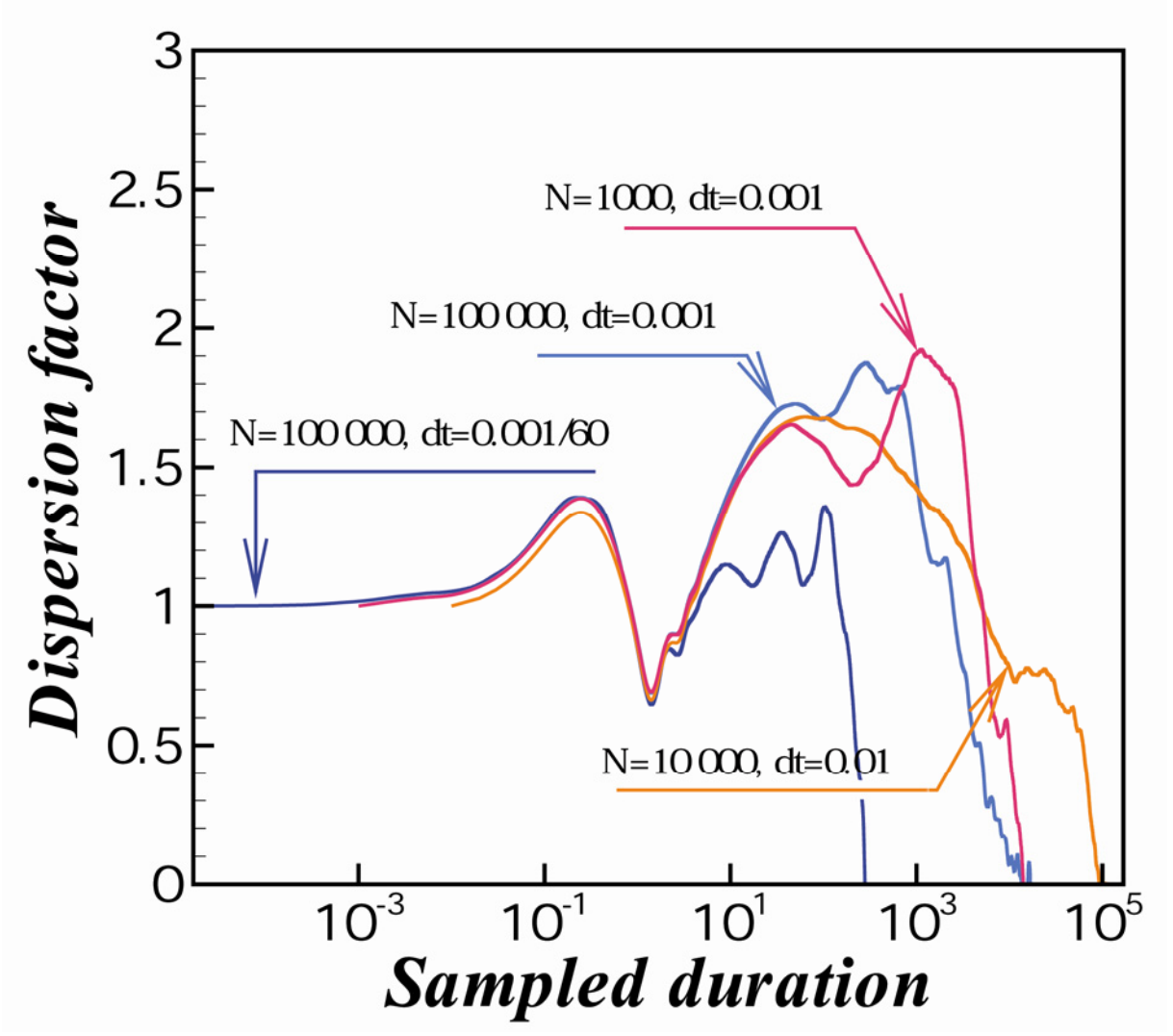

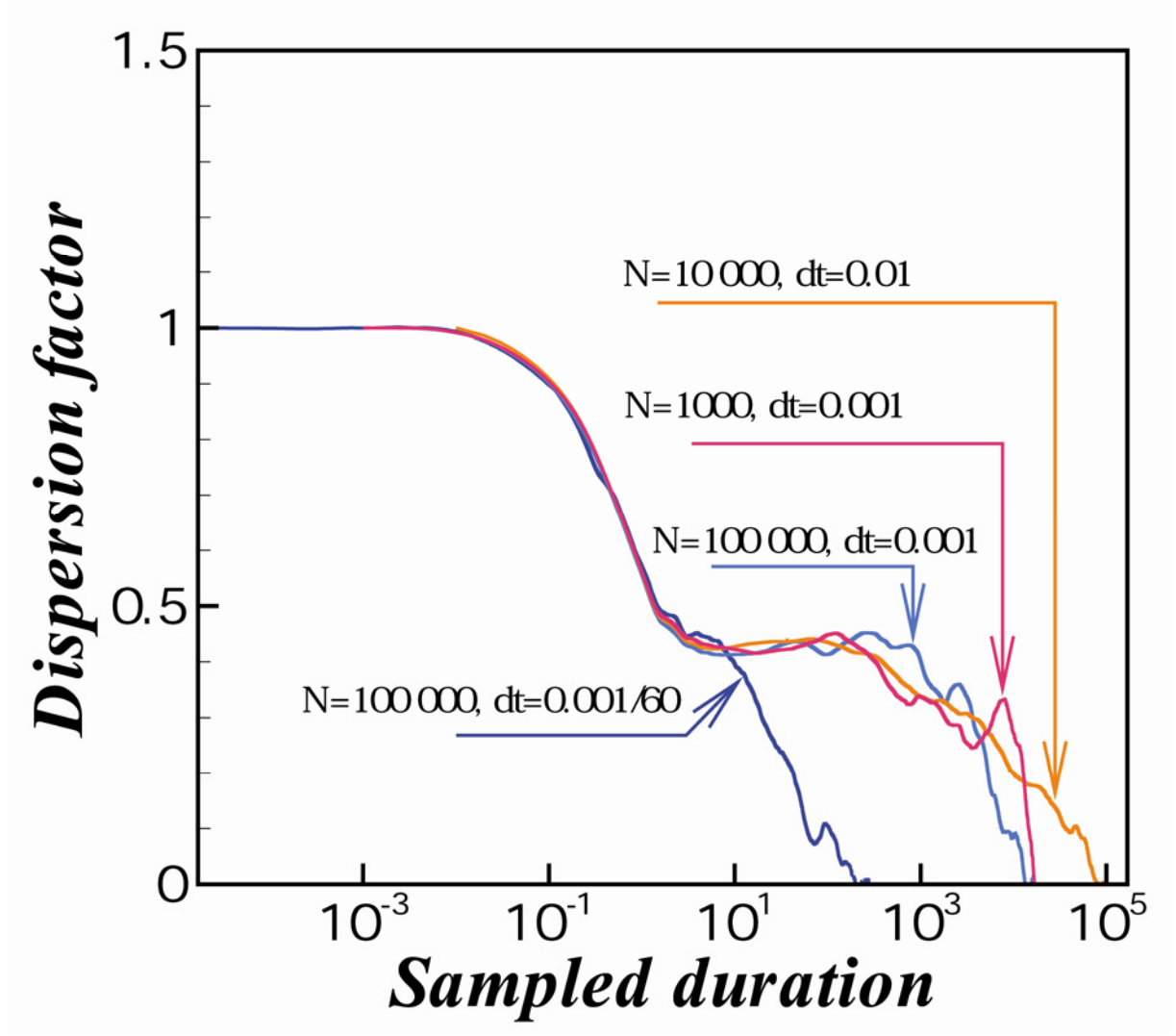

**Рис. 4.6-3**. Зависимость поправочного коэффициента дисперсии от длительности осреднения для давления на холодную пластиной.

**Рис. 4.6-4**. Зависимость поправочного коэффициента дисперсии от длительности осреднения для теплового потока на холодную пластину.

Итак, если реальная дисперсия давления на холодную пластину оказывается на 70 % выше оценки, не учитывающей автокорреляции, то реальная дисперсия энергии, наоборот, оказывается на 60 % ниже оценки. Этому можно найти простое наглядное объяснение. Если вследствие случайных флуктуаций пластина отражает частицы обратно в поток с повышенной скоростью (за счёт этого повысится и давление на пластину), то это приведёт к повышению давления и температуры газа вблизи пластины. Вследствие повышения давления, газ ещё некоторое время будет давить на пластину сильнее, пока флуктуация давления не релаксирует. Это увеличивает чувствительность к флуктуациям давления. Для теплового потока картина противоположная. Если вследствие случайных флуктуаций пластина сообщила дополнительную энергию газу, повышая энергию отражаемых частиц, то температура газа вблизи пластины вырастет. Вследствие нагрева, газ будет поставлять на пластину более энергичные частицы, стремясь вернуть избыток энергии обратно. Это приводит к ослаблению флуктуаций теплового потока.

Выше были проанализированы поправки на дисперсию параметров, измеряемых на холодной пластине. Однако, бо́льший интерес представляет поправка на дисперсию величин, получаемых объединением результатов от холодной и горячей пластины.

Но прежде, необходимо ввести некоторые обозначения. Итак, для простоты будем полагать, что сбор данных не стенке производится интервалами длительности $\delta\tau$. При этом известны математические ожидания приращений энергии $\delta\mathcal{E}^h$ (для горячей стенки) и $\delta\mathcal{E}^c$ (для холодной), а также их дисперсии $D^h(\delta\tau)$ и $D^c(\delta\tau)$ соответственно. Считаем, что длина выборок составляет $K=\frac{\tau}{\delta\tau}$ элементов. Математическое ожидание полного баланса составляет: $\mathcal{E}^h = K\cdot\delta\mathcal{E}^h$ и $\mathcal{E}^c = K\cdot\delta\mathcal{E}^c$ , дисперсии соответственно обозначим $D^h(\tau)$ и $D^c(\tau)$ , которые ищутся по формуле: $D^h(\tau) = K\cdot D^h(\delta\tau)\cdot\left[1+2\sum_{k'=1}^{K-1}\mathrm{corr}\left[\delta\mathcal{E}^h{}_{k_0},\delta\mathcal{E}^h{}_{k_0+k'}\right]\cdot\left(1-\frac{k'}{K}\right)\right]$ (для горячей пластины), аналогично для холодной. Определим также «перекрёстную дисперсию»: $D^X(\tau) = K\cdot\sqrt{D^c(\delta\tau)\cdot D^h(\delta\tau)}\cdot\sum_{k'=-(K-1)}^{K-1}\mathrm{corr}\left[\delta\mathcal{E}^h{}_{k_0},\delta\mathcal{E}^c{}_{k_0+k'}\right]\cdot\left(1-\frac{|k'|}{K}\right)$. Математическое ожидание объединённой выборки составляет: $\mathcal{E}(\beta)=\beta\cdot\mathcal{E}^c-(1-\beta)\cdot\mathcal{E}^h$, здесь $\beta$ – коэффициент взвешенного среднего. Дисперсия объединённой выборки: $D(\tau,\beta)=\beta^2 D^c(\tau)+2\beta(1-\beta)\cdot D^X(\tau)+(1-\beta)^2\cdot D^h(\tau)$ . Значение $\beta$ , при котором дисперсия минимальна, составляет: $\beta_{\mathrm{MIN}}=\frac{D^h(\tau)-D^X(\tau)}{D^c(\tau)+D^h(\tau)-2D^X(\tau)}$, величина минимальной дисперсии $D^{\mathrm{MIN}}(\tau)=\frac{D^c(\tau)\cdot D^h(\tau)-\left[D^X(\tau)\right]^2}{D^c(\tau)+D^h(\tau)-2D^X(\tau)}$. Если же при объединении выборок используется коэффициент $\beta_{\mathrm{IND}}=\frac{D^h(\delta\tau)}{D^c(\delta\tau)+D^h(\delta\tau)}$, то дисперсия объединённой выборки составит $D^{\mathrm{U}}(\tau)=\frac{\left[D^h(\delta\tau)\right]^2\cdot D^c(\tau)+2D^c(\delta\tau)\cdot D^h(\delta\tau)\cdot D^X(\tau)+\left[D^c(\delta\tau)\right]^2\cdot D^h(\tau)}{\left[D^c(\delta\tau)+D^h(\delta\tau)\right]^2}$ . Если же все приращения энергии действительно независимы, то в этом случае $D^{\mathrm{MIN}}(\tau)=D^{\mathrm{U}}(\tau)=D^{\mathrm{IND}}(\tau)$, где $D^{\mathrm{IND}}(\tau)=K\cdot\frac{D^c(\delta\tau)\cdot D^h(\delta\tau)}{D^c(\delta\tau)+D^h(\delta\tau)}$.

На **Рис. 4.6-5** показаны зависимости поправочных коэффициентов для давления. **Голубая** и **красная** кривые показывают поправочные коффициенты $\frac{D^c(\tau)}{K\cdot D^c(\delta\tau)}$ и $\frac{D^h(\tau)}{K\cdot D^h(\delta\tau)}$ соответственно для холодной и горячей пластин. Как видно, давление на горячую пластину сходится лучше. Это, по-видимому, связано с тем, что газ у горячей пластины более разрежен, и флуктуации в нём затухают быстрее. Кроме того, излишек тепла может не только возвращаться на горячую пластину, но и рассасываться в сторону холодной пластины.

**Чёрная** кривая показывает отношение $D^{\mathrm{U}}(\tau)/D^{\mathrm{IND}}(\tau)$ – поправку к дисперсии величины, полученной объединением результатов для двух пластин. Для объединённой выборки, поправочный коэффициент достигает $D^{\mathrm{U}}(\tau)/D^{\mathrm{IND}}(\tau)\approx 2.7$ – довольно большой величины. Это объясняется значительной взаимной корреляцией давления на двух пластинах. Для сравнения, зависимость для наименьшей дисперсии – отношение $D^{\mathrm{MIN}}(\tau)/D^{\mathrm{IND}}(\tau)$ – представлена на **Рис. 4.6-5** **коричневой** кривой. Практически всюду эта кривая совпадает с **чёрной** кривой, что говорит о том, что получить сколько-либо заметный выигрыш использованием $\beta=\beta_{\mathrm{MIN}}$ не удаётся, и $\beta_{\mathrm{IND}}$ является весьма хорошим приближением.

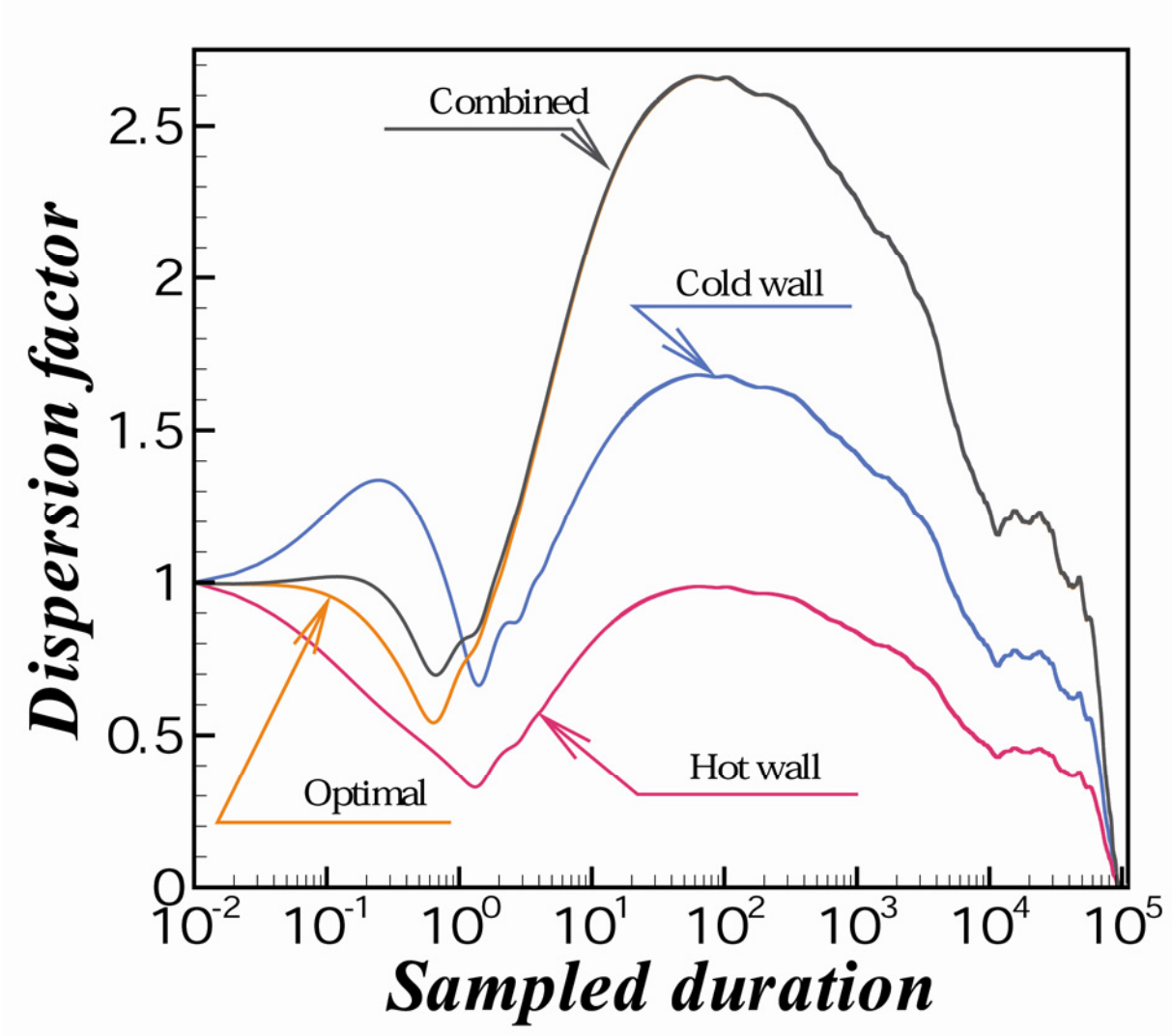

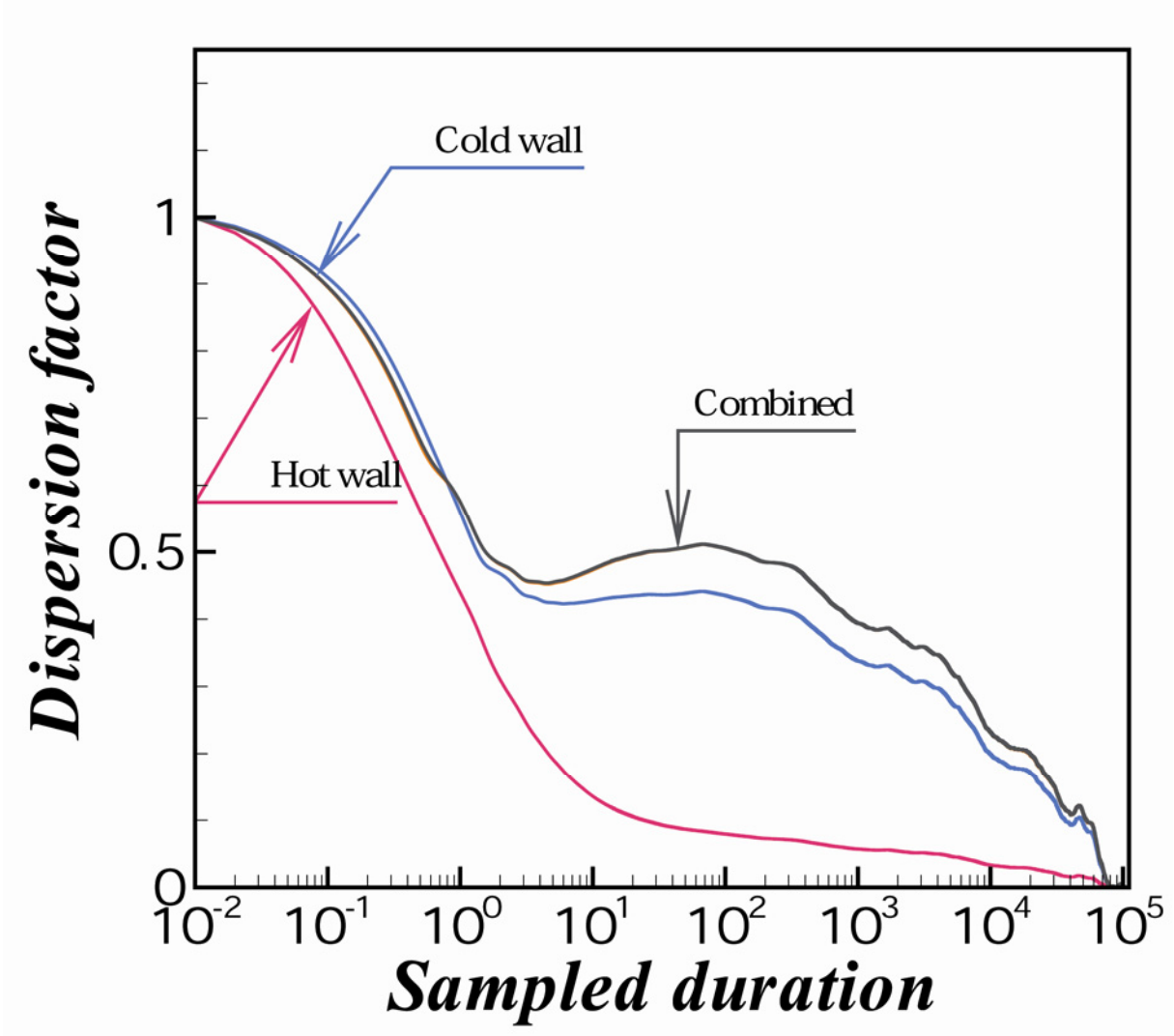

**Рис. 4.6-5**. Поправочные коэффициенты дисперсии для давления.

**Рис. 4.6-6**. Поправочные коэффициенты дисперсии для теплового потока.

Аналогично, на **Рис. 4.6-6** показаны зависимости поправочных коэффициентов дисперсии для теплового потока. Разница между холодной и горячей пластинами гораздо более выражена, чем для давления. Поправочный коэффициент для объединённой выборки сходится к $D^{\mathrm{U}}(\tau)/D^{\mathrm{IND}}(\tau) \approx 0.5$. Оптимальный подбор коэффициентов взвешенного среднего также не позволяет улучшить дисперсию.

Итак, реальная дисперсия теплового потока оказалась на 50 % меньше, чем посчитанная без учёта корреляционных свойств выборок. Реальная дисперсия давления, наоборот, оказалась в 2.7 раз выше. Полученные значения поправочных коэффициентов относятся к числу Кнудсена $Kn_0$=0.025.

Поправочные коэффициенты для $Kn_0$=0.1 составляют: $D^{\mathrm{U}}(\tau)/D^{\mathrm{IND}}(\tau) \approx 1.3$ для давления и $D^{\mathrm{U}}(\tau)/D^{\mathrm{IND}}(\tau) \approx 1.0$ для теплового потока.

# 5 Схема временных множителей

Схемы весовых множителей позволяют избирательно увеличивать плотность расчетных частиц, однако, их недостатком являются как трудности с выполнением законов сохранения, так и повышенное влияние повторных столкновений на точность макропараметров.

Альтернативный подход заключается в применении временных множителей, который, в общем случае, может использоваться совместно с весовыми множителями. При использовании временных множителей, время для разных компонент и/или в различных областях пространства течет с разной скоростью. Очевидно, этот подход применим только для стационарных задач.

Применение разного временного шага в разных частях расчетной области при решении стационарных задач предлагалось и раньше (19). Однако, при этом в пределах ячейки временной шаг оставался одинаковым для всех частиц.

В настоящей работе описан и исследован подход с использованием разного временного шага для разных расчетных частиц, находящихся в одной ячейки. Например, может использоваться разный временной шаг для различных компонент смеси.

Пусть $F_1$ и $F_2$ – фиксированные компонентные весовые множители, характеризующие количество реальных молекул, приходящихся на одну расчетную частицы. Введём $\mathrm{T}_1$ и $\mathrm{T}_2$ –временные множители, показывающие, какую долю от глобального расчетного времени «ощущает» частица. Функции распределения расчетных частиц двух сортов обозначим $f_1$ и $f_2$. Запишем уравнения Больцмана в этих переменных:

$$\frac{1}{\mathrm{T}_1}\cdot\frac{\partial f_1}{\partial t}+v_i\frac{\partial f_1}{\partial x_i}=F_1\cdot I_{11}[f_1,f_1]+F_2\cdot I_{12}[f_1,f_2],$$

$$\frac{1}{\mathrm{T}_2}\cdot\frac{\partial f_2}{\partial t}+v_i\frac{\partial f_2}{\partial x_i}=F_2\cdot I_{22}[f_2,f_2]+F_1\cdot I_{21}[f_2,f_1].$$

В стационарном случае, временные производные обращаются в ноль, и временные множители не должны изменить решение системы уравнений.

Эти же уравнения можно переписать в виде:

$$\frac{\partial f_1}{\partial t}+(\mathrm{T}_1 v_i)\frac{\partial f_1}{\partial x_i}=\frac{F_1}{\mathrm{T}_1}\cdot\mathrm{T}_1\mathrm{T}_1\cdot I_{11}[f_1,f_1]+\frac{F_2}{\mathrm{T}_2}\cdot\mathrm{T}_1\mathrm{T}_2\cdot I_{12}[f_1,f_2],$$
$$\frac{\partial f_2}{\partial t}+(\mathrm{T}_2 v_i)\frac{\partial f_2}{\partial x_i}=\frac{F_2}{\mathrm{T}_2}\cdot\mathrm{T}_2\mathrm{T}_2\cdot I_{22}[f_2,f_2]+\frac{F_1}{\mathrm{T}_1}\cdot\mathrm{T}_1\mathrm{T}_2\cdot I_{21}[f_2,f_1].$$

Отсюда можно заключить, что теперь роль весовых множителей при оценке макропараметров по пересечению границы (в т.ч. потоков частиц, импульса и энергии через поверхности, включая логику запуска частиц) будут играть соотношения $\frac{F_i}{\mathrm{T_i}}$. Это же касается и логики парных столкновений, с тем дополнением, что в вероятность столкновения пары дополнительно войдёт множитель $\mathrm{T_i T_j}$. При перемещении частиц, длительность временного шага учитывается с дополнительным коэффициентом $\mathrm{T_i}$. Макропараметры, оцениваемые по «фотографиям», по-прежнему, вычисляются с весами $F_i$.

Итак, если раньше частота столкновения каждой пары частиц вычислялась по формуле: $\nu_{ij}=\max(F_i,F_j)\cdot\frac{\sigma_{ij}(\mathrm{c}_{ij})\cdot\mathrm{c}_{ij}}{V}$, и параметры $i-$й частицы после столкновения изменялись с вероятностью $\frac{F_i}{\max(F_i,F_j)}$, то теперь частота вычисляется следующим образом: $\nu_{ij}=\mathrm{T}_i\mathrm{T}_j\cdot\max\left(\frac{F_i}{\mathrm{T}_i},\frac{F_j}{\mathrm{T}_j}\right)\cdot\frac{\sigma_{ij}(\mathrm{c}_{ij})\cdot\mathrm{c}_{ij}}{V}$, а вероятность изменения параметров частицы составляет $\frac{\frac{F_i}{\mathrm{T}_i}}{\max\left(\frac{F_i}{\mathrm{T}_i},\frac{F_j}{\mathrm{T}_j}\right)}$.

Отношения $\frac{F_i}{\mathrm{T_i}}$ будет называть ***комбинированными весами***.

В частном случае $\frac{F_i}{\mathrm{T_i}}=\frac{F_j}{\mathrm{T_j}}$, обе частицы всегда изменяют свои параметры после столкновения, такое сочетание компонентных и временных множителей будет называться сбалансированным.

Весовые и временные множители могут зависеть от координат. Если при этом всюду сохраняется сбалансированность, т.е. $\frac{F_i}{\mathrm{T_i}}=const$, то не требуется применения логики рождения и уничтожения частиц при их перемещении в пространстве. Так как уменьшение временного множителя после пересечения частицей какой-либо границы повлечет за собой уменьшение её эффективной скорости перемещения, то, в силу сохранения потока частиц, их плотность в области, где время замедлено, будет пропорционально выше.

Итак, если для всех пар частиц во всей расчетной области выполняется соотношение сбалансированности $\frac{F_i}{\mathrm{T_i}}=\frac{F_j}{\mathrm{T_j}}=const$, схема остается консервативной.

Впрочем, консервативность при использовании временных множителей имеет некоторую особенность. Физические величины сохраняются теперь только в среднем, и лишь после того, как достигнут стационарный режим. Например, если временной шаг зависит от координат, то, при сохранении количества расчетных частиц, физическая масса газа сохраняться не будет, так как расчетная частица представляет различное число реальных молекул, в зависимости от того, в какой части расчетной области она находится. Тем не менее, сохраняющиеся физические величины не подвержены случайным блужданиям, как в неконсервативных схемах, и будут колебаться вокруг равновесных значений, после того как достигнут стационарный режим.

Очевидный мотив применения временных множителей – избавление от неконсервативных схем весовых множителей, когда концентрации компонент сильно различаются. Однако, именно в этом случае схема может не дать выигрыша, так как скорость сбора статистики зависит от комбинированного веса, о чём будет подробно сказано ниже.

Другой, менее очевидный мотив – уменьшение числа расчетных частиц. Например, если в случае бинарной смеси сечение столкновения $\sigma_{12} \gg \sigma_{11}$, то, без применения временных множителей, количество расчетных частиц компоненты 1 при наличии любого количества компоненты 2 необходимо увеличить в $\frac{\sigma_{12}}{\sigma_{11}}$ раз по сравнению с требующимся для расчета только чистого газа 1, иначе вероятность повторного столкновения частиц компоненты 2 с одной и той же частицей компоненты 1 будет велика. С другой стороны, концентрация компоненты 2 может быть значительно ниже концентрации компоненты 1. В этом случае, имеет смысл использовать временные множители для замедления компоненты 2: $\mathrm{T}_1 = 1, \mathrm{T}_2 \sim \frac{\sigma_{11}}{\sigma_{12}}$. При этом повышения числа частиц компоненты 1 не требуется, т.е. не требуется уменьшать $F_1$. Если пропорционально увеличить число расчетных частиц компоненты 2, положив $F_2 = F_1 \cdot \frac{\sigma_{11}}{\sigma_{12}}$, то схема временных множителей будет сбалансированной, в то время как общее число частиц вырастет незначительно. Такой подход может быть полезен при исследовании движения примеси крупных органических молекул в составе легкого несущего газа, а также при исследовании движения газа, содержащего кластеры.

Впрочем, из-за того, что относительные скорости частиц пары в фазах перемещения и столкновения отличаются, особенно при ненулевой средней поступательной скорости потока, ход рассуждений, использованный ранее при выводе критерия $N_c\mathrm{Kn}_c$, не применим в первоначальном виде. Если принять предположение, что функция распределения близка к равновесной, а средняя поступательная скорость равна нулю для обоих компонент, то можно привести критерий к виду:

$$\frac{F_2\sigma_{12}h}{V} \ll 1 \qquad \rightarrow \qquad \sqrt{\frac{1+\frac{m_1}{m_2}}{1+\frac{m_1}{m_2}\cdot\left(\frac{\mathrm{T}_2}{\mathrm{T}_1}\right)^2}} \cdot \frac{F_2\sigma_{12}h}{V} = \sqrt{\frac{1+\frac{m_1}{m_2}}{\frac{1}{\mathrm{T}_2{}^2}+\frac{m_1}{m_2}\cdot\frac{1}{\mathrm{T}_1{}^2}}} \cdot \frac{W_2\sigma_{12}h}{V} \ll 1$$

Случай $\frac{m_1}{m_2} = \mathrm{M} \gg 1$ (тяжелая частица сталкивается с легкой): $\sqrt{\frac{1+\frac{1}{\mathrm{M}}}{\frac{1}{\mathrm{T}_1{}^2}+\frac{1}{\mathrm{M}}\cdot\frac{1}{\mathrm{T}_2{}^2}}} \cdot \frac{W_2\sigma_{12}h}{V} \ll 1$

Случай $\frac{m_1}{m_2} = \frac{1}{\mathrm{M}} \ll 1$ (легкая частица сталкивается с тяжелой): $\sqrt{\frac{1+\frac{1}{\mathrm{M}}}{\frac{1}{\mathrm{T}_2{}^2}+\frac{1}{\mathrm{M}}\cdot\frac{1}{\mathrm{T}_1{}^2}}} \cdot \frac{W_2\sigma_{12}h}{V} \ll 1$

Из приведенных выше выражений видно, что при сбалансированной схеме уменьшению вероятности повторных столкновений между молекулами тяжелой и лёгкой компоненты способствует прежде всего уменьшение временного множителя для тяжелой компоненты (с сопутствующим уменьшением статистического веса и увеличением числа тяжелых расчетных частиц), в то время как влияние временного множителя лёгкой компоненты слабое.

Если же сбалансированности не требуется, то, при одних и тех же величинах $F_i$, опять же, изменение отношения временных множителей оказывает более сильное влияние на вероятность повторного столкновения тяжелой молекулы о лёгкую, а не наоборот.

В общем случае, вероятность повторного столкновения выбранной частицы 1 с частицей 2 при использовании временных множителей следующая: $\frac{F_2\sigma_{12}h}{V}\cdot\frac{\|\vec{v_1}-\vec{v_2}\|}{\|\vec{v_1}\mathrm{T}_1-\vec{v_2}\mathrm{T}_2\|}$, откуда видно, что наименее благоприятным при $\mathrm{T}_1 \neq \mathrm{T}_2$ является такое сочетание скоростей, при котором $\vec{v_1}\mathrm{T}_1 \approx \vec{v_2}\mathrm{T}_2$. Вследствие этого, следует проявлять осторожность в выборе временных множителей при расчете течений с сильным скольжением компонент, чтобы случайно не попасть в «резонанс», при котором значительная доля физически движущихся с разной скоростью молекул будут в расчете длительное время оставаться в непосредственной близости друг от друга, двигаясь параллельно, будучи склонными к повторным столкновениям. Последствия такого совпадения специально не исследовались, но самопроизвольно не возникло ни разу при испытаниях и использовании метода. Подобные подозрения можно высказать на тему изменения условий устойчивости стационарного решения модифицированной системы уравнений. Во всех выполненных тестах временные множители давали решение, близкое к решению, полученному без их использования.

Принципиальным недостатком временных множителей является неприменимость к нестационарным задачам. Другим важным недостатком является уменьшение скорости сходимости к стационарному решению. Так как для некоторых компонент время течет медленнее, то и стационарное решение, скорее всего, будет достигнуто позднее. Говоря проще, замедленная частицы должна испытать то же количество столкновений, а для этого требуется больше временных шагов.

Вернёмся к скорости набора статистики. При замедлении времени плотность расчетных частиц пропорционально вырастает (если схема сбалансирована), и это увеличивает размер выборки. Однако, увеличивается время, за которое параметры частиц существенно меняются (что составляет порядка времени между столкновениями). В итоге, при кратно увеличившемся числе расчетных частиц, скорость статистической сходимости макропараметров оказывается практически такой же, как и без использования временных множителей. Статистическая сходимость параметров, оцениваемых по пересечению, так же остаётся прежней. Схема компонентных весовых множителей, например, не обладает подобным недостатком – повышение плотности частиц какой-либо компоненты за счет снижения её статистического веса, как правило, ускоряет статистическую сходимость её макропараметров как при оценке по «фотографиям», так и по оценке по пересечениям.

Последний недостаток несущественен, если сечение столкновения замедляемой компоненты с остальными велико, либо если её масса очень мала (и велика тепловая скорость), так как при отсутствии временных множителей пришлось бы уменьшать временной шаг для всей системы, а не только для отдельной компоненты.

## 5.1 Качественное испытание временных множителей

Численное испытание временных множителей на примере задачи обтекания плоской пластины смесью гелий + ксенон (4) обнаружило отличное совпадение как макропараметров обеих компонент, так и энергетического спектра молекул ксенона, достигших пластины, при условии, что критерий $N_c\mathrm{Kn}_c \gg 1$.

Краткое описание задачи: двумерная постановка, плоская пластина установлена перпендикулярно набегающему сверхзвуковому потоку газовой смеси гелий + ксенон. Сверхзвуковой поток имеет температуру 29.27 K и скорость 672.5 м/с, что соответствует числу Маха M = 5 для чистого гелия и температуре торможения $T_0$ = 273.15 K. Для столкновений использована модель VSS с параметрами из (1). Варьировались состав смеси, число Кнудсена набегающего потока $\mathrm{Kn}_\infty$, коэффициент поглощения тяжелой компоненты на пластине. Число Кнудсена определяется как отношение длины свободного пробега в невозмущенном потоке в ширине пластины.

Для демонстрации преимущества схемы временных множителей были выбраны следующие условия: 1) $\mathrm{Kn}_\infty = 3$, $C_{Xe}/C_{He} = 0.1$ и 2) $\mathrm{Kn}_\infty = 0.3$, $C_{Xe}/C_{He} = 0.025$, при этом в обоих случаях отсутствует поглощение ксенона на пластине. Число частиц было выбрано таким, чтобы значение критерия $N_c\mathrm{Kn}_c$ для столкновений гелий-гелий составляло единицу. Вычисления проводились тремя методами: стандартным DSMC, DSMC с временным множителем для ксенона 0.15 и таким же компонентным весовым множителем (сбалансированная схема), и DSMC с применением только компонентного весового множителя для ксенона 0.15. На **Рис. 5.1-1** приведены полученные профили макропараметров в плоскости симметрии. **Чёрные** кривые соответствуют решению, полученному при $N_c\mathrm{Kn}_c \gg 1$, **синие** – стандартной схеме DSMC, **красные** – DSMC с применением временных множителей, **зеленые** – DSMC с применением компонентных весовых множителей. Сравнение макропараметров гелия показывает, что временные множители не дают преимущества перед весовыми множителями в определении макропараметров гелия, хотя оба дают лучший результат, чем стандартная схема DSMC. Однако, на макропараметрах ксенона различие более существенное: можно видеть, что **красные** кривые (результат временных множителей) заметно лучше соответствует **чёрной** (образцовой) кривой, чем **зеленая** кривая (весовые множители). Кроме того, в области невозмущенного потока зеленая кривая наглядно демонстрирует артефакты, характерные для весовых множителей (завышение температуры ксенона).

На **Рис. 5.1-2** приведены энергетические спектры частиц ксенона, столкнувшихся с пластиной. Снова можно видеть, что схема временных множителей даёт лучший результат. Так, для режима $\mathrm{Kn}_\infty = 3$, **зеленая** кривая показывает несколько смещенное положение максимума, а **синяя** кривая, кроме того, показывает завышенный поток высокоэнергетичных частиц при уменьшении потоков частиц низких и средних энергий.

Объяснить результаты численных экспериментов можно следующим образом. При использовании стандартной схемы DSMC, значения критерия $N_c\mathrm{Kn}_c$ составляют 0.33 для столкновений гелий-ксенон и 0.17 для столкновений ксенон-ксенон. При введении только компонентных весовых множителей, $N_c\mathrm{Kn}_c$ для столкновений ксенон-ксенон повышается до 1.1, для столкновения гелия с ксеноном – до 2.2, но для столкновений ксенона с гелием – остаётся по-прежнему 0.33. Если же используются ещё и временные множители, то значение критерия для столкновений ксенона с гелием тоже повышается до 2.2, т.е. доля повторных столкновений ксенона с одной и той же частицей гелия уменьшается. Этого нельзя достигнуть только уменьшением весового множителя ксенона (т.е. увеличения числа расчетных частиц примеси). Из приведенных соображений можно понять, почему применение временных множителей, по сравнению с компонентными весовыми множителями, уменьшило отклонение макропараметров ксенона, но почти не повлияло на макропараметры гелия. Так как в данном случае роль играет уменьшенная частота повторных столкновений ксенона с гелием, то, улучшение макропараметров ксенона происходит за счёт более точного описания его взаимодействия с несущим газом. Однако, точность описания несущего газа осталась прежней, и параметры несущего газа могут измениться только косвенным образом – через изменение параметров ксенона, которое не особо велико.

В Табл. 5-1 представлены измеренные значения коэффициента сопротивления и теплового потока, которые также демонстрирует наилучшее согласие именно при использовании схемы временных множителей.

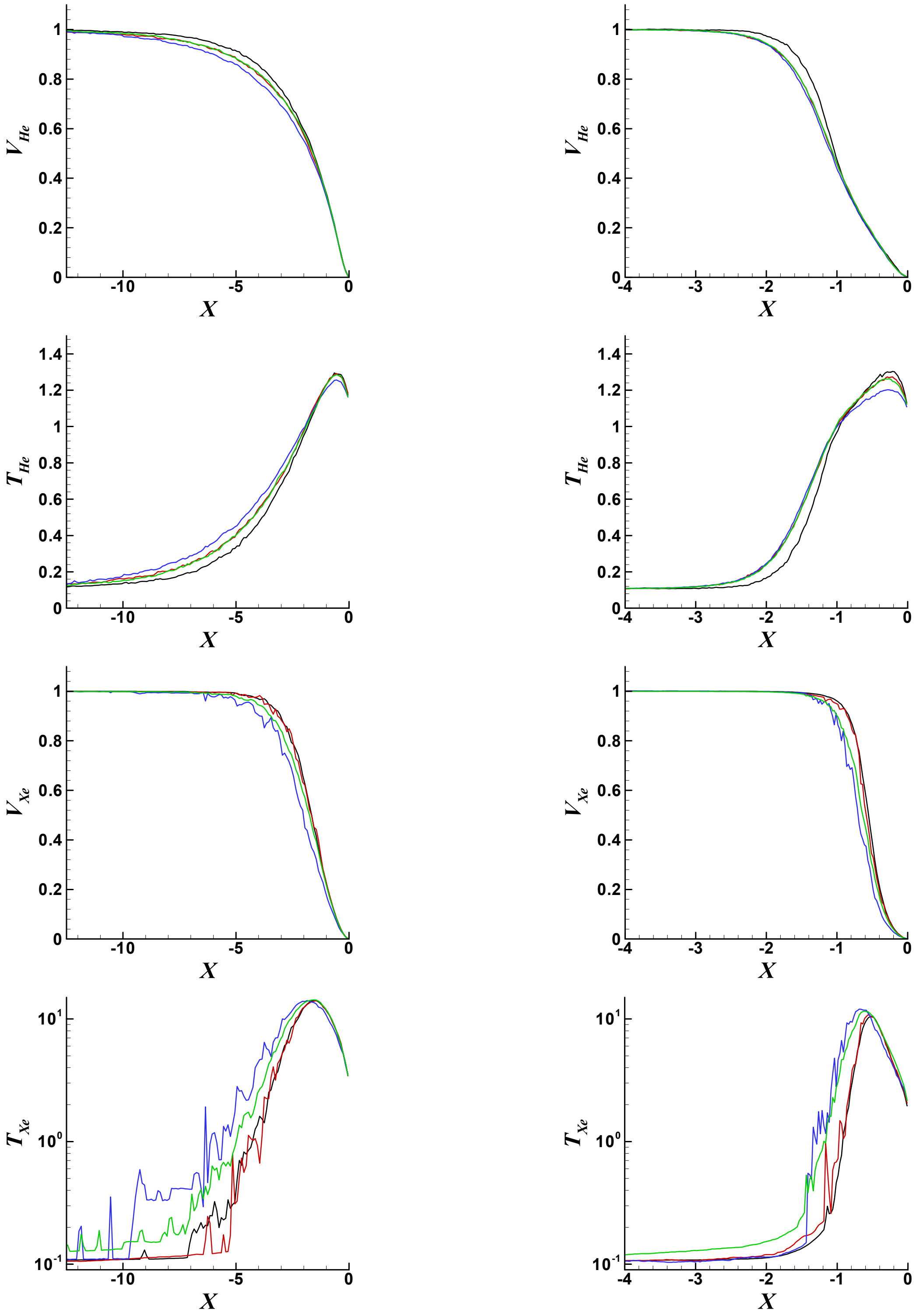

**Рис. 5.1-1**. Сравнение макропараметров (скорости и температуры обеих компонент) в плоскости симметрии, полученных тремя модификациями метода DSMC. **Черная** кривая – эталонный расчет, **красная** – использована сбалансированная схема временных множителей, **зеленая** – использованы компонентные весовые множители, **синяя** – стандартный DSMC.
Слева – режим Kn∞=3, $C_{Xe}/C_{He} = 0.1$, справа – Kn∞=0.3, $C_{Xe}/C_{He} = 0.025$.

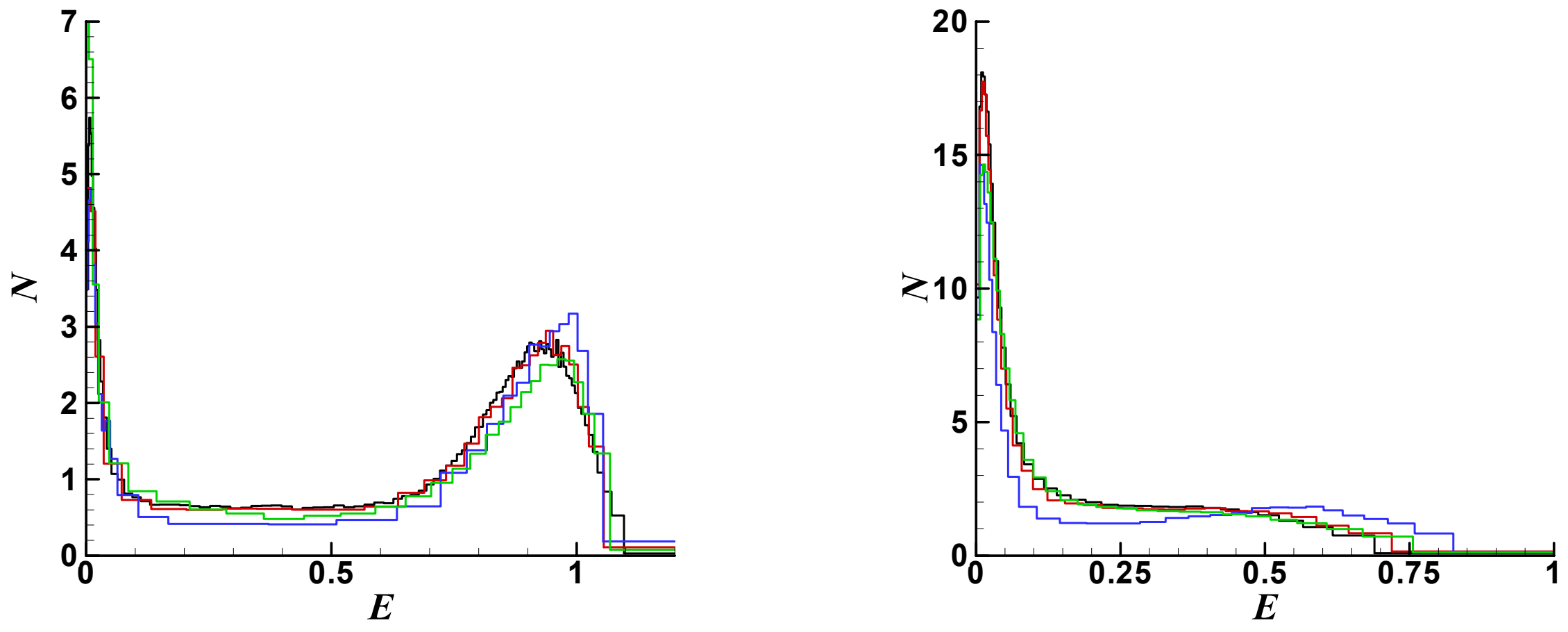

**Рис. 5.1-2**. Сравнение энергетических спектров молекул ксенона, столкнувшихся с пластиной, полученных тремя модификациями метода DSMC. **Черная** кривая – эталонный расчет, **красная** – использована сбалансированная схема временных множителей, **зеленая** – использованы компонентные весовые множители Бёрда, **синяя** – стандартный DSMC.
Слева – режим Kn∞=3, $C_{Xe}/C_{He} = 0.1$, справа – Kn∞=0.3, $C_{Xe}/C_{He} = 0.025$.

Табл. 5-1. Сравнение коэффициентов сопротивления и теплового потока, полученных тремя модификациями метода DSMC.

| | Kn∞=3, $C_{Xe}/C_{He} = 0.1$. | | Kn∞=0.3, $C_{Xe}/C_{He} = 0.025$. | |
|---|---|---|---|---|
| | $C_x$ | $Q$ | $C_x$ | $Q$ |
| Эталонный расчет, $N_c \mathrm{Kn}_c \gg 1$. | 2.10 | 0.65 | 1.83 | 0.26 |
| $\frac{F_2}{F_1} = \frac{\mathrm{T}_2}{\mathrm{T}_1} = 0.15, \mathrm{T}_1 = 1.$ | 2.11 | 0.66 | 1.87 | 0.27 |
| $\frac{F_2}{F_1} = 0.15, \mathrm{T}_2 = \mathrm{T}_1 = 1.$ | 2.13 | 0.66 | 1.87 | 0.27 |
| $\frac{F_2}{F_1} = \mathrm{T}_2 = \mathrm{T}_1 = 1.$ | 2.15 | 0.68 | 1.88 | 0.31 |

## 5.2 Количественное испытание при замедлении легкой компоненты

Для такого испытания идеально подходит задача Фурье о теплопередачи между двумя параллельными пластинами. Как известно, добавление сравнительно небольшого количества примеси легкой компоненты к тяжелому газу существенно повышает теплопроводность смеси. В качестве модельного газа в данном исследовании использовалась смесь 10 % He + 90 % Xe, при этом теплопроводность такой смеси примерно в 1.5 раза выше теплопроводности ксенона. Число Кнудсена, оцениваемое по сечению столкновения He-He, средней плотности смеси и расстоянию между пластинами, составляет 0.1 (≈0.017 по сечению Xe-Xe). Использовались VSS параметры Бёрда для моделирования столкновений, при этом температуры пластин отличаются в 4 раза и составляют 137 K и 546 K.

Тепловая скорость гелия в ≈5.7 раз превышает тепловую скорость ксенона. Без использования временных множителей, необходимо либо использовать очень маленький временной шаг, чтобы частицы гелия не пролетали несколько ячеек за временной шаг, либо мириться с ошибкой, связанной со слишком высоким временным шагом (хотя частично это можно компенсировать кратным повышением общего числа частиц). Если же замедлить легкий газ в 2-3 раза, применив временные множители, то разница в тепловой скорости уменьшится, и можно использовать увеличенный временной шаг при той же величине ошибки.

На **Рис. 5.2-1** приведены зависимости ошибки теплового потока от временного шага. Как можно видеть, замедление легкой компоненты в 1.7 раз заметно снижает ошибку при том же временном шаге (значение выбиралось таким, чтобы частоты столкновений для обоих компонент стали близки). Ещё сильнее эффект проявляется при замедлении в 3 раза. При замедлении гелия в 3 раза и малых временных шагах преимущество временных множителей не наблюдается. Однако, последнее может быть лишь следствием статистического разброса, так как абсолютная величина ошибки очень мала, что затрудняет численное исследование.

Замедление гелия в 1.7 раз позволяет использовать на 25 % больший временной шаг, оставаясь в рамках 1 % погрешности по тепловому потоку, а замедление в 3 раза позволяет увеличить временной шаг уже на 50 %. При этом, в первом случае общее число частиц увеличивается всего на 7 %, а в последнем – на 20 %, причём общее количество столкновений в единицу расчетного времени не меняется с увеличением числа частиц. При одном и том же временном шаге, замедление гелия в 1.7 раз уменьшает ошибку теплового потока на 25 %, а замедление в 3 раза – на 35 %.

Проведенные исследования показывают, что использование временны множителей для замедления легкой компоненты может быть оправдано при моделировании смесей, в которых легкая компонента составляет малую примесь.

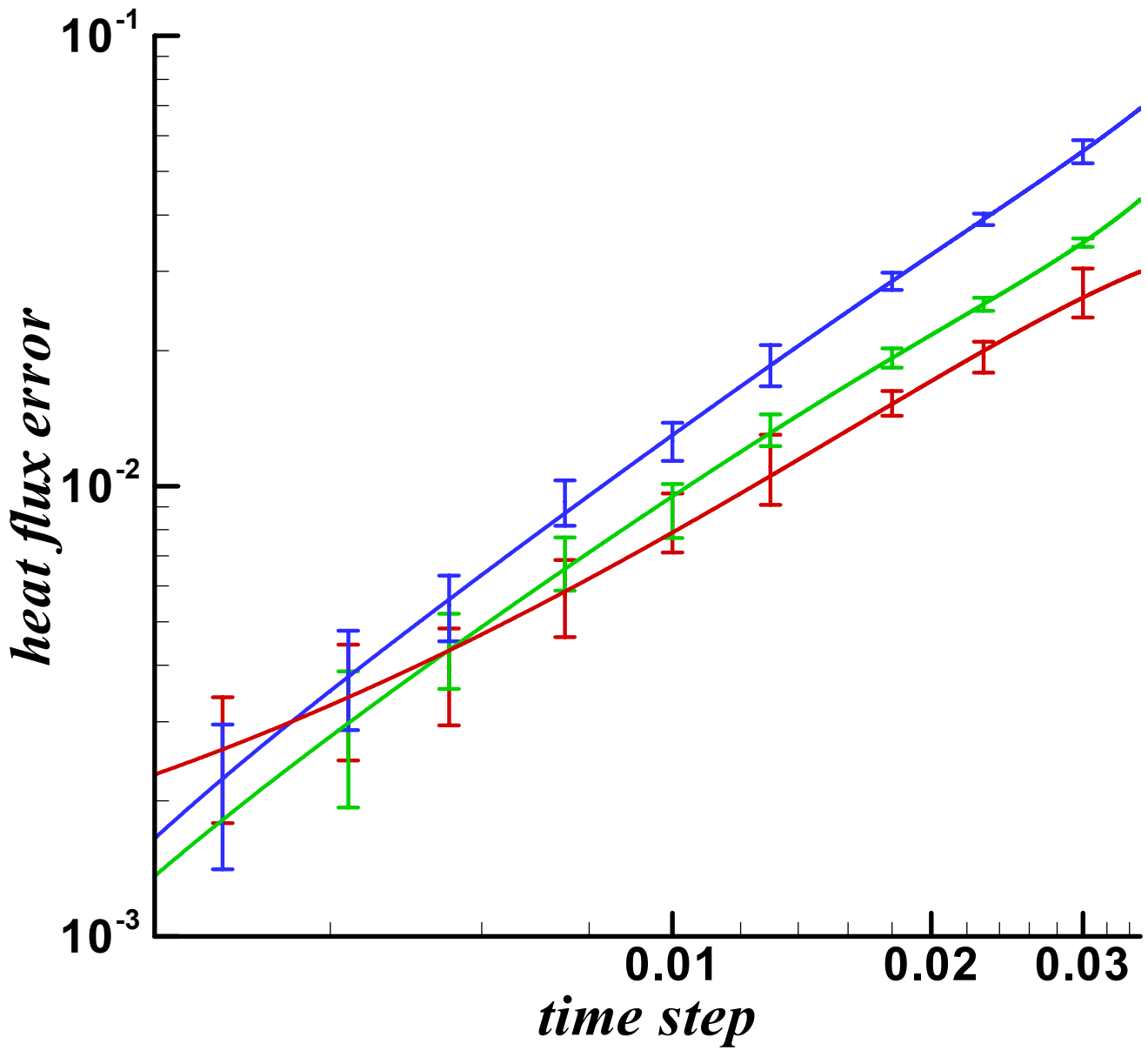

**Рис. 5.2-1.** Ошибка теплового потока между пластин с изменением временного шага. **Синяя** кривая – временные множители не используются, **зеленая** кривая – гелий замедлен в 1.7 раза, **красная** кривая – гелий замедлен в 3 раза.

## 5.3 Количественное исследование при замедлении тяжелой компоненты

Поставить задачу, в которой небольшая примесь тяжелого газа существенно меняет какой-либо легко измеримый и при этом чувствительный к ошибкам счета параметр, не так просто. В качестве основы взята задача о движении тяжелой примеси через сжатый слой. При этом измеряемым параметром является средняя энергия частиц тяжелой примеси, достигающей пластины. Сжатый слой моделируется как газ, заключенный между двумя изотермическими пластинами. Для несущего газа установлено диффузное отражение на обеих пластинах. Тяжелая примесь диффузно отражается от левой пластины и полностью поглощается на правой. Однако, при поглощении тяжелой частицы на правой пластине, с левой пластины в сжатый слой запускается тяжелая частица, имеющая скорость, соответствующую предельно достижимой скорости (M=∞) гелия. Такой подход позволяет держать общее количество тяжелых частиц в расчетной области постоянным. Несмотря на несколько условную постановку граничных условий и почти двукратное увеличение числа столкновений тяжелых частиц со сжатым слоем по сравнению с двумерной моделью, данная постановка сохраняет многие свойства исходной задачи, например, слабое влияние концентрации тяжелой примеси на их среднюю энергию при достижении мишени. При этом простая одномерная постановка задачи и фиксированное число частиц в расчетной области позволяют эффективно исследовать зависимость ошибки величины средней энергии от полного числа частиц.

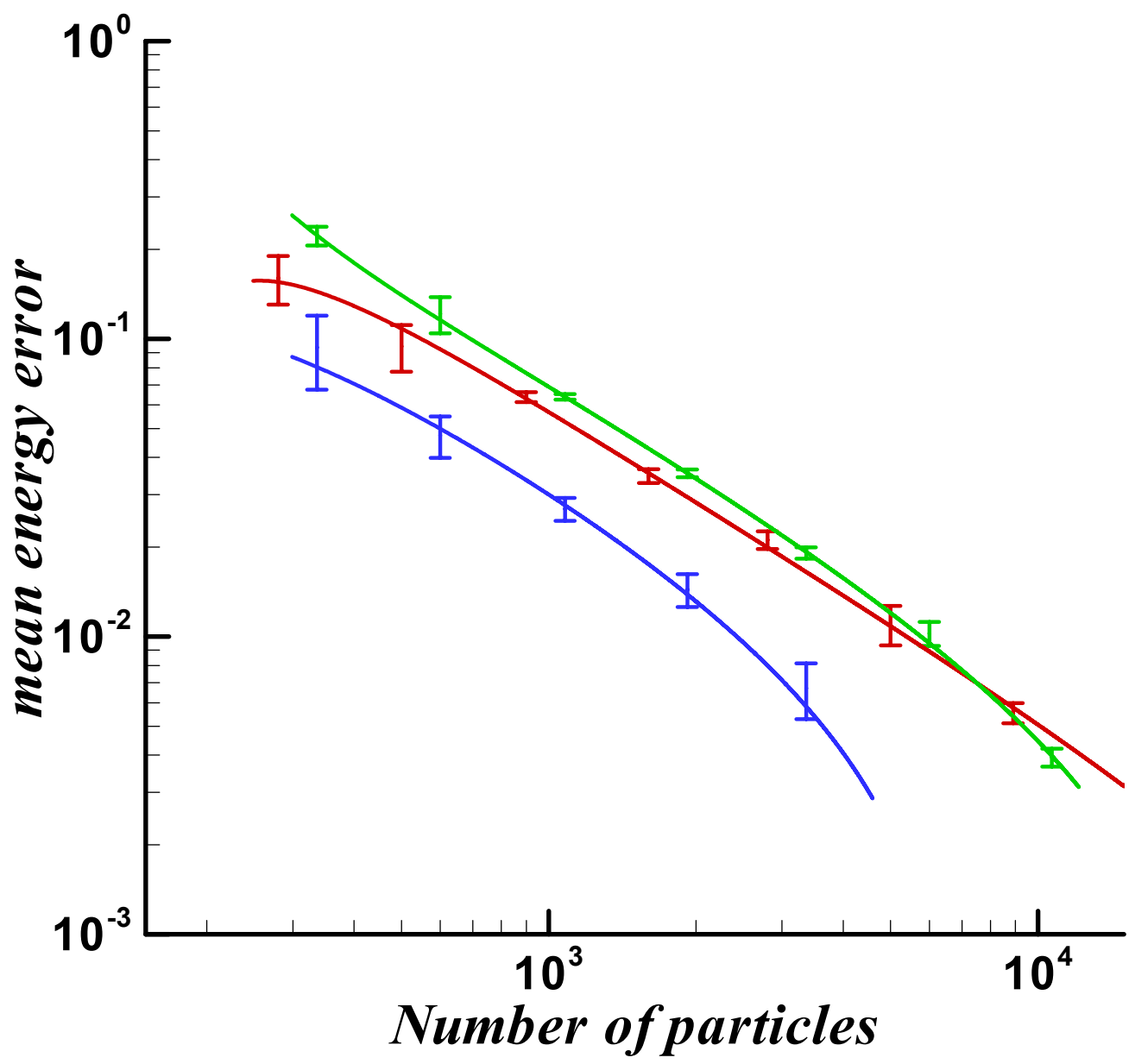

**Рис. 5.3-1**. Зависимость ошибки средней энергии от числа частиц в расчетной области. **Красная** кривая – временные и компонентные множители не используются, **зеленая** кривая – использованы компонентные весовые множители Бёрда, **синяя** – использованы сбалансированные временные множители.

Использовалась модель столкновений VSS с параметрами Бёрда. Температура пластин – 273 K. Концентрация ксенона – 10 %. Число Кнудсена по средней плотности смеси, сечению He-He и расстоянию между пластинами составляет 0.04. Расчет проводился, используя 3 подхода: без использования компонентных и весовых множителей, с использованием компонентных весовых множителей Бёрда при 3-кратном увеличении числа частиц ксенона, и с использованием сбалансированной схемы с 3-кратным увеличением числа частиц ксенона и таким же его замедлением.

На **Рис. 5.3-1** показана зависимость ошибки от числа частиц в расчетной области для 3 перечисленных подходов. При использовании временных множителей либо компонентных множителей Бёрда, в обоих случаях число частиц ксенона повышается 3-кратно, а общее число частиц – на 20 %.

Сравнение кривых показывает, что применение компонентных весовых множителей Бёрда не даёт какого-либо преимущества. Результат вполне предсказуем, так как их применение не уменьшает число повторных столкновений молекул ксенона с одной и той же молекулой гелия, но именно эти повторные столкновения отрицательно сказываются на точности расчета потери энергии в сжатом слое. Кроме того, схема весовых множителей Бёрда не консервативна, что может также давать вклад в погрешность.

Однако, замедление ксенона уменьшает ошибку величины энергии более чем в 2 раза. Аналогично, ту же самую точность значения энергии можно получить при более чем 2-кратной экономии числа частиц.

Следует отметить, что сечение столкновения He-Xe приблизительно в 3 раза превышает сечение He-He. При ещё больших сечениях столкновения между несущим и примесным газом, например, при расчете сложных органических молекул либо кластеров, ожидаемый выигрыш в числе расчетных частиц ещё выше.

# 6 Схемы весовых множителей для осесимметричных задач

*<u>Схема радиальных весовых множителей</u>* предназначена для расчета осесимметричных течений, цель использования – повысить плотность расчетных частиц в приосевой зоне. Классическая схема весовых множителей сопоставляет частице вес, связанный с расстоянием от оси. После того, как частица удалилась от оси – она либо уничтожается, либо увеличивает свой вес. Если же частица движется к оси, то её необходимо размножить – превратить в несколько частиц. Однако, параметры этих размножившихся частиц будут совпадать, что является недостатком схемы.

## 6.1 Особенности классической схемы Бёрда

Остановимся подробнее на тестовой задаче (1) обтекания цилиндра, ориентированного торцом в сторону сверхзвукового потока. Для этой задачи, Бёрд предоставил и реализацию расчетной программы на фортране, в виде файла DSMC2A, прилагающегося к его монографии, что позволяет самостоятельно повторить расчёт.

Цилиндр радиусом 1 см обтекается с левого торца потоком аргона, имеющим скорость 1 км/с, плотность $10^{21}$ м$^{-3}$, температуру 100 K. Расчетная область имеет радиус 3 см и простирается на 2 см в обе стороны от торца цилиндра. Размер ячейки сетки составляет 0.5 мм, временной шаг составляет 1 мкс. Число расчетных частиц лежит в диапазоне 30 000 – 40 000 шт. Температура окружности цилиндра составляет 300 K, температура торца – 1000 K, что довольно близко к температуре торможения потока. Используется модель VHS с параметрами (1).

В тестовой программе реализована схема весовых множителей. До радиуса 4 мм веса всех частиц одинаковы, на бо́льших радиусах вес пропорционален расстоянию до оси. Используется дополнительный буфер для задержки копий, ёмкостью 200 частиц. Каждая ячейка разделена на 2x2 подъячеек.

Изучая результат расчёта в виде изолиний плотности и температуры (**Рис. 6.1-1**), легко заметить излом изолиний в сжатом слое перед цилиндром. Излом располагается как раз приблизительно на расстоянии 4 мм от оси (см. изолинию плотности F).

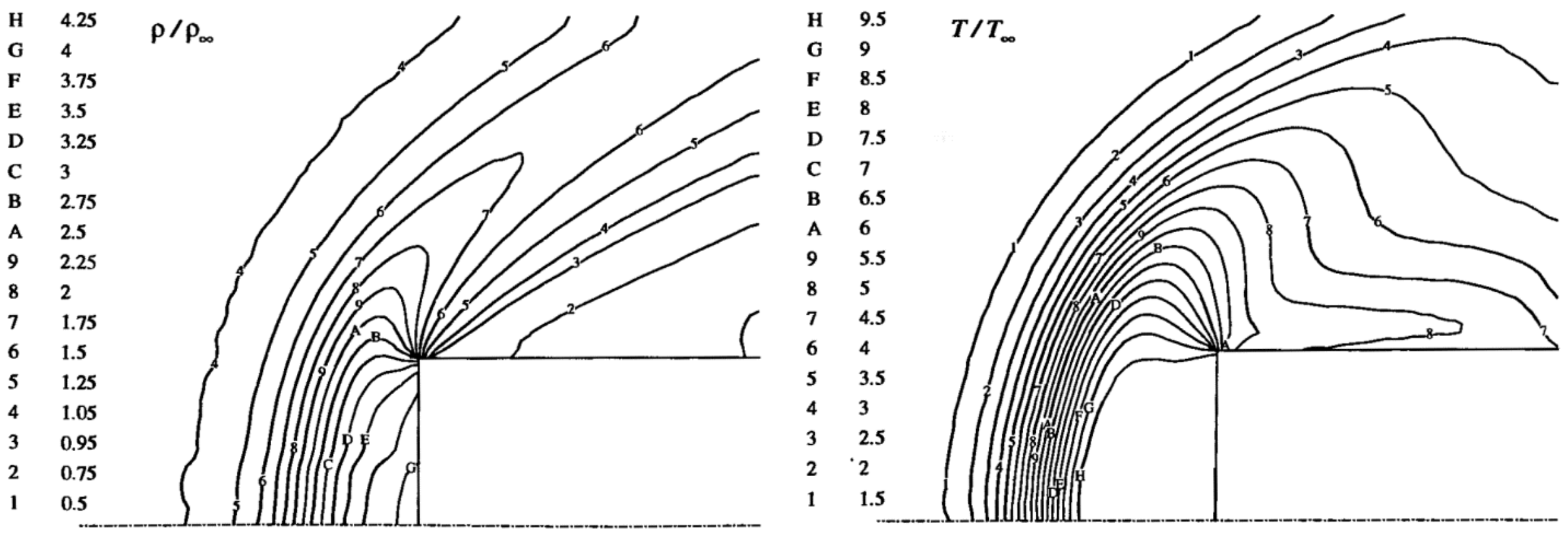

**Рис. 6.1-1.** Изолинии плотности (слева) и температуры (справа) по данным (1)

В работе (23) изучается влияние весовых множителей на примере этой же задачи. При этом весовые множители постоянны в пределах каждой ячейки. Авторы обнаружили аналогичный излом изолиний. Кроме того, если вводить весовые множители вплоть до оси, то вблизи оси развиваются ярко выраженные артефакты счета (**Рис. 6.1-2**). Из этого авторы делают вывод, что применение радиальных весовых множителей не даёт никакого выигрыша.

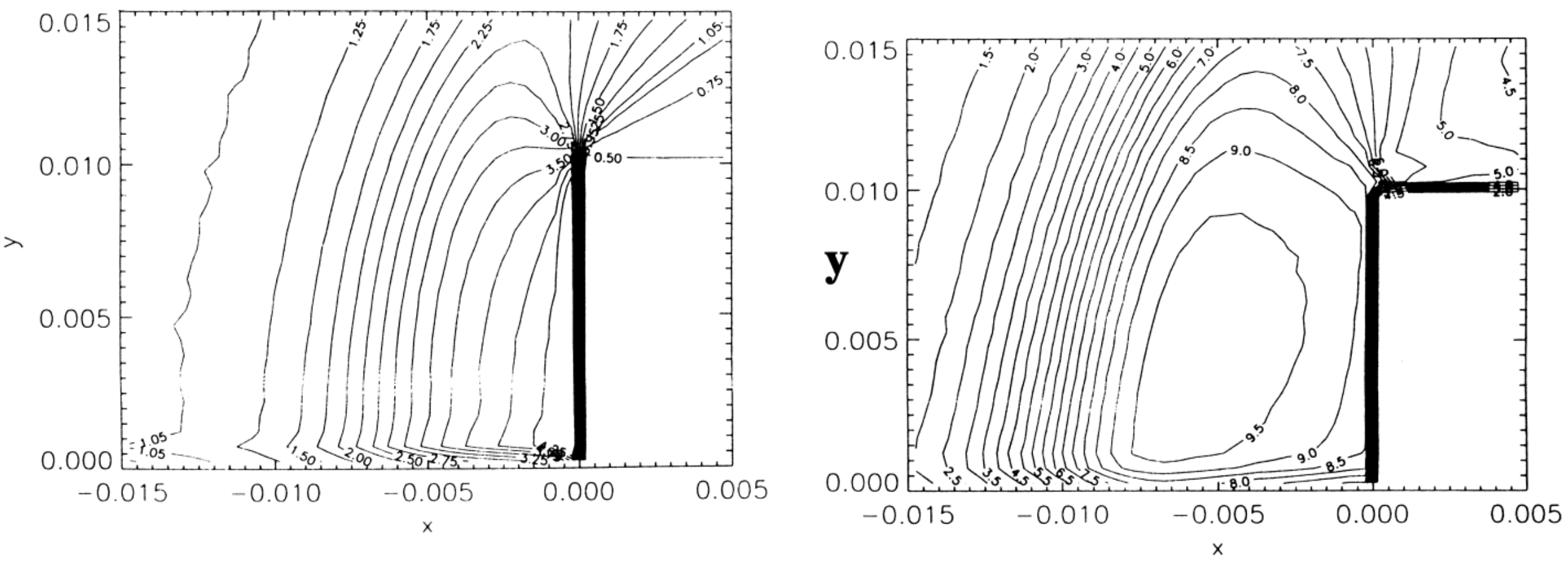

**Рис. 6.1-2.** Изолинии плотности (слева) и температуры (справа) по данным (23) при использовании логики пространственных весовых множителей во всём пространстве.

Мотивацией к поиску источника артефактов весовых множителей, реализующихся в расчётах Бёрда, послужило обнаружение несовпадения с результатами, полученных методом мажорантной частоты.

Метод мажорантной частоты не генерировал подобные артефакты, поэтому результаты не совпадали. Это можно увидеть на **Рис. 6.1-3**, где **красные** изолинии соответствуют результатам, полученным при использовании схемы MFS, **коричневые** изолинии соответствуют результатам, полученным по схеме NTC.

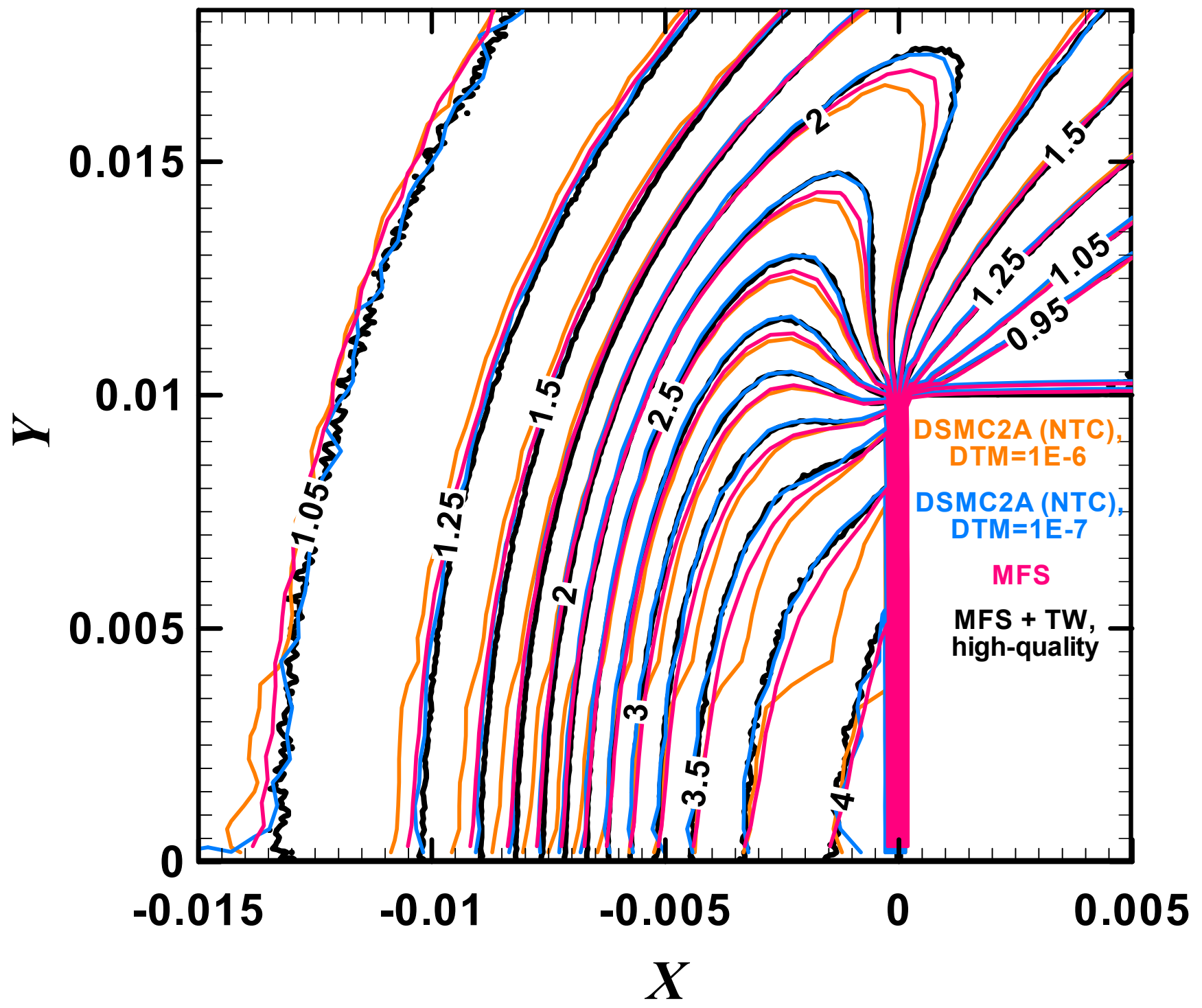

**Рис. 6.1-3**. Сравнение изолиний плотности. **Красная** кривая – результат схемы MFS, **коричневая** кривая – результат схемы NTC, **голубая** кривая – результат схемы NTC при уменьшенном временном шаге, **чёрная** кривая – результат повышенной точности, полученный при заведомо избыточной дискретизации.

В связи с этим, в процессе разработки новых схем весовых множителей пришлось подробно изучить природу этих артефактов. Для этого в программе Бёрда варьировались различные параметры. Отключение логики подъячеек, отключение защиты от выбора одинаковых частиц, увеличение размера буфера копий до 2000 частиц – всё это практически не влияло на результат. Однако, после уменьшения временного шага в 10 раз (до 0.1 мкс), артефакты счёта в виде изломов изолиний исчезли (**Рис. 6.1-3**, **голубые** изолинии), и результаты, полученные программой Бёрда, хорошо совпали с результатами, полученными методом мажорантной частоты при шаге 1 мкс. Следует заметить, что временной шаг 1 мкс слишком велик: среднее время между столкновениями при максимальной плотности (в сжатом слое) составляет около 0.65 мкс.

Замена схемы столкновений с NTC на MFS – в данном случае, фактически, единственное существенное отличие логики программы. Это говорит в пользу того, что схема MFS намного более устойчива к артефактам при использовании весовых множителей, по сравнению со схемой NTC.

С целью определить влияние выбора параметров на возможное расхождение результатов, был проведен также расчёт при заведомо корректных параметрах: размером ячейки 0.08 мм (около 1/6 от минимальной длины свободного пробега), временным шагом 70 нс (около 1/9 от минимального среднего интервала между столкновениями) и числом частиц порядка 635 000, что соответствует $N_C Kn_C > 100$.. Кроме того, при этом использовалась схема **tw** (комбинированные радиальные множители, будет описана далее), лучше всего подходящая для решения стационарных осесимметричных задач. На **Рис. 6.1-3** этому результату соответствуют **чёрные** изолинии. Как видно, дальнейшее повышение дискретизации лишь незначительно сказалось на форме изолиний.

Может показаться, что улучшение дискретизации вовсе не даёт никакого выигрыша. Однако, сравнение теплового потока на торец цилиндра показывает снижение на 12 %, что может быть существенным увеличение точности для некоторых приложений.

## 6.2 Размножение частиц при столкновениях

Альтернативой размножению частиц при перемещении может быть размножение частиц во время столкновения, предлагающееся автором настоящей работы.

Рассмотрим простейшую из таких схем. Каждой столкновительной ячейке присваивается характерный вес, связанный с ее расстоянием от оси. Частицам тоже присваивается вес. При перемещении частиц в ячейку с бóльшим весом, она уничтожается, либо приобретает вес ячейки, аналогично классической схеме радиальных весовых множителей. Таким образом, вес каждой частицы в ячейке всегда больше или равен целевому для ячейки значению: $p_c$. Если веса частиц в ячейке различны, то, в столкновительной фазе, частица большего веса пропорционально имеет больше шансов участвовать в столкновениях:

$$\nu_{ij} = \frac{p_i p_j \sigma_{ij}(c_{ij}) \cdot c_{ij}}{p_c V/F}$$

В результате столкновения образуется пара частиц, имеющих вес, равный $p_c$, при этом вес обоих исходных частиц тоже уменьшается на $p_c$. Если после столкновения пары вес какой-либо частицы станет меньше $p_c$, то частица снова либо уничтожается, либо приобретает вес ячейки.

Хотя в этом методе увеличиваются затраты на расчет столкновений, при этом несколько уменьшаются флуктуации числа частиц. Это достигается за счет того, что частицы, приблизившиеся к оси и затем снова удалившиеся от нее, не испытав при этом столкновений, не испытывают уничтожения или умножения. В остальном метод практически эквивалентен классическому методу размножения частиц при перемещении – в данном случае одна частица большого веса ведёт себя аналогично набору тождественных частиц с весом $p_c$, которые образовались бы при размножении в фазе перемещения.

## 6.3 Пакетное размножение частиц

Автором настоящей работы предложен ещё один подход к размножению частиц. Он позволяет уменьшить число столкновительных тестов по сравнению с предыдущим подходом, увеличив эффективность. Частота столкновений пары вычисляется по формуле:

$$\nu_{ij} = \frac{\max(p_i,\ p_j)\sigma_{ij}(\mathrm{c}_{ij}) \cdot \mathrm{c}_{ij}}{V/F}$$

При этом в результате столкновения должны, исходя из общих соображений, образоваться две частицы с весом $\min(p_i,\ p_j)$ каждая, и на эту же величину должен быть уменьшен вес исходных частиц, в результате чего, всегда одна из исходных частиц уничтожается. Однако, вместо образования одной новой пары частиц с весом $p_* = \min(p_i,\ p_j)$, производится разыгрывание столкновения несколько раз подряд, при этом образуются несколько частиц меньшего веса. Так как процесс столкновений стохастичен, размножаемые таким образом частицы обладают меньшей взаимосвязью параметров, чем при прямом размножении какой-либо частицы.

Алгоритм многократных столкновений заключается в следующем. Сначала вычисляется число частиц, которые должны появиться в результате столкновений. Пусть в результате столкновения частицы должны иметь веса $\acute{p}_\iota \le p_i$ и $\acute{p}_J \le p_j$. Тогда ожидаемое количество новых частиц должно составлять $N_i = p_*/\acute{p}_\iota, N_j = p_*/\acute{p}_J$. Кроме того, избыточный вес исходной частицы (пусть для определенности это будет частица $i$, в то время как частица $j$ полностью теряет свой вес и уничтожается) после уменьшения на $p_*$ составляет $\delta p$. Также введем величину $\delta p_* = \acute{p}_\iota \cdot \mathrm{frac}\, N_i$, характеризующую «дробную» часть веса $p_*$, по которой и будем определять, в большую или меньшую сторону округлять $N_i$. Рассмотрим четыре случая:

- Случай $\delta p = 0$. Исходная частица безусловно уничтожается. В результате, частица с «дробным» весом $\delta p_* < \acute{p}_\iota$ может породить 0 или 1 частицу. В этом случае дробную часть $N_i$ следует интерпретировать как вероятность округления $N_i$ в бóльшую сторону.
- Случай $\delta p \ge \acute{p}_\iota$. Исходная частица безусловно выживает, имея при этом вес $\delta p$. В остальном, этот случай полностью аналогичен предыдущему
- Случай $\delta p + \delta p_* \le \acute{p}_\iota$. В этом случае из рассматриваемых двух частиц «дробного» веса могут породиться в сумме 0 или 1 частица, причем вероятность последнего $\frac{\delta p + \delta p_*}{\acute{p}_\iota}$. При этом, вероятности выживания каждой из двух частиц должны соотноситься так же, как их веса.
- Случай $\acute{p}_\iota < \delta p + \delta p_* \le 2\acute{p}_\iota, \delta p < \acute{p}_\iota$. Порождаются 1 или 2 частицы. Вероятность последнего $\frac{\delta p + \delta p_*}{\acute{p}_\iota} - 1$.Случай, когда порождается одна частица, в остальном аналогичен предыдущему.

Алгоритм выглядит сложным, но, технически, его можно свести лишь к двум случаям:

- Случай $\delta p \ge \acute{p}_\iota$. Исходная частица безусловно выживает, имея при этом вес $\delta p$. К дробной величине $N_i$ прибавляется равномерно распределенное в интервале 0..1 случайное число, после чего отбрасывается дробная часть.
- Случай $\delta p < \acute{p}_\iota$. Сперва определим, выживет ли исходная частица. Вероятность этого равна $\frac{\delta p}{\acute{p}_\iota}$. Введем переменную $k$, которая равна 0, если исходная частица выживает, либо 1, если исходная частица уничтожается. Теперь разыграем вероятность выживания второй частицы. Если $\left(k - \frac{\delta p}{\acute{p}_\iota}\right) \cdot \mathrm{RF} < \frac{\delta p_*}{\acute{p}_\iota} + k - 1$, то вторая частица выживает, здесь $\mathrm{RF}$ – случайное число 0..1. От $N_i$ сперва отбрасывается дробная часть, после чего, если вторая частица выжила, к $N_i$ прибавляется единица.

$N_j$ всегда округляется прибавлением случайного числа и отбрасыванием дробной части.

После определения целочисленных $N_i$ и $N_j$ вычисляется число розыгрышей столкновений $N_c = \max(N_i,\ N_j)$. При каждом столкновении, соответствующие результирующие частицы запускаются в поток с вероятностями $N_i/N_c$ и $N_j/N_c$ и весами $\acute{p}_\iota$ и $\acute{p}_J$ соответственно, причем после каждого принятого и осуществленного запуска соответствующее $N_i$ либо $N_j$ уменьшается на единицу, $N_c$ также уменьшается на единицу.

Смысл это сложной процедуры заключается в уменьшении случайного блуждания числа частиц за счет введения искусственной взаимосвязи между событиями их выживания. Так, в частном случае $\acute{p}_i = p_i$ и $\acute{p}_j = p_j$, эта процедура сводится к описанной выше схеме Бёрда компонентных весовых множителей, сохраняющей число частиц.

Недостаток этой схемы в том, что, если при приближении к оси частицы не успевают размножаться, что неизбежно происходит, если ожидаемый послестолкновительный вес существенно меняется на длине свободного пробега (например, у оси), и при этом размер ячеек не позволяет достичь желаемой редкости повторных столкновений при весе прибывающих частиц, то вблизи оси начинают проявляться счетные артефакты. Теоретически, эту проблему можно обойти, если использовать несколько вложенных сеток. При этом каждому уровню сетки соответствует свой диапазон весов частиц, и частицы с разных уровней сетки трактуются как разные компоненты. Таким образом, при столкновении двух частиц из разных диапазонов весов, используется наиболее грубая из двух сеток.

Однако, такая схема будет довольно сложна в реализации, и, кроме того, в определенных условиях увеличивает эффективный размер столкновительной ячейки. В то же время, можно получить сходный результат и упрощенным вариантом алгоритма, просто подстраивая шаг сетки у оси к локальной длине свободного пробега. В общем случае увеличение шага сетки у оси невыгодно – пространственные весовые множители, вообще говоря, используются именно для борьбы с подобной необходимостью. Причём, в значительной доле осесимметричных расчетов вблизи оси присутствуют как области повышенной плотности, так и области пониженной плотности.

## 6.4 Сбалансированные радиальные множители

Идея (19) использовать разный временной шаг в разных частях расчетной области с целью управления количеством расчетных частиц (и, соответственно, локальным значением критерия $N_c\mathrm{Kn}_c$) естественным образом наталкивает на разработку схемы сбалансированных радиальных временных множителей, при которой время для приосевых частиц замедляется. Схема имеет ограниченную практическую ценность – недостатки временных множителей были рассмотрены выше. Тем не менее, она позволяет моделировать стационарные осесимметричные задачи при разумном числе частиц в расчетной области, не прибегая к использованию неконсервативных схем. Так, размножение-уничтожение частиц может приводить к флуктуациям плотности (11) в некоторых условиях.

Кто касается расчета частоты столкновений, то изменение алгоритма незначительно: $\nu_{ij} = \frac{p_i p_j \sigma_{ij}(\mathrm{c}_{ij}) \cdot \mathrm{c}_{ij}}{V/W}$. Здесь $p_i = \mathrm{T}_i = r_i / r_{max}$ – временные множители частиц, жёстко привязанные к расстоянию от оси, $W$ – всюду постоянный комбинированный вес, в данном случае соответствующий классическому весу частиц на максимальном расстоянии от оси. Так как диапазон $p_i$ в пределах одной ячейки небольшой, то изменение алгоритма принятия-непринятия касается лишь использования в качестве мажоранты величины $\left[p_i p_j \sigma_{ij}(\mathrm{c}_{ij}) \cdot \mathrm{c}_{ij}\right]_{max}$ вместо прежней при расчете максимальной вероятности столкновений.

Непростая деталь этого метода заключается в зависимости локального временного шага от расстояния до оси, которое само меняется в процессе перемещения частиц. Удобной является следующая последовательность действий. 1) перевести величину глобального временного шага $\delta t$ в величину максимального локального шага $\delta T$. Под локальным временем понимается время, в котором законы движения не зависят от расстояния до оси. 2) Перемещение частицы в течение локального времени $\delta T$, с обработкой сопутствующих пересечений траекторий частиц с границами расчетной области. 3) Перевод фактически использованного интервала $\acute{\delta T}$ в глобальный временной интервал $\acute{\delta t}$ и, если остаток $\acute{\delta t} < \delta t$,

т.е. если произошло пересечение границ, то обработать пересечение, а затем аналогично обработать перемещение частицы в течение остатка глобального времени.

Приведенные ниже формулы получены в предположении отсутствия действующих на частицы внешних сил. Первая операция производится следующим образом:

$$\delta T = Y \cdot \left[1 - \frac{(|V_\perp| - V_r)Y}{2(r_0 + |V_\perp|Y)}\right],\ Y = \frac{\mathrm{Exp}\left\{\beta \frac{r_0}{r_{max}} \delta t\right\} - 1}{\beta},\ \beta = \frac{|V_\perp|}{r_0},\ |V_\perp| = \sqrt{V_r^2 + V_\varphi^2}.$$

Здесь $V_r, V_\varphi, V_z$ – соответствующие проекции скорости частицы, $r_0$ – начальная радиальная координата (расстояние до оси). При вычислении величины $Y$ следует проявлять осторожность. Во-первых, при малых значениях аргумента экспоненты уместно использовать разложение в ряд Тейлора, иначе возможно переполнение снизу при вычитании близких величин (underflow). Во-вторых, следует избегать слишком больших значений аргумента экспоненты. Последнее обычно случается, если в конце траектории частица должна оказаться на очень большом расстоянии от оси, далеко за пределами расчётной области (чего не происходит – пересечение границы расчетной области произойдёт раньше). В таком случае, значение аргумента достаточно ограничить.

Третья операция производится так:

$$\acute{\delta t} = \frac{r_{max}}{r_0} \frac{\mathrm{Log}\left\{1 + \beta \acute{Y}\right\}}{\beta},\ \acute{Y} = \frac{2 r_0 \acute{\delta T}}{r_0 + r_1 - |V_\perp| \acute{\delta T}}$$

Здесь $r_1$– радиальная координата в конце траектории. При вычислении логарифмического выражения, аналогично, следует позаботиться о возможности вычитания близких величин.

## 6.5 Комбинированная схема радиальных множителей

Недостатком радиальных весовых множителей с размножением частиц при столкновениях является чувствительность к длине свободного пробега: при слишком большой длине свободного пробега, частицы не успевают размножаться, приближаясь к оси.

Объединив схему весовых множителей с размножением при столкновениях и схему временных множителей, можно воспользоваться преимуществами обоих методов.

Идея заключается в том, чтобы, при столкновении двух частиц с больши́ми комбинированными весами $W_i$ (и потому сильно замедленных), разыгрывать столкновения несколько раз и запускать в поток частицы с меньшими $\acute{W_i} \sim \frac{r_i}{r_{max}}$. По мере того, как частица с малым комбинированным весом удаляется от оси, её комбинированный вес увеличивается (так, чтобы $\mathrm{T}_i \le 1$ всюду), либо частица уничтожается с определенной вероятностью. Однако, при движении частицы с больши́м весом в сторону оси, $W_i$ остаётся неизменным, а $\mathrm{T}_i$ уменьшается.

Фактически, при столкновении частиц частота рассчитывается по формуле: $\nu_{ij} = \frac{F_i F_j \sigma_{ij}(\mathrm{c}_{ij}) \cdot \mathrm{c}_{ij}}{V \cdot \min(\acute{W_i}, \acute{W_j})}$, где $F_i$включает в себя линейную зависимость от расстояния до оси. В случае успеха, для частицы $i$ с вероятностью $\frac{\acute{W_i}}{\max(\acute{W_i}, \acute{W_j})}$ в поток запускается частица с комбинированным весом $\acute{W_i}$, и на ту же величину уменьшается вес $W_i$ исходной частицы.

При перемещении частицы, необходимо дополнительно отслеживать момент достижения частицей $\mathrm{T}_i = 1$ при удалении от оси (пусть этому соответствует радиус $r_*$), и отключать пересчёт временного интервала при дальнейшем удалении, включая вместо него обычную логику выбора между уничтожением частицы и увеличением её веса.

Локальное время, когда частица достигает критического расстояния $r_*$, рассчитывается по формуле:

$\acute{\delta T} = \frac{{r_*}^2 - {r_0}^2}{r_0 V_r + \sqrt{(r_0 V_r)^2 + |V_\perp|^2({r_*}^2 - {r_0}^2)}}$, в случае $V_r > 0$;

$\acute{\delta T} = \frac{\sqrt{(r_0 V_r)^2 + |V_\perp|^2({r_*}^2 - {r_0}^2)} - r_0 V_r}{|V_\perp|^2}$, в случае $V_r < 0$.

Определять макропараметры по «фотографиям» следует: 1) непосредственно перед фазой столкновения, после перемещения частиц и удаления частиц со слишком малым весом, 2) непосредственно после фазы столкновения и размножения. При этом используется вес частицы $F_i$, а не комбинированный вес.

Если же применяются компонентные весовые $\widetilde{F}_l$ и/или временные $\widetilde{\mathrm{T}}_l$ множители, то изменения, вносимые в схему, вполне естественны и несущественны, поэтому здесь опускаются.

Преимущество использования представленной схемы в том, что, при приближении к оси, частицы скапливаются не только за счет размножения при столкновениях, но и за счет уменьшения временного множителя. Основные же недостатки унаследованы от временных множителей: замедление сходимости к стационарному решению, неприменимость к нестационарным задачам.

Испытание данного метода подтвердило его работоспособность и показало повышенную скорость работы процедуры обработки столкновений по сравнению с подходом размножения частиц при столкновениях. Однако, обработка перемещения частиц, наоборот, стала заметно медленнее.

К сожалению, данная схема лишь частично избавляет от проблем, ради борьбы с которыми она разрабатывалась. Так, в случае смесей газов сильно отличающихся масс, вероятность повторного столкновения тяжелой молекулы с легкой, если последняя сильно замедлена, остаётся велика. Действительно, согласно эвристической оценки критерия для случая временных множителей, при столкновении тяжелой частицы с лёгкой, получаем: $\sqrt{\frac{1+M}{1+\mathrm{M}\cdot{\mathrm{T}_2}^2}} \cdot \frac{F_2 \sigma_{12} h}{V} \ll 1$. В случае $M \gg 1$ и $\mathrm{T}_2 \ll 1$, значение критерия может оставаться довольно высоким – в $\sqrt{1+M}$ раз выше по сравнению со случаем $\mathrm{T}_2 \sim 1$. Однако, при отсутствии замедления частиц большого комбинированного веса, положение было бы ещё хуже – значение критерия вырастает в количество раз, соответствующее отношению фактического веса легкой частицы к целевому весу для её радиуса, что составляет величину порядка $1/\mathrm{T}_2 \gg 1$ раз – заведомо бóльшую и ничем не ограниченную.

# 7 Практическое сравнение схем решения осесимметричных задач

Итак, в настоящей работе описаны 3 схемы решения осесимметричных задач, основанные на размножении/уничтожении частиц при их приближении к оси / удалении от оси соответственно. Введём следующие обозначения этих схем:

- **wmax**: схема радиальных множителей с пакетным размножением частиц, в которой, при столкновении пары частиц, происходит размножение частиц путём многократного розыгрыша результатов столкновения и замены пары большого веса несколькими парами частиц меньшего веса.
- **wpp**: схема, описанная в начале раздела 6Вероятность столкновения пропорциональна произведению весов. Перед столкновением, каждая исходная частица пары логически расщепляется на две, одна из которых имеет вес ячейки и участвует в столкновении.
- **tw**: Комбинированная схема радиальных множителей, сочетающая в себе логику размножения частиц при столкновениях с логикой замедления времени для частиц, приближающихся к оси.

Кроме того, в сравнении также участвуют ещё два подхода:

- **nw**: размножения/уничтожения частиц, связанного с изменением расстояния до оси, не происходит. Проблема недостатка частиц у оси попросту игнорируется. «Обычная» консервативная схема ПСМ.

- **tnw**: схема сбалансированных радиальных множителей. Все частицы имеют одинаковый комбинированный вес, но локальный временной шаг пропорционален расстоянию от оси.

Результаты расчета по разным схемам сравниваются как между собой, так и с результатом **x10** – расчета, полученного по схеме **wpp**, но при 10-кратном увеличении количества частиц. Для сравнения используется задача о конвергентом сверхзвуковом течении смеси He + Xe (24), так как в ней имеет место трансзвуковое движение газа по направлению к оси (что считается неблагоприятным условием для схем с размножением частиц).

Краткое описание задачи: звуковое сопло в виде кольцевой щели, создающее поток газа, направленный в сторону оси кольца. Газ – смесь гелий + ксенон. Варьируется концентрация ксенона, отношение диаметра кольца к ширине щели, число Кнудсена $Kn_*$. Последнее вычисляется как отношение длины свободного пробега при условиях торможения к ширине щели.

## 7.1 Случай сильно несбалансированных весов компонент

Для начала сравним результаты расчетов, выполненных при следующих условиях: смесь 100 % гелия с пренебрежимо малой примесью ксенона (для расчета ксенона используются компонентные весовые множители), Отношение диаметра источника к ширине щели $D_0/h_* = 4$, число Кнудсена по параметрам торможения и ширине щели $Kn_*$ = 0.02. Над кольцевой щелью расположен канал с зеркальными плоскими стенками, имеющий разницу радиусов $h_*$ – для того, чтобы более мягко перейти от жёстких граничных условий к формированию звуковой поверхности. На максимальном удалении от оси, значение критерия $N_C Kn_C$ = 7.54 для столкновений He-He (3.7 для столкновений Xe-He).

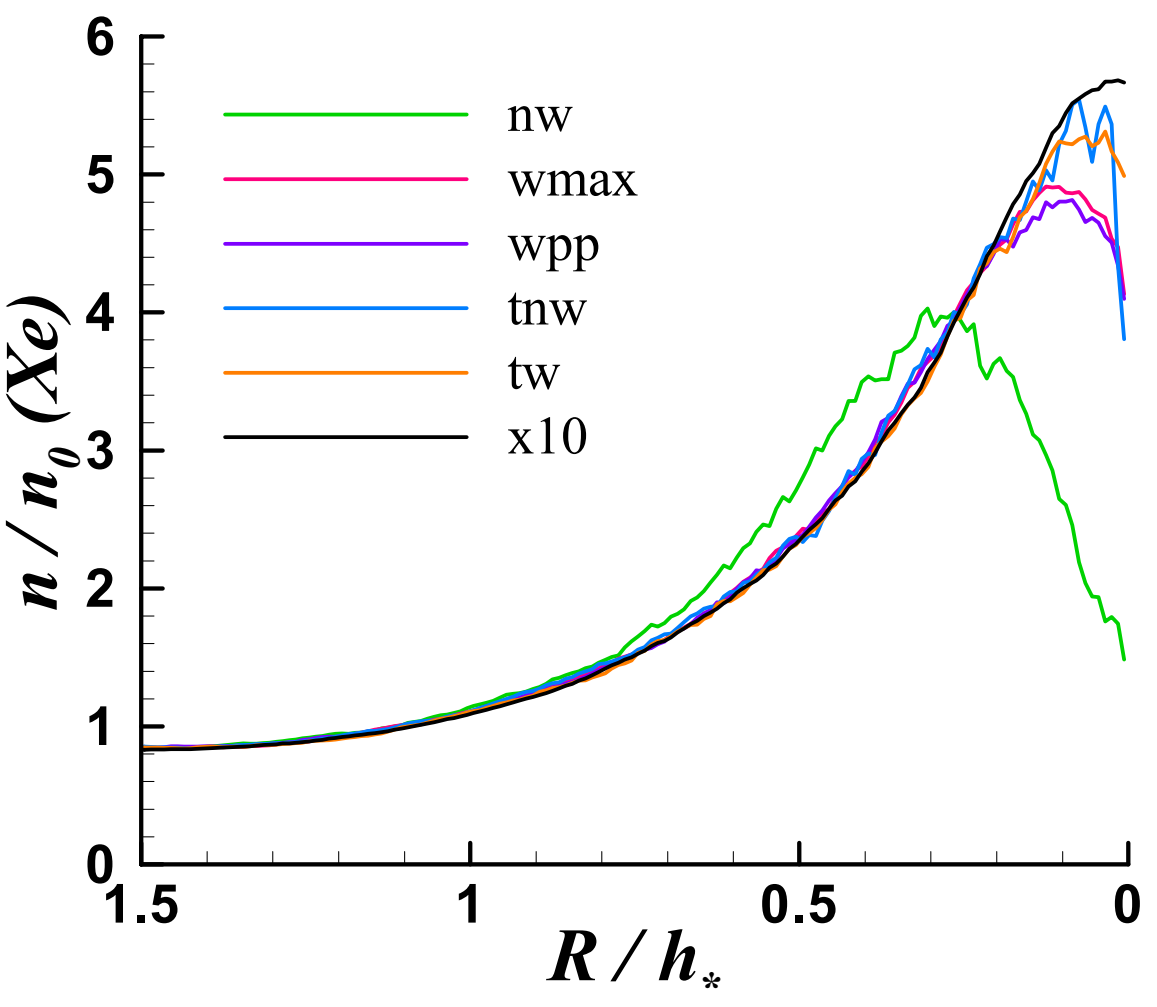

**Рис. 7.1-1**. Профиль плотности тяжелой компоненты (ксенона) в окрестности оси в плоскости симметрии для разных численных подходов.

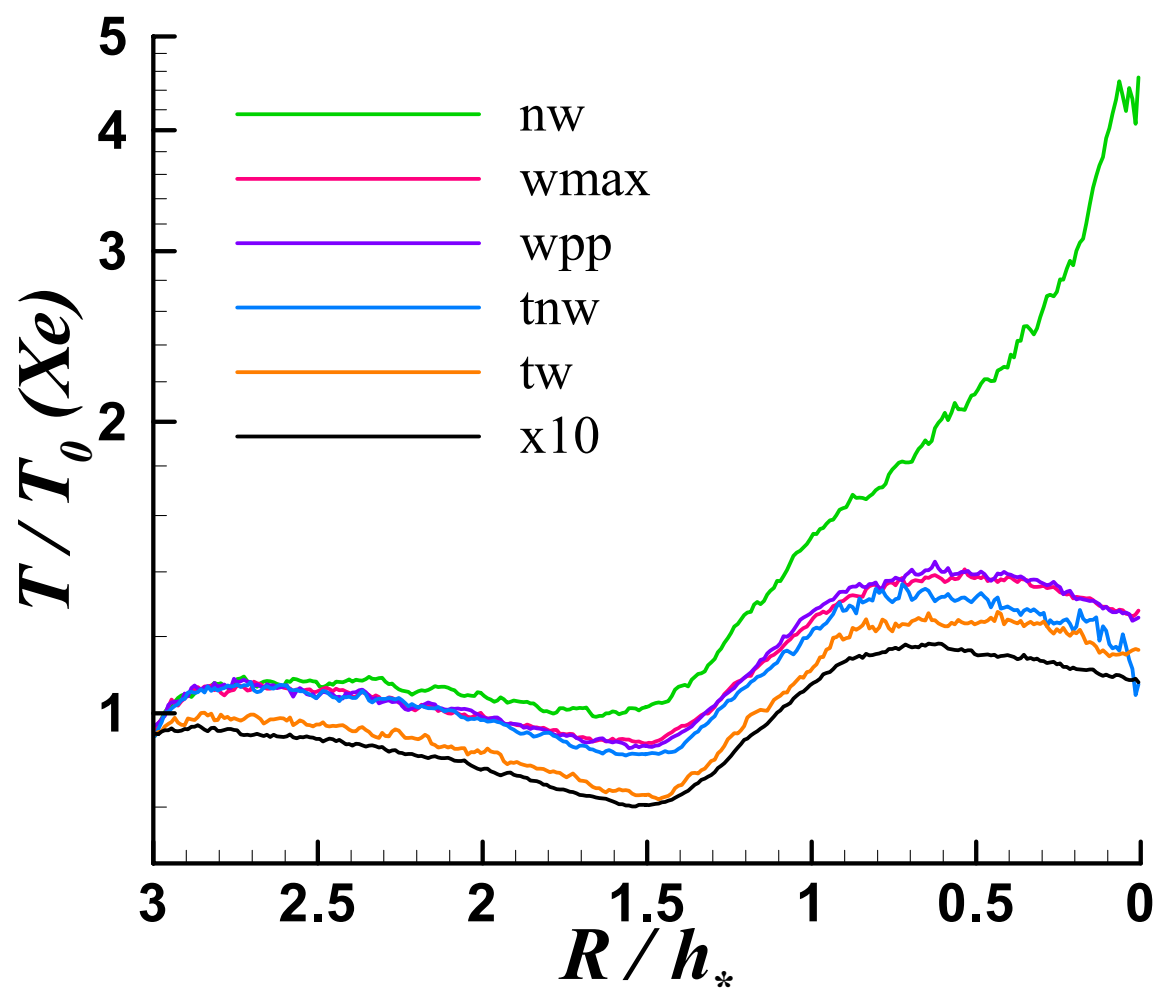

**Рис. 7.1-2**. Профиль температуры тяжелой компоненты в плоскости симметрии для разных численных подходов.

На **Рис. 7.1-1** показано распределение плотности Xe (за 1 взята плотность в камере торможения) в плоскости симметрии, полученное при различных расчетах. Легко можно видеть, что наиболее точное значение плотности вблизи оси получилось в расчетах **tw** и **tnw**, причём схема **tnw** демонстрирует ухудшенную статистическую сходимость у оси. Ошибки приосевой плотности при расчете схемами **wpp** и **wmax** примерно вдвое больше и сопоставимы между собой. Приосевая ошибка схемы **nw**, в которой радиальные веса не используются, в разы превышает ошибку всех остальных вариантов, причём значительные отклонения начинают проявляться в 4 раза дальше от оси.

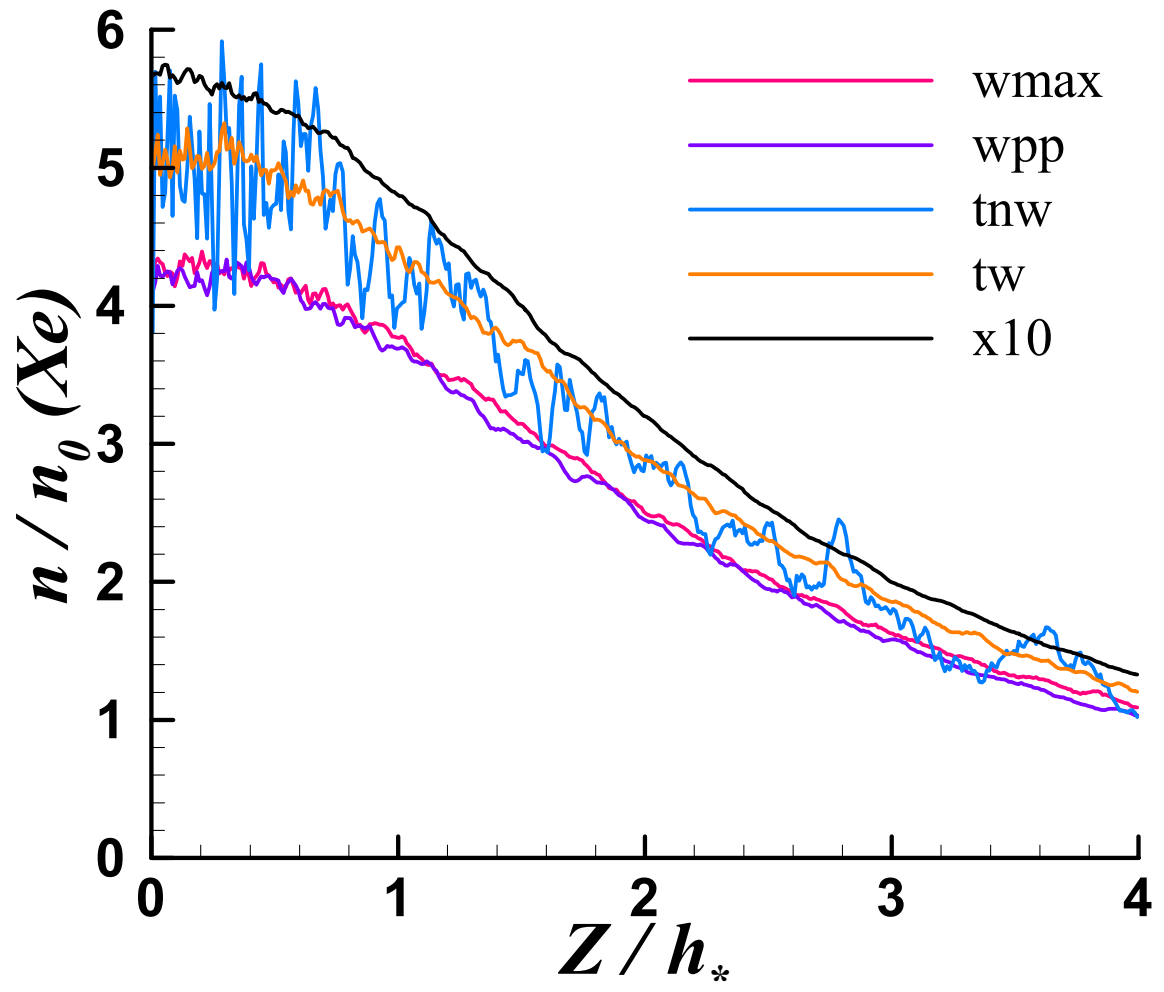

**Рис. 7.1-3**. Профиль плотности тяжелой компоненты (ксенона) у оси для разных численных подходов.

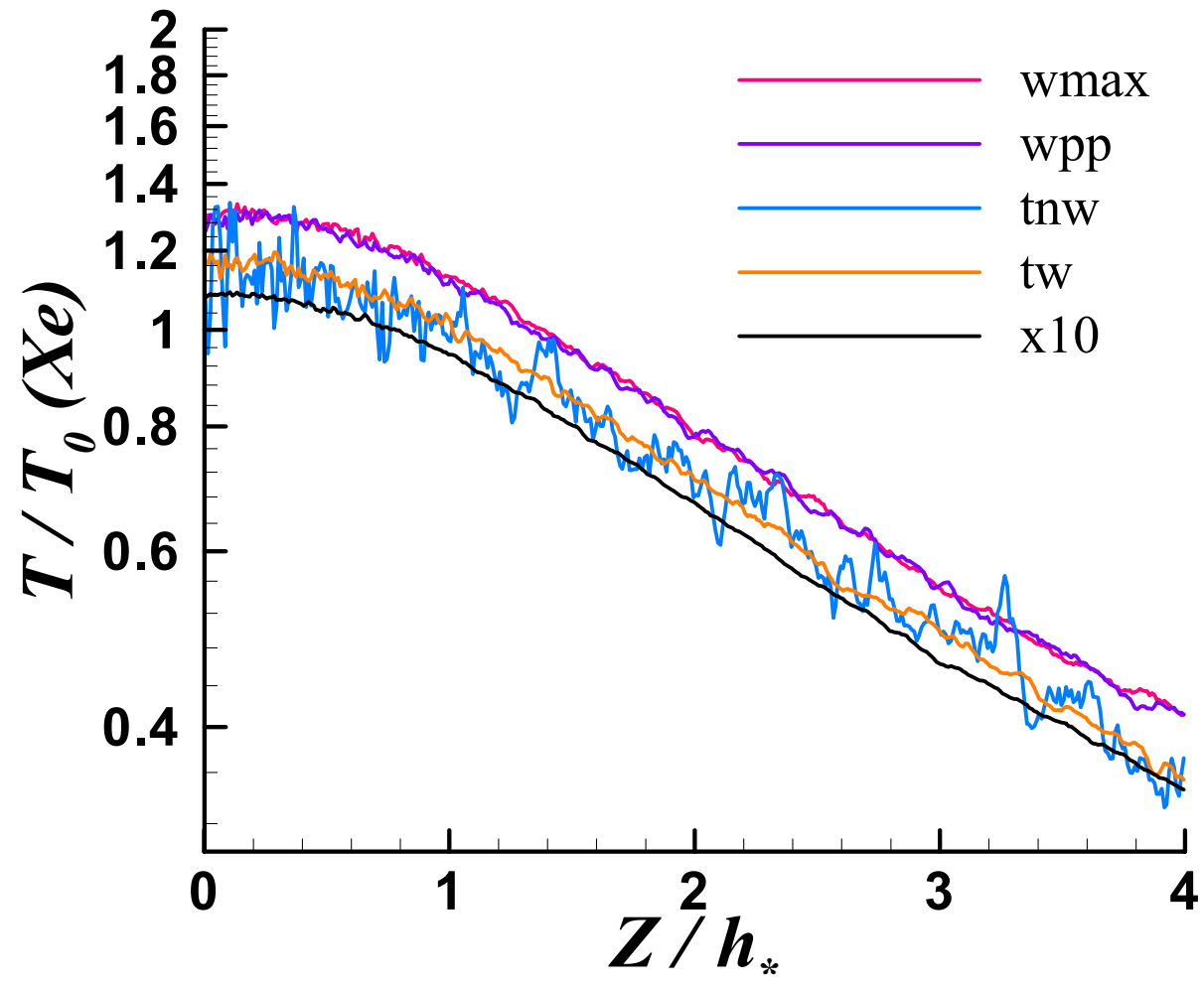

**Рис. 7.1-4**. Профиль температуры тяжелой компоненты у оси для разных численных подходов.

На **Рис. 7.1-2** показано распределение температуры Xe, также в плоскости симметрии. Как можно видеть, во всех случаях отличие от эталонного результата **x10** начинает проявляться практически в непосредственной близости у границы расчетной области, с которой истекает газ. Основной источник отклонения – нарушение статистики обмена энергии-импульса в результате повторных столкновений Xe-He при их неконсервативности. Как и следовало ожидать, схема **nw** даёт наименее адекватные значения температуры, особенно вблизи оси. Ошибка схемы **tnw** на большом расстоянии от оси примерно совпадает с ошибкой схем **wpp** и **wmax**. Однако, вблизи оси ошибка схем **wpp** и **wmax** вдвое больше. По-видимому, увеличение последней связано с тем, что вблизи оси частицы не успевают размножаться, и эффективное значение $N_C Kn_C$ уменьшается. Ошибка схемы **tw** вблизи оси практически совпадает с ошибкой **tnw**. Однако, вдали от оси схема **tw** демонстрирует аномально низкую ошибку температуры тяжелой примеси, в то время ожидаемая величина ошибки должно лежать между величинами **tnw** и **wpp**. По-видимому, причина этого заключается в случайном совпадении благоприятных для схемы обстоятельств – например, ксенон оказался в среднем сильнее замедлен по сравнению с гелием. При расчете режима $D / h_* = 15$, например, такого аномального поведения схемы **tw** не наблюдается.

На **Рис. 7.1-3** и **Рис. 7.1-4**, аналогично, показаны профили плотности и температуры соответственно, но вдоль оси в её непосредственной близости. В данном случае, течение газа переходит от центростремительного к похожему на обычное струйное расширение. Как видно, на оси ошибки **tw** и **tnw**, по-прежнему, практически совпадают по величине и имеют меньшее значение по сравнению с **wpp** и **wmax**. Из этих же рисунков можно оценить приосевую ошибку. Для схем **tw** и **ntw**, ошибка составляет 10 – 15 % для плотности и около 10 % для температуры. Для схем **wpp** и **wmax** ошибка имеет величину 25 – 35 % для плотности и 20 % для температуры. Как можно видеть, величина артифактов поступательной температуры тяжелой примеси имеет примерно тот же порядок, что и, например, ошибка плотности.

С увеличением $N_C Kn_C$, температурные артефакты ослабевают, так же, как и другие погрешности. Вообще, не следует предвзято относиться к нефизичной разнице поступательных температур компонент при их равновесии. Да, градиенты температуры в неравновесном потоке обычно меньше по сравнению, например, с градиентами плотности и давления, поэтому погрешности температуры более заметны. Во-вторых, по разнице поступательных температур часто оценивается выраженность поступательной неравновесности, чему артефакты мешают. Если «консервативные» ошибки, прежде всего, проявляются как отклонение вязкостных сил, диффузионных и тепловых потоков от истинного значения при сохранении общей динамики макропараметров, то артефакты неконсервативности имеют более глубокую природу и искажают саму функцию распределения.

Главный вывод данного подраздела: при использовании несбалансированных компонентных весовых множителей, переход на предложенную в настоящей работе схему **tw** выгоден, так как последняя, при прочих равных, может давать вдвое меньшую ошибку вблизи оси. Схемы **wpp** и **wmax**, тем не менее, всё ещё могут быть использованы (прежде всего, для нестационарных задач), но с бóльшим числом частиц. Что касается схемы **nw**, то её поведение при несбалансированных весах компонент неадекватно, поэтому использование её малопродуктивно.

Следует отметить, что характерные для неконсервативных схем ошибки сильнее всего выражены тогда, когда веса компонент существенно отличаются – как, например, в рассмотренном здесь случае. Если веса компонент близки друг к другу, то обычно наблюдаются более слабые искажения равновесной функции распределения.

## 7.2 Случай близких весов компонент

Если компонентные весовые множители сбалансированы, то столкновения консервативны. Однако, при применении неконсервативной схемы радиальных весов, веса частиц будут отличаться от локальных целевых значений, хотя и имеют тенденцию приближаться к ним. Из рассматриваемых схем, только схемы **tnw** и **nw** ведут себя заведомо консервативно.

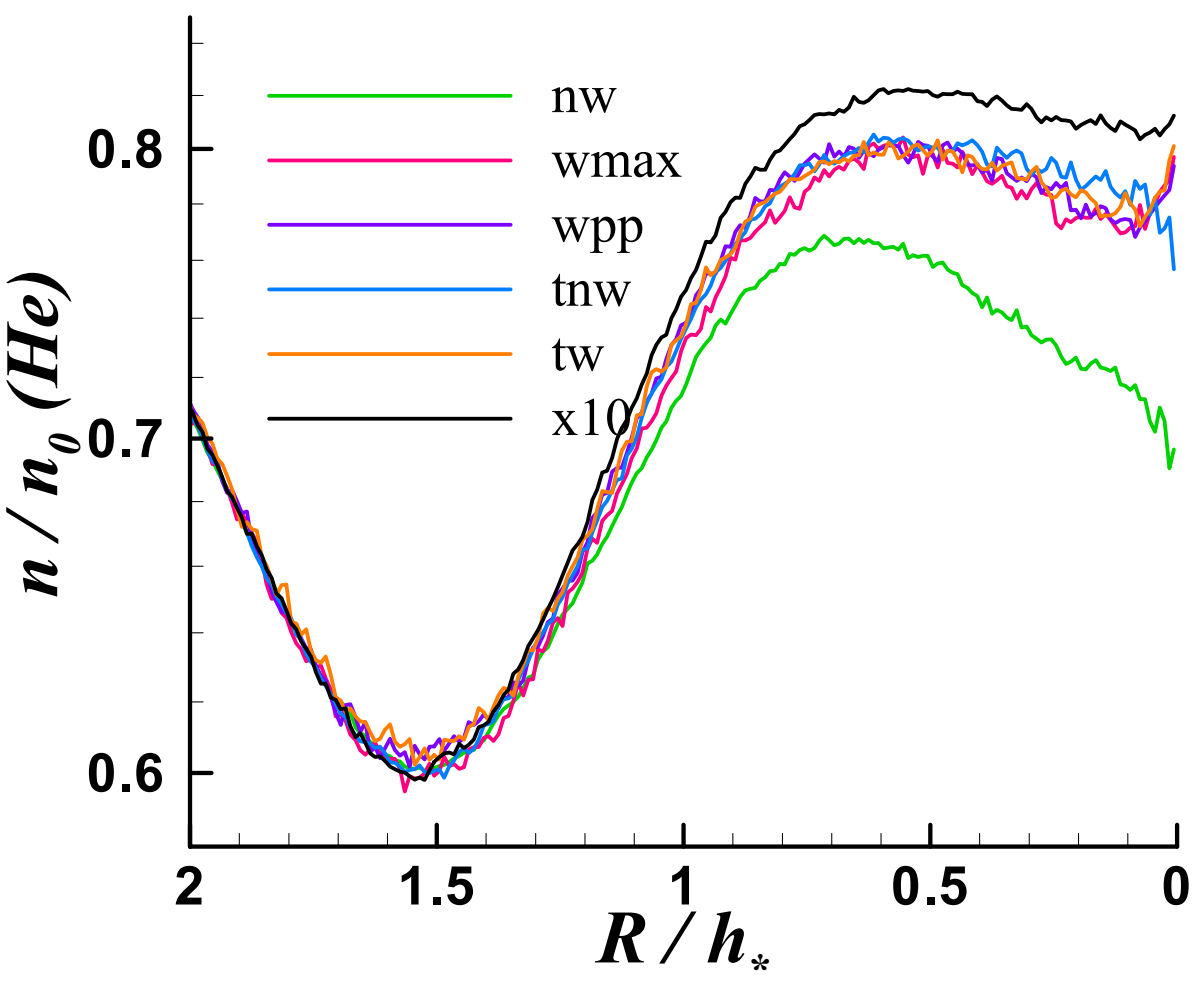

Рис. 7.2-1. Профиль плотности гелия в окрестности оси в плоскости симметрии для разных численных подходов (случай близких компонентных весов).

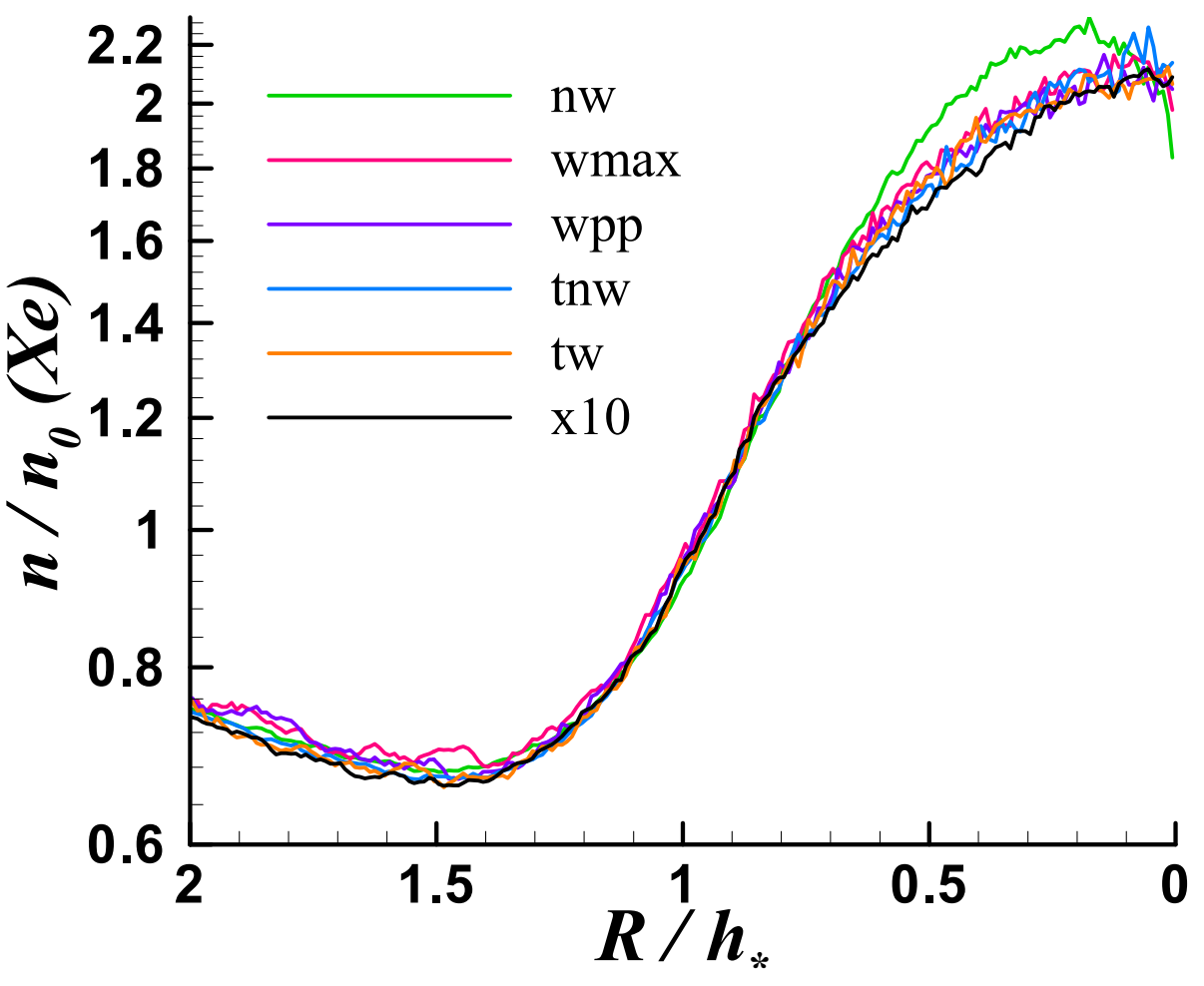

Рис. 7.2-2. Профиль плотности гелия в окрестности оси в плоскости симметрии для разных численных подходов (случай близких компонентных весов).

Проведем теперь расчеты для смеси 90 % He + 10 % Xe. Как следствие, становится существенным влияние ксенона на течение несущего газа. Значения критерия $N_C Kn_C$ следующие: 7.54 для He-He, 3.7 для Xe-He, 1.8 для Xe-Xe. Это довольно низкие значения, особенно для Xe-Xe, в реальных расчетах обычно выдерживались значения не менее 20.

Рассматривая зависимости плотности гелия (Рис. 7.2-1.) и ксенона (Рис. 7.2-2) в плоскости симметрии, можно заключить, что погрешность примерно одинакова при всех схемах, кроме **nw.** Консервативная схема **nw** вновь «забраковывается» из-за очень сильных артефактов вблизи оси.

Анализ зависимостей средней температуры смеси (Рис. 7.2-3) и температуры ксенона (Рис. 7.2-4) показывает следующее. Схема **tw** более не даёт такого же превосходства по сравнению с **wpp** и **wmax**, как в случае сильно отличающихся компонентных весов. Все три неконсервативных схемы показывают примерно одинаковое завышение температуры ксенона по сравнению со средней температурой. Схема **tnw**, благодаря своей консервативности, полностью устраняет эффект завышения температуры ксенона.

Выводы следующие. Схема **tnw** позволяет решать осесимметричные задачи при полном отсутствии атрефактов неконсервативности. Схема **tw** не показала превосходства в случае сбалансированных весов компонент, но, тем не менее, работает не хуже схем **wpp** и **wmax** и потому может считаться довольно универсальным инструментом для решения осесимметричных задач как при сбалансированных, так и не при сбалансированных весах компонент.

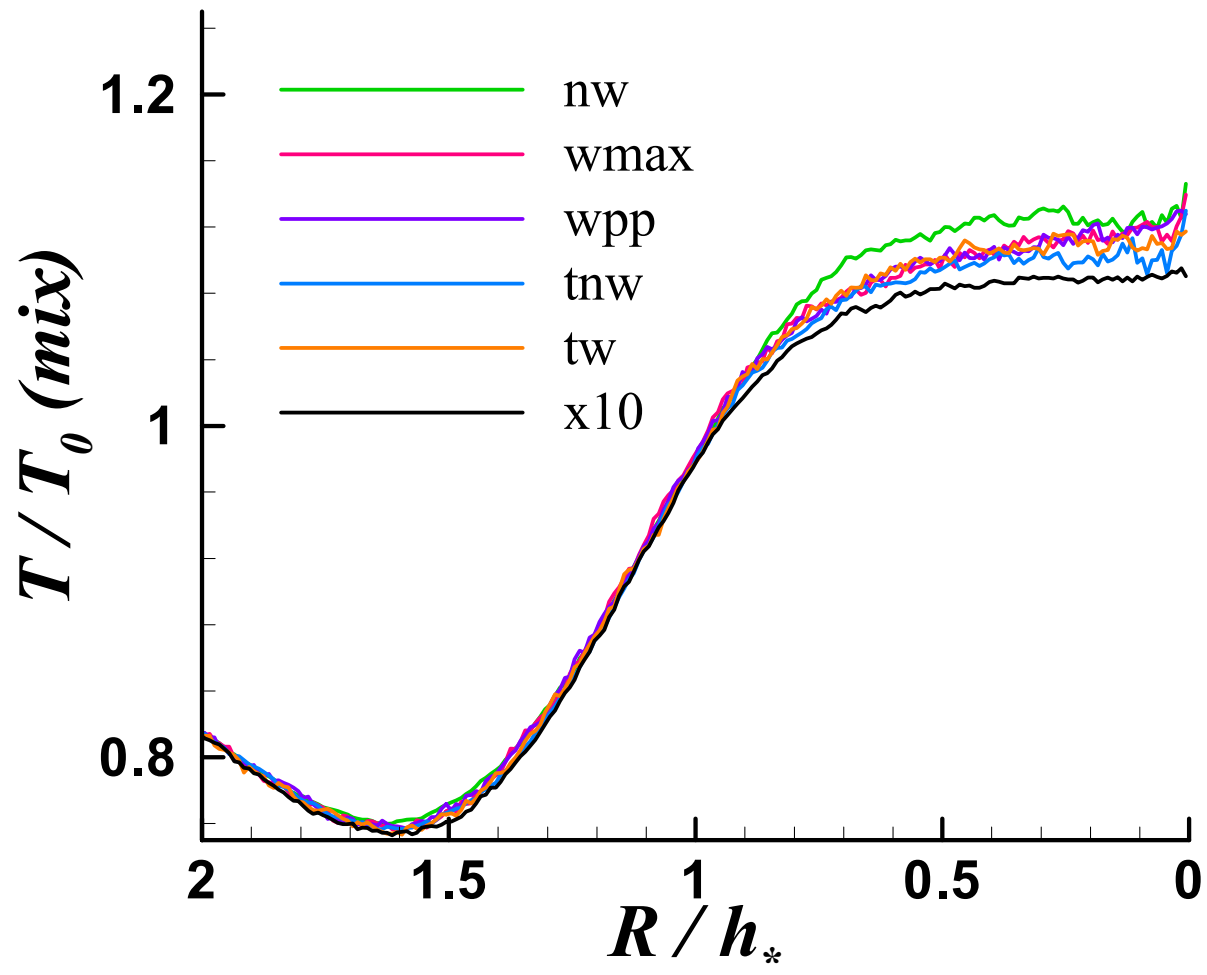

Рис. 7.2-3. Профиль средней температуры смеси в окрестности оси для разных численных подходов (случай близких компонентных весов).

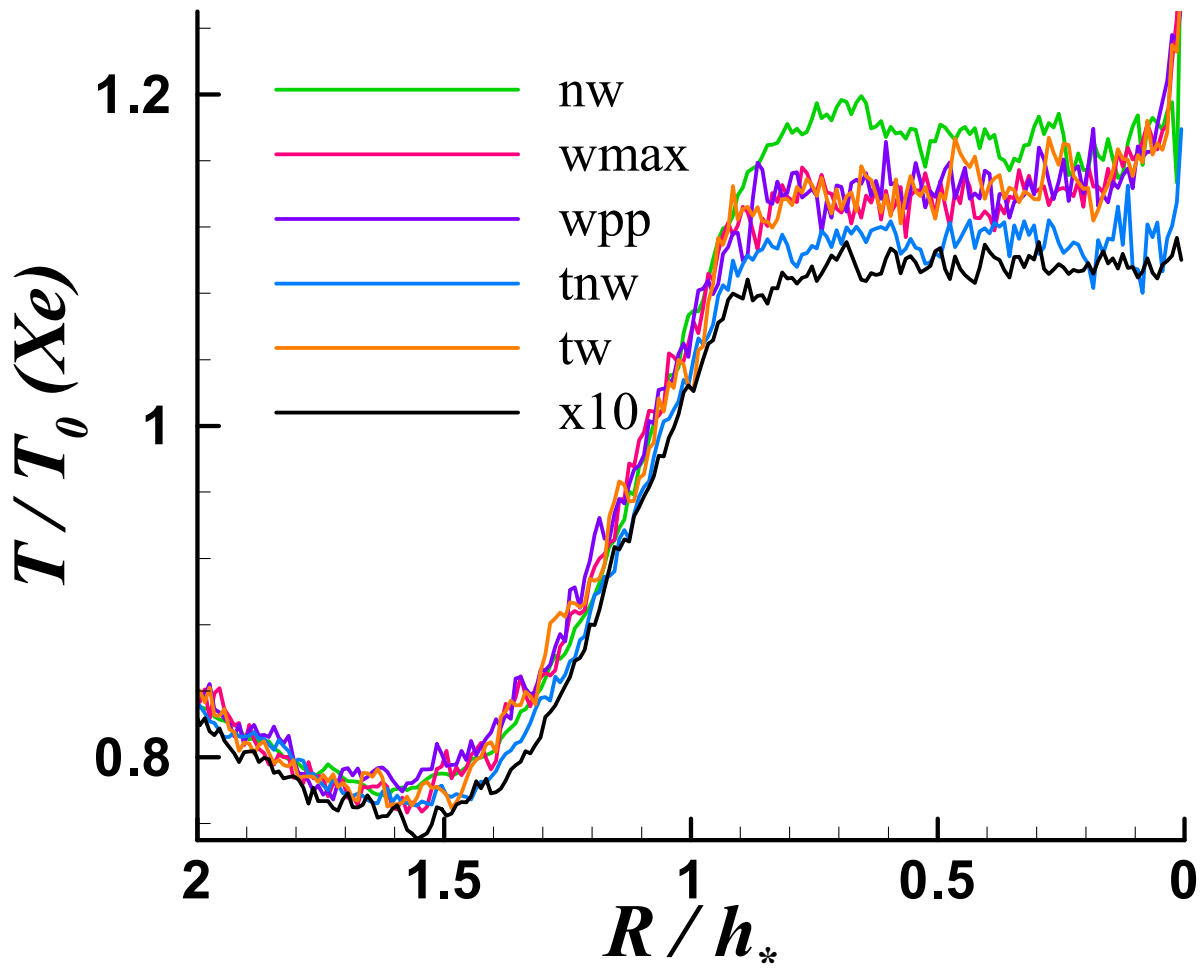

Рис. 7.2-4. Профиль средней температуры ксенона в окрестности оси для разных численных подходов (случай близких компонентных весов).

# 8 Подавление флуктуаций числа частиц

Известный недостаток схем пространственных весовых множителей заключается во флуктуациях числа частиц (11), приводящих к флуктуациям плотности. Реальную неприятность это, однако, представляет только для узкого класса задач, в которых граничные условия не способствуют погашению флуктуаций. Прежде всего, это процессы в замкнутых объёмах, а также крайне медленные дозвуковые течения. Устранение этого недостатка позволит расширить класс доступных задач. Кроме того, уменьшенный разброс числа частиц может способствовать некоторому снижению погрешности при расчете трансзвуковых течений.

Предлагаемый способ решения проблемы флуктуаций основан на простой идее: уничтожать частицы следует в момент превышения ею величины веса, при котором она образовалась.

## 8.1 Автомодельный процесс размножения/уничтожения частиц при перемещении

Рассмотрим для начала процесс уничтожения. Процесс уничтожения реализуется при условии, что частица движется в направлении увеличения своего пространственного веса. Классическая формулировка алгоритма следующая. Пусть частица, изначально имевшая вес

Сначала обратим внимание на классический процесс уничтожения частиц. Пусть частица, имевшая вес $W_0$, переместилась в другую точку, которой соответствует вес $W_f > W_0$. Частица либо выживает и приобретает вес $W_f$ – с вероятностью $\frac{W_0}{W_f}$, либо, в противном случае, уничтожается. Заметим, что, если частица всё это время движется в направлении увеличения веса, её окончательная вероятность выживание не изменяется в зависимости от того, на сколько промежуточных этапов разбит процесс. Так, если перемещение частицы произошло в два этапа: $W_0 \to W_2 \to W_f$, причём $W_0 < W_2 < W_f$, то вероятность выживание останется прежней: $\frac{W_0}{W_2} \cdot \frac{W_2}{W_f} = \frac{W_0}{W_f}$. Это позволяет разбить процесс уничтожения на произвольное число промежуточных этапов, вплоть до непрерывного процесса, без изменения конечного результата. Назовём такое свойство процесса автомодельностью.

В качестве отступления, можно упомянуть, что процесс розыгрыша столкновений по схеме мажорантной частоты обладает свойством автомодельности по времени, а по схеме NTC – нет.

Для процесса уничтожения частиц можно заметить то обстоятельство, что, вместо многократного розыгрыша выживания с вероятностью $\frac{W_0}{W_f}$ при каждом перемещении частицы, можно сразу разыграть вес частицы, при котором она уничтожится: $W_{MAX} = \frac{W_0}{RF}$. Здесь $RF$ – случайное число в диапазоне 0..1. Эту величину необходимо сгенерировать и запомнить для каждой частицы и считать её действительной до тех пор, пока частица не завершит движение в направлении повышения веса, и не станет двигаться в направлении его понижения. Теперь, до тех пор, пока вес частицы в конце фазы перемещения не превышает максимальный, т.е. $W_f$<$W_{MAX}$, частица выживает, в противном случае – уничтожается.

Стандартная процедура размножения частиц следующая. На этот раз $W_f < W_0$. Число новых частиц, которые следует запустить в поток, определяется по формуле: $N_+ = \frac{W_0}{W_f} - RF$ и округляется в меньшую сторону. Как можно заметить, стандартная процедура размножения не обладает свойством автомодельности – разброс в конечном числе частиц будет увеличиваться при внесении промежуточных этапов.

Построим автомодельный процесс размножения частиц, отталкиваясь от непрерывного процесса. Считаем, что все копии исходной частицы движутся синхронно с оригиналом, имея тот же самый вес $W$. Вес частиц уменьшается непрерывно. Запишем вероятности $p_n(W)$ для каждого возможного числа копий $n$ как функции от $W$. Пусть вес частиц снизился на малую величину $\delta W$ и составляет $W - \delta W$. Каждая из частиц удвоится с вероятностью $\frac{\delta W}{W}$. Вероятность того, что из $n$+1 частиц удвоится какая-то одна, составляет $\approx (n+1)\frac{\delta W}{W}$. Это позволяет записать систему дифференциальных уравнений: $\frac{dp_0}{dW} = -\frac{p_0}{W}$, $\frac{dp_{n+1}}{dW} = -(n+1)\frac{p_n}{W} + (n+2)\frac{p_{n+1}}{W}$.

При условии, что в начале процесса не было ни единой копии, вероятности будут следующие: $p_n(W) = \frac{W}{W_0}\left(1 - \frac{W}{W_0}\right)^n$, т.е. представляют собой экспоненциальное распределение.

Помимо непосредственно числа копий, интерес представляет также функция распределения значений $W_n$, при которых были рождены копии. Это позволяет впоследствии разыграть не только число копий, но и величины $W_n$. Исследуя, как меняется число копий: $-\frac{d}{dW}\sum np_n(W)\Big|_{W=W_n}$, можно найти плотность вероятности: $\sim\frac{1}{{W_n}^2}$.

Таким образом, для реализации автомодельного процесса размножения необходимо разыграть число копий согласно экспоненциальному распределению. Это можно сделать следующим образом. Разыгрываем величину $W_{MAX} = \frac{W_0 W_f}{W_f + RF \cdot W_0}$, представляющую собой вес, при котором произошла копия. Если $W_{MAX} < W_f$ (неравенство выполняется, если $RF \cdot W_0 > W_0 - W_f$), то копия в поток не запускается, и цикл размножения для данной исходной частицы прерывается. В противном случае, одна копия исходной частицы помещается в поток, и цикл размножения продолжается (разыгрывается следующая копия).

## 8.2 Описание подхода

По отдельности, ключевые идеи уже были сформулированы выше. Осталось лишь сформулировать алгоритм с целом.

- Задан пространственный вес частицы как функция от координат. Для удобства, его можно вычислить один раз для каждой частицы в процедуре перемещения и затем хранить в массиве, чтобы не вычислять повторно в процедуре столкновений и в последующем шаге перемещения, но это не обязательно.
- Для каждой частицы хранится максимально допустимый вес.
- Во время запуска в поток новой частицы, её максимальный вес разыгрывается по формуле $W_{MAX} = \frac{W_0}{RF}$, где $W_0$ – пространственный вес запускаемой частицы.
- После перемещения частицы, вычисляется новый пространственный вес частицы $W_f$ в конце траектории, также при этом определяется пространственный вес в начале траектории $W_0$. Если $W_f > W_{MAX}$, то частица уничтожается. Если $W_f$<$W_0$, то запускается алгоритм автомодельного размножения исходной частицы. При этом величины $W_{MAX}$ для частиц-копий разыгрываются согласно процедуре размножения, а исходная частица-оригинал сохраняет прежнее значение $W_{MAX}$.
- В процедуре столкновения, частицы не размножаются: либо частица меняет свои параметры при столкновении, либо нет, но остаётся в единственном числе. При этом частица сохраняет свою величину $W_{MAX}$.

Стоит пояснить, почему предложенный подход, действительно, должен уменьшить флуктуации числа частиц. При стандартном подходе, как уже было сказано, никакой предыстории частиц не сохраняется. Это означает, что, в среднем, каждая частица может как размножиться, так и уничтожиться. При предложенном подходе, уже в момент запуска в расчётную область, некоторые частицы имеют такое большое значение $W_{MAX}$, что заведомо не уничтожатся, пока не покинут расчетную область. Доля таких частиц составляет около 50 % при однородном распределении плотности газа. Если плотность газа выше на периферии – доля неуничтожаемых частиц будет выше. Если же выше плотность вблизи оси, то доля неуничтожаемых частиц будет, наоборот, ниже. Количество частиц в расчетной области ограничено снизу числом неуничтожаемых частиц. Именно неуничтожаемые частицы будут в итоге основными прародителями большинства размноженных частиц. Все частицы-потомки имеют ограниченные значения $W_{MAX}$ и потому рано или поздно будут уничтожены.

## 8.3 Испытание эффективности подхода

Испытание проводилось при условиях работы (11). Расчетная область представляет собой трубу диаметром 8 единиц и длиной 10 единиц. Размер ячейки составляет 0.25 единиц. Одна единица длины соответствует длине свободного пробега при единичной плотности и единичной температуре. Все три границы расчетной области представляют собой стенки, имеющие единичную температуру, отражение на них полностью диффузное. На правом торце трубы, дополнительно, задана 3 % вероятность поглощения частицы. На левом торце трубы, дополнительно, задан поток входящих молекул, соответствующих плотности 3 % от единичной, нулевой скорости потока, единичной температуры. При таких условиях, реализуется крайне медленное дозвуковое течение с левого торца трубы в сторону правого.

Число модельных частиц выбрано таким, чтобы при единичной плотности и отсутствии пространственных множителей в объёме расчетной области помещалось около 1250 частиц (точное значение: 1256.64 частиц). Реально использовались весовые множители, зависящие от радиуса по линейному закону и меняющиеся в диапазоне 0.025..1. Это соответствует ожидаемому числу частиц 2333.9.

Для проведения сравнения была дополнительно реализована стандартная схема Бёрда пространственных весовых множителей с размножением при перемещении.

Расчёт проводился в течении продолжительного времени, при этом вычислялись средние по временным шагам параметры, а также их среднеквадратические отклонения: число частиц, средняя по сечению (площади расчетной области) плотность газа, средняя по объёму плотность газа.

Результаты сравнения следующие:

| | Стандартный подход | Представленный новый подход |
|---|---|---|
| Число частиц | 2117.7 ± 527.1 | 2335.2 ± 93.8 |
| Средняя по сечению плотность | 0.908 ± 0.227 | 1.001 ± 0.044 |
| Средняя по объёму плотность | 0.905 ± 0.225 | 0.998 ± 0.033 |

Как видно, флуктуации средней плотности уменьшились в 7.5 раз. Кроме того, получено весьма точное значение средней плотности, в то время как при стандартном алгоритме плотность оказалась в среднем заниженной на 10 %. Результаты данного испытания можно считать положительными.

Производилось тестирование и на других задачах.

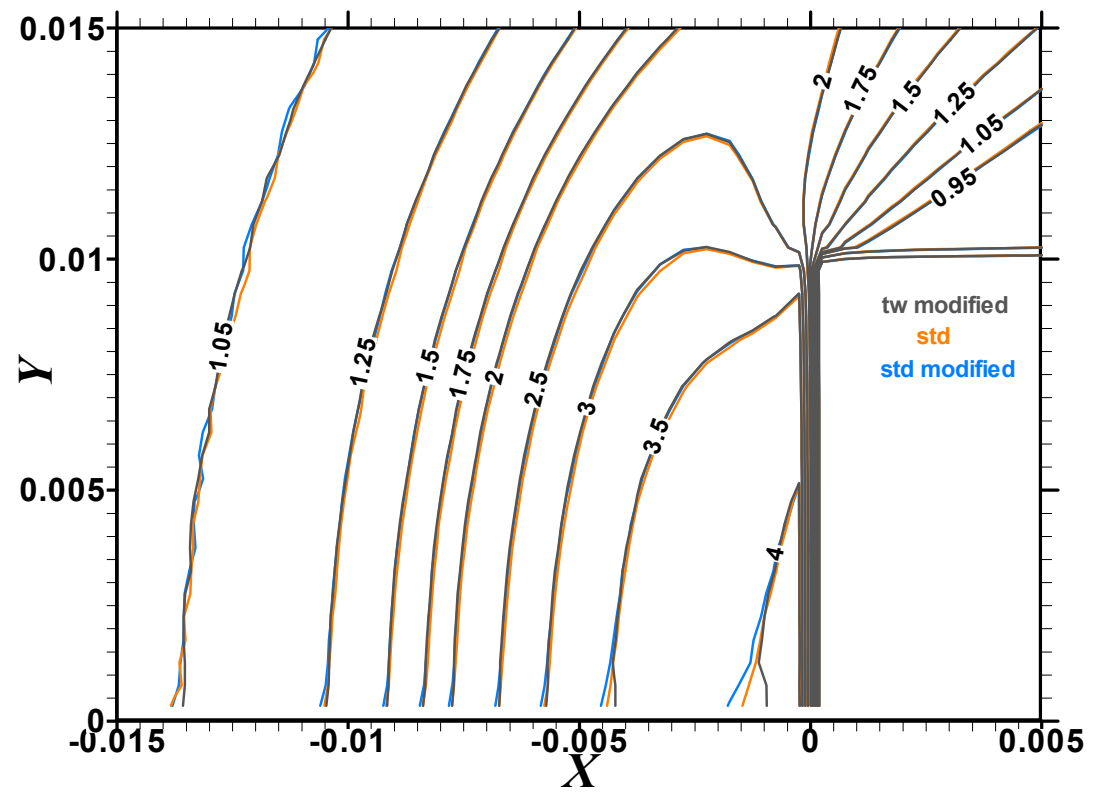

**Рис. 8.3-1.** Сравнение изолиний плотности, полученных тремя схемами: стандартной схемой весовых множителей (**коричневая** кривая), доработанной стандартной схемой (**голубая** кривая), доработанной схемой **tw** (**чёрная** кривая).

Например, одним из тестов являлось решение задачи обтекания цилиндра, описанной в разделе 6Результаты для стандартной схемы весовых множителей и предложенной схемы с подавлением флуктуаций числа частиц дали слабо отличимый результат. В этом можно убедиться, сравнив показанные на **Рис. 8.3-1** изолинии плотности, полученные в трёх расчетах: расчет по классической схеме размножения частиц при перемещении, расчет по доработанной схеме с размножением частиц при перемещении, а также расчет по доработанной схеме **tw** (детали доработки описаны в следующем параграфе). Сравнение говорит в пользу работоспособности нового алгоритма.

## 8.4 Применение подхода в схемах с размножением частиц в столкновениях

Как было сказано ранее, нюансом подхода с размножением частиц в столкновениях является максимальная отсрочка внесения разброса в число частиц. Это достигается сохранением информации о весе каждой частицы в момент, когда её параметры последний раз изменились (т.е. при последнем столкновении или при запуске новой частицы в поток), что и позволяет произвести процедуру размножения в произвольный момент времени.

Пусть частица имеет в момент столкновения вес $W_0$, тогда после столкновения её средний вес должен составлять $W_0 - W_1$, и дополнительно должна быть создана послестолкновительная частица с весом $W_1$. Выше был описан автомодельный случайный процесс размножения при перемещении. Количество частиц, получившихся из одной исходной частицы, описывается экспоненциальным распределением. Следуя духу этого случайного процесса, можно построить процесс размножения при столкновении следующим образом. С вероятностью $\frac{W_1}{W_0}$ исходная частица уничтожается, а послестолкновительная частица наследует её максимально допустимый вес. В противном случае (с вероятностью $1 - \frac{W_1}{W_0}$) исходная частица выживает и сохраняет как свой текущий вес $W_0$, так и максимально допустимый вес, а максимально допустимый вес послестолкновительной частицы разыгрывается по формуле: $W_{MAX} = \frac{W_1 W_0}{W_1 + RF(W_0 - W_1)}$, где $RF$ – случайное число в интервале 0..1. Назовём такой алгоритм Способом I.

Описанная процедура может быть использована в схемах **wpp** и **tw**. Доработка схемы **rmax** автором не проводились, так как, при наличии схемы **tw**, применение схемы **rmax** оправдано в основном для нестационарных задач, которые в рамках настоящей работы не решались.

Однако, существует и другой способ построить процесс размножения при столкновении, опираясь на минимизацию разброса числа частиц. Алгоритм состоит в следующем. Сперва определяется, будет ли максимальный вес исходной частицы передан послестолкновительной частице (с вероятностью $\frac{W_1}{W_0}$), или же, напротив, останется за исходной частицей. В последнем случае, максимальный вес послестолкновительной частицы разыгрывается по формуле: $W_{MAX} = \frac{W_1 W_0}{W_1 + RF(W_0 - W_1)}$ . Если же исходная частица передала максимальный вес послестолкновительной, то её новый максимальный вес рассчитывается по формуле: $W_{MAX} = \frac{W_0(W_0 - W_1)}{W_0 - RF \cdot W_1}$. Далее рассматриваются три случая:

- $W_0 - W_1 \geq W_1$. В этом случае, исходная частица после столкновения приобретает вес $W_0 - W_1$.
- $W_0 - W_1 < W_1$, за исходной частицей остался прежний максимальный вес. В этом случае, вес исходной частицы после столкновения будет составлять $W_1$.
- $W_0 - W_1 < W_1$, максимальный вес исходной частицы передан послестолкновительной частице. В этом случае, если разыгранный новый максимальный вес исходной частицы составляет не менее $W_1$, то частица выживает и приобретает вес $W_1$. В противном случае, исходная частица уничтожается.

Последний вариант алгоритма назовём Способом II.

Испытания подхода на схеме **tw** показали, что, при использовании Способа II, погрешности у оси слегка более заметны, по сравнению со Способом I. Тем не менее, различие весьма незначительно и обнаруживается лишь при заведомо некорректном задании параметров счета: временной шаг сравним со временем между столкновениями или превышает его, причём частицы пролетают за временной шаг несколько ячеек сетки. Если временной шаг уменьшить до 1/3 от времени между столкновениями, погрешности становятся необнаружимы в обоих случаях. Следовательно, можно рекомендовать пользоваться Способом I как более простым в реализации и требующим меньше операций, и только в случае обнаружения заметных искажений параметров вблизи оси попробовать воспользоваться Способом II.

# 9 Детектирование повторных столкновений

С некоторой вероятностью, неизбежны повторные столкновения, когда при стохастическом выборе сталкивающейся пары оказывается, что эти две молекулы уже сталкивалась между собой ранее (и после этого обе не сталкивались с другими молекулами).

Для модели твёрдых сфер угол рассеяния при столкновении изотропен, поэтому серия повторных столкновений практически идентична одному столкновению, за исключением лишь того, что, между повторными столкновениями, возможно некоторое смещение частиц. В этом случае наличие повторных столкновений можно грубо рассматривать просто как занижение частоты столкновений молекул, т.е. как нехватку столкновений. Для более сложных моделей столкновений, это уже не так.

Простейший способ избежать повторных столкновений – для каждой молекулы запоминать последнего партнера и выбирать другую пару, если столкновение повторное. Это, однако, возможно лишь в том случае, если молекул в ячейке больше двух. При любом небольшом среднем числе частиц в ячейке, вероятность наличия в ячейке всего двух молекул остаётся конечной. Но, при не очень малом среднем числе частиц в ячейке, метод может быть довольно эффективен. При том же значении $N_C Kn_C$, точность расчета может повыситься. В настоящей работе такой подход систематически не исследовался.

Если же используется неконсервативная схема с размножением частиц при столкновении, то обнаружение повторных столкновений существенно усложняется. В частности, за счёт столкновения одной и той же пары может образоваться сразу несколько молекул, которые нежелательно более сталкивать друг с другом. В схемах с размножением частиц при перемещении, детектирование нисколько не проще, так как логически разные размноженные частицы могут иметь совпадающие параметры. Но именно неконсервативные схемы генерируют малоприятные артефакты при повторных столкновениях.

В целях борьбы с этими артефактами был опробован подход, предполагающий детектирование и последующее игнорирование повторных столкновений, без попытки выбрать другую пару.

Заранее стоит сказать, что, вместо того, чтобы просто игнорировать повторные столкновения, можно, по-прежнему, выбирать другую пару. Конечно, кроме проблемы детектирования повторных столкновений, возникает задача определить, возможны ли в ячейке не повторные столкновения вообще. Если нет, то цикл выбора другой пары никогда не остановится. Простейший выход из ситуации может заключаться в прерывании цикла, если несколько (например, три) подряд выбранных пар частиц определены как повторные. Никаких испытаний подходов с выбором альтернативной пары, опять же, автором настоящей работы не проводилось, в том числе и для неконсервативных схем.

Предлагаемый алгоритм детектирования повторных столкновений следующий. Для каждой частицы теперь дополнительно хранятся три 16-битных числа: *hash*, *hashA*, *hashB*, всего 6 байт на частицу. Число *hash* – это «уникальный» номер молекулы, а по сути – случайное число 0÷65535, которое при каждом столкновении генерируется заново для каждой частицы, порождённой в столкновении. Числа *hashA*, *hashB* – номера *hash* двух партнеров исходной пары, при столкновении которой образована новая частица. При введении новой частицы в поток через границу области, всем трём переменным присваивается одно и то же случайное число.

После выбора пары, шесть чисел (по три на частицу) сравниваются между собой, совпадение говорит о повторности столкновения. Опробованы три «уровня глубины» детектирования. Итак, пусть сталкиваются две исходных молекулы ($A$, $B$), после столкновения которых генерируется пара молекул ($A'$, $B'$). Уровни контроля задаются следующим образом:

0) Контроль повторных столкновений отключён.

1) Если для молекулы А хотя бы одно из её чисел *A.hashA*, *A.hashB* совпадает с числом *B.hash* (молекулы *B*), то столкновение является повторным для молекулы *A*. В этом случае вес молекулы *A* не уменьшается, молекула *A′* в поток не запускается. Аналогично проверяется повторность столкновения относительно молекулы *B*.

2) Столкновение отменяется сразу для обеих частиц, если выполнено хотя бы одно из условий: *A.hashA==B.hash, A.hashB==B.hash, A.hash==B.hashA, A.hash==B.hashB*. Т.е., согласно уровню 1), столкновение является повторным хотя бы для одной частицы.

3) Столкновение считается повторным согласно уровню 2), а также в случаях: *A.hashA==B.hashA, A.hashA==B.hashB, A.hashB==B.hashA, A.hashB==B.hashB.*

Уровень 1) оказался непригоден – его асимметричность может порождать дополнительные артефакты, особенно при использовании схем **tw** и **tnw**. Уровень 2) уменьшает величину артефактов (прежде всего расхождение температур компонент) примерно вдвое по сравнению с 0) и не порождает таких новых артефактов, как уровень 1). При использовании уровня 3) отменяется примерно вдвое больше столкновений по сравнению с уровнем 2), однако, сколько-либо заметного дополнительного уменьшения артефактов не наблюдается. Кроме того, уровень 3) требует в два раза больше проверок.

Испытание подхода показало, что введение детектирования уровня 2) позволяет заметно ослабить артефакты неконсервативных схем. При этом дополнительно требуется 6 байт на молекулу. Вероятность ложной детекции невелика и составляет около $\approx {}^{1}/_{16384}$.

Для демонстрации преимуществ выбрана задача о конвергентном течении. Намеренно использовалось не очень большое число частиц, чтобы подчеркнуть артефакты. Выбран диаметр источника $D/H_* = 15$, число Кнудсена 0.02, состав смеси 90 % He + 10 % Xe. Для ксенона использован компонентный весовой множитель 0.25. Для столкновений He – He, значение $N_C Kn_C \approx 8$.

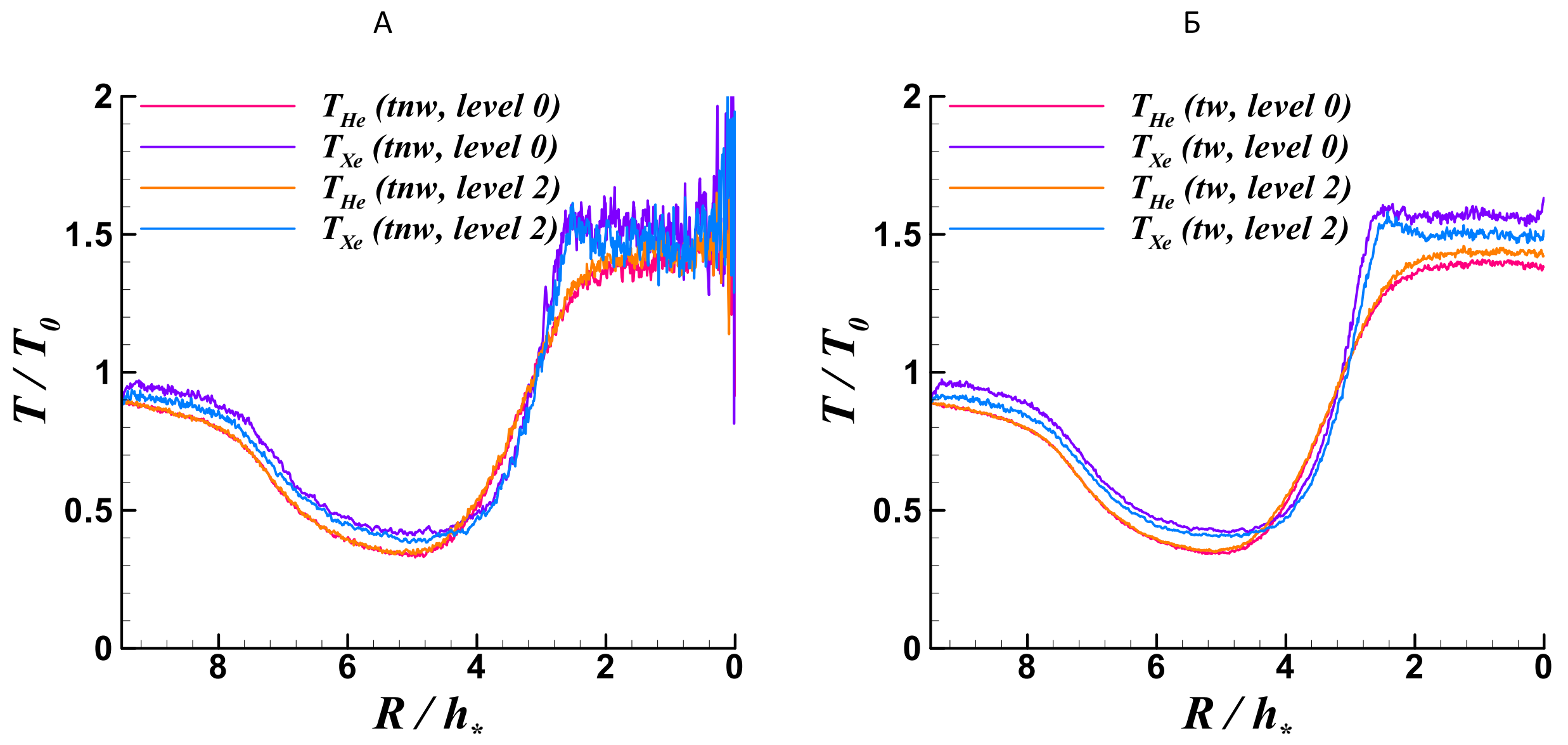

**Рис. 8.4-1.** Температуры в плоскости симметрии конвергентного источника, рассчитанные с контролем повторных столкновений (уровень 2) и без него (уровень 0). Использованы схемы **tnw** (А) и **tw** (Б).

На **Рис. 8.4-1** (А, Б) показано сравнение зависимостей температур гелия и ксенона в плоскости симметрии, рассчитанные схемой **tnw** (А) и схемой **tw** (Б). В данном случае, зоны $R/h_* < 1.5$ и $R/h_* > 7$ являются околоравновесными, температуры компонент в них не должны сильно отличаться, и разница температур компонент в этих зонах может считаться артефактом. Схема **tw** демонстрирует снижение разницы температур компонент почти в два раза в области $R/h_* > 7$ и в 3 раза в области $R/h_* < 1.5$. Схема **tnw** обладает низкой сходимостью статистики у оси, поэтому в зоне $R/h_* < 1.5$ оценить разницу затруднительно, а в зоне $R/h_* > 7$ наблюдается разница того же порядка, как и для схемы **tw**.

# 10 Заключение

Выше были описаны доработки метода ПСМ, повышающие его эффективность для современных задач газовой динамики.

- Предложен критерий подобия, позволяющий оценить требуемое для расчета число частиц. Показано, что критерий не зависит от локальной плотности.
- Предложена и испытана схема временных множителей, особенно полезная для расчета смесей газов с сильно отличающимися сечениями столкновения компонент смеси.
- На основе схемы временных множителей создана новая схема решения осесимметричных задач, в которой недостаток статистически независимых расчетных частиц вблизи оси подавляется при помощи уменьшения временного шага для частиц, приближающихся к оси и при этом длительное время не испытывающих столкновений.
- Предложен подход для борьбы с флуктуациями числа частиц при их размножении-уничтожении.
- Предложен метод отбраковки повторных столкновений.
- Предложены некоторые рекомендации по технической оптимизации алгоритма.

Данные подходы, совместно с другими известными доработками, такими как схема мажорантной частоты и пространственно-зависимый временной шаг, превращают метод ПСМ в мощный инструмент моделирования течений газовых смесей, для которых характерны большие перепады плотности и большие отношения сечений столкновения между компонентами смеси.

# 11 Благодарности

Автор выражает благодарность Институту Теплофизики СО РАН за поддержку при выполнении настоящей работы. Особые  благодарности Михаилу Ю. Плотникову и Алексею А. Морозову за обсуждение результатов работы, ценные комментарии и помощь в подготовке рукописи.

# 12 Приложение: Новый алгоритм выборки из гамма-распределения

Для числа степеней свободы менее 2 используется классический подход (25). Для числа степеней свободы более 2 был разработан собственный алгоритм выборки из гамма-распределения, основанный на принципе принятия-отклонения. Охватывающая функция распределения строится из двух частей: левая часть представляет собой степенную функцию, правая – затухающую экспоненту. Коэффициенты двух кривых и точка раздела между ними вычисляются по эмпирическим формулам. Пример охватывающей функции представлен на **Рис. 12.2-1**.

Алгоритм выборки из гамма-распределения включает в себя около десятка констант, рассчитываемых для заданного числа степеней свободы. Алгоритм разделен на две части. Первая часть (функция ***GammaGen::init***) заполняет структуру с набором констант. Вторая часть (функция ***GammaGen::get***) получает эту структуру в качестве аргумента.

Реализация алгоритма генерирует около 8 млн. чисел в секунду на одном ядре процессора Core2Duo E6600 2.4 GHz. Сравнение скорости выборки из бета-распределения (используется в модели Боргнакке-Ларсена), основанного на новом алгоритме, с классическим представлено на **Рис. 12.2-2**. Как видно, классический алгоритм оказывается примерно вдвое быстрее при 2-4 внутренних степенях свободы, при 10 степенях свободы быстродействия двух алгоритмов практически совпадают, а при дальнейшем увеличении числа внутренних степеней свободы, быстродействие классического алгоритма катастрофически падает, в то время как быстродействие нового алгоритма стабильно.

## 12.1 Код алгоритма на C++

```
struct GammaGen
{
	double alpha;
	double beta;
	double gamma;
	double lnx0;
	double l_p1;
	double l_p2;
	double r_p1;
	double r_p2;
	double r_p3;
	double bp1;

	void init(double shape)
	{
		if (shape > 1)
		{
			double ialpha = sqrt(shape);
			alpha = 1/ialpha;
			double x0 = (shape-ialpha)*exp(alpha);
			double s = shape-x0;
			beta = 0.5*(s+sqrt(4*x0+s*s));
			double ibeta = 1 / beta;
			gamma = shape-1;
			lnx0 = log(x0);
			l_p1 = shape - ialpha;
			l_p2 = l_p1 * (1-log(l_p1));
			r_p2 = 1-ibeta;
```

```
                r_p3 = gamma * (log(gamma/r_p2)-1);
                double logb = r_p3 + l_p2 - ialpha * lnx0 -
                     x0 * ibeta + log(beta * ialpha);
                bp1 = 1+exp(logb);
                r_p1 = x0 + beta * logb;
           } else
           {
                double ialpha = shape;
                alpha = 1 / ialpha;
                double x0 = sqrt(shape);
                double ibeta = beta = 1;
                gamma = shape-1;
                lnx0 = log(x0);
                l_p1 = 0;
                l_p2 = 0;
                r_p2 = 0;
                r_p3 = gamma * lnx0;
                double s = ialpha/x0;
                bp1 = 1+s*exp(-x0);
                r_p1 = log(s);
           }
      }

      double get()
      {
           for (;;)
           {
                double p = bp1 * get_random_real();
                double r = log(get_random_real());
                double x,w;
                if (p <= 1)
                {
                     double a = lnx0 + alpha*log(p);
                     x = exp(a);
                     w = l_p1*a-x+l_p2;
                } else
                {
                     x = r_p1 - beta * log(p-1);
                     w = gamma*log(x) - r_p2*x-r_p3;
                }
                if (r < w) return x;
           }
      }
};
```

## 12.2 Тестирование алгоритма.

Тестирование производилось следующим образом. При помощи тестируемой функции генерировалось несколько выборок из гамма-распределения, затем для них на основе методики (26) вычислялись статистики критериев согласия с теоретической функцией распределения. Использовались следующие статистики: Колмогорова $S_K$, Смирнова $S_m$, Крамера–Мизеса–Смирнова $S_\omega$, Андерсона–Дарлинга $S_\Omega$. Каждый из этих критериев суть мера отличия полученной экспериментальной функции распределения (построенной по выборке) от предполагаемой теоретической, особым образом нормированная на размер выборки. Если выборка соответствует распределению, то значение каждого из этих критериев случайно, но подчиняется некоторой известной образцовой функции распределения (например, статистика Колмогорова $S_K$ подчиняется распределению Колмогорова). Если же выборка не соответствует предполагаемой теоретической функции распределения, то величины статистик более не укладываются в образцовые функции распределения и быстро нарастают с ростом размера выборки.

В качестве примера, в **Табл. 12-1** показаны значения статистик при тестировании функции выборки из гамма-распределения при количестве степеней свободы 0.5, 3 и 10. Каждый из тестов был повторен 9 раз и производился на выборке размером 100 000 000. Для сравнения, представлены значения статистик в случае, когда сравнение производится с несколько искаженными функциями распределения: для числа степеней свободы 0.5005 и 3.003 и 10.01 (на 0.1 % выше исходного). Можно видеть, что, даже при таком небольшом несоответствии распределений, значения статистик существенно возрастают. Для 3 и 10 степеней свободы была протестирована функция из текста программ Бёрда. Если для 3 степеней соответствие оказалось в пределах разумного, то уже для 10 степеней свободы искажение функции распределения налицо.

Для вычисления образцовой функции распределения использовалась библиотека численного анализа (27)

**Табл. 12-1.** Результаты тестирования алгоритма выборки из гамма-распределения

| Число степеней свободы при генерации выборки | Число степеней свободы сверочной функции распределения | Вычисленные значения статистик | | | |
|---|---|---|---|---|---|
| | | $S_K$ | $S_m$ | $S_\omega$ | $S_\Omega$ |
| 0.5 | 0.5005 | 3.689641 | 54.45380 | 7.240794 | 39.19554 |
| | | 4.228223 | 71.51149 | 8.350878 | 45.46036 |
| | | 4.645225 | 86.31247 | 9.8037 | 53.30613 |
| | | 4.4522 | 79.28832 | 9.734659 | 51.6677 |
| | | 3.605921 | 52.01065 | 6.068401 | 33.87501 |
| | | 3.857159 | 59.51070 | 7.498696 | 43.54092 |
| | | 2.996927 | 35.92628 | 4.782118 | 28.27664 |
| | | 3.570501 | 50.99391 | 4.820722 | 26.25529 |
| | | 4.349745 | 75.68112 | 10.1795 | 55.94271 |
| 0.5 | 0.5 | 0.854846 | 2.923046 | 0.085123 | 0.64998 |
| | | 0.929491 | 3.455814 | 0.148519 | 0.907524 |
| | | 0.999021 | 0.736596 | 0.189666 | 1.231211 |
| | | 0.853583 | 2.914418 | 0.103284 | 0.724369 |
| | | 1.008742 | 4.070244 | 0.211939 | 1.254673 |
| | | 0.815194 | 2.658168 | 0.105332 | 0.603958 |
| | | 0.448005 | 0.802835 | 0.034838 | 0.302223 |
| | | 0.511888 | 0.215562 | 0.04301 | 0.375158 |
| | | 0.941176 | 3.543251 | 0.131994 | 0.722955 |
| 3 | 3.003 | 5.675351 | 128.8384 | 18.54901 | 98.8144 |
| | | 6.079349 | 147.8339 | 19.84912 | 106.727 |
| | | 5.494808 | 120.7717 | 15.61166 | 83.36814 |
| | | 6.261752 | 156.8381 | 21.85383 | 115.2126 |
| | | 5.761025 | 132.7576 | 18.24985 | 95.15996 |
| | | 5.42536 | 117.7381 | 15.79119 | 85.35641 |
| | | 5.960689 | 142.1192 | 18.87967 | 100.3314 |
| | | 5.702442 | 130.0714 | 17.96158 | 93.63478 |
| | | 6.228521 | 155.1779 | 21.47791 | 116.3242 |
| 3 | 3 | 0.956549 | 1.049887 | 0.189627 | 0.89525 |
| | | 1.303129 | 6.792585 | 0.377277 | 1.729676 |
| | | 1.167986 | 5.456765 | 0.172275 | 0.882733 |
| | | 0.960795 | 3.692511 | 0.180477 | 0.975291 |
| | | 0.855415 | 2.816303 | 0.217142 | 1.262431 |
| | | 0.857568 | 2.941689 | 0.088226 | 0.453152 |
| | | 0.857103 | 2.938502 | 0.138324 | 0.730843 |
| | | 0.958112 | 3.671915 | 0.188492 | 1.030596 |
| | | 0.791987 | 2.508974 | 0.093627 | 0.699842 |
| 3 (код Бёрда) | 3 | 2.061543 | 16.99984 | 1.366306 | 24.94822 |
| | | 1.919349 | 14.7356 | 0.839222 | 23.53953 |
| | | 2.137955 | 18.28341 | 1.667305 | 26.92417 |
| | | 1.711401 | 11.71557 | 1.155739 | 23.05634 |
| | | 1.736262 | 12.05842 | 0.790997 | 21.35872 |
| | | 1.756274 | 12.338 | 0.869213 | 21.53808 |
| | | 1.837666 | 13.50807 | 0.789766 | 23.038 |
| | | 1.8152 | 13.17981 | 1.290437 | 25.59783 |
| | | 1.908628 | 14.57145 | 0.844811 | 24.31927 |
| 10 | 10.01 | 9.300668 | 346.0097 | 44.63087 | 232.1158 |
| | | 8.770661 | 307.698 | 46.11035 | 252.1073 |
| | | 9.89668 | 391.7771 | 48.48008 | 250.9043 |
| | | 9.658494 | 373.146 | 48.7174 | 256.0664 |
| | | 9.932141 | 394.5897 | 52.79102 | 273.9844 |
| | | 9.178863 | 337.0061 | 48.46185 | 257.4466 |
| | | 9.593468 | 368.1385 | 53.76679 | 285.4175 |
| | | 9.772626 | 382.0169 | 51.49459 | 266.905 |
| | | 9.230357 | 340.7979 | 47.07379 | 252.4219 |
| 10 | 10 | 0.573069 | 1.313634 | 0.072472 | 0.678899 |
| | | 0.989985 | 3.920284 | 0.239868 | 1.428942 |
| | | 0.631982 | 0.13088 | 0.076777 | 0.650964 |
| | | 1.048939 | 0.293291 | 0.135537 | 1.124916 |
| | | 0.718409 | 1.953424 | 0.093302 | 0.533794 |
| | | 1.209206 | 0.448888 | 0.337763 | 1.800649 |
| | | 0.729103 | 2.126366 | 0.0955 | 0.617484 |
| | | 1.059261 | 0.290167 | 0.176194 | 1.149087 |
| | | 0.762857 | 2.327803 | 0.11333 | 0.790154 |
| 10 (код Бёрда) | 10 | 292.5306 | 342296.6 | 28477.28 | 275894.8 |
| | | 292.527 | 342288.2 | 28574.03 | 276436.7 |
| | | 292.5281 | 342290.8 | 28535.01 | 276199 |
| | | 292.5282 | 342290.9 | 28488.54 | 275775.9 |
| | | 292.5276 | 342289.6 | 28443.21 | 275703.6 |
| | | 292.5303 | 342295.9 | 28394.46 | 275583.7 |
| | | 292.5281 | 342290.7 | 28364.69 | 275201.5 |
| | | 292.5269 | 342288 | 28580.2 | 276525.9 |
| | | 292.5272 | 342288.6 | 28635.09 | 276870.4 |
| 2000 | 2002 | 125.6203 | 63121.86 | 9151.74 | 47960.98 |
| | | 126.0314 | 63535.62 | 9175.865 | 48007.43 |
| | | 126.412 | 63919.97 | 9199.891 | 48093.29 |
| | | 126.6074 | 64117.71 | 9270.717 | 48465.51 |
| | | 126.1731 | 63678.57 | 9174.696 | 47996.91 |
| | | 125.3847 | 62885.3 | 9122.127 | 47761.86 |
| | | 126.2336 | 63739.72 | 9203.424 | 48129.84 |
| | | 126.4675 | 63976.14 | 9225.494 | 48223.23 |
| | | 125.9191 | 63422.47 | 9146.907 | 47844.74 |
| 2000 | 2000 | 0.622973 | 0.208317 | 0.068829 | 0.472358 |
| | | 0.699993 | 0.367599 | 0.105044 | 0.964541 |
| | | 0.692793 | 1.919847 | 0.089294 | 0.645757 |
| | | 0.654269 | 1.075807 | 0.049229 | 0.303368 |
| | | 0.859795 | 2.956992 | 0.189296 | 1.194068 |
| | | 0.836351 | 0.927202 | 0.145153 | 0.753433 |
| | | 0.895596 | 1.030379 | 0.089315 | 0.722913 |
| | | 0.79266 | 0.574676 | 0.065028 | 0.427412 |
| | | 0.804569 | 2.589322 | 0.17783 | 1.087046 |

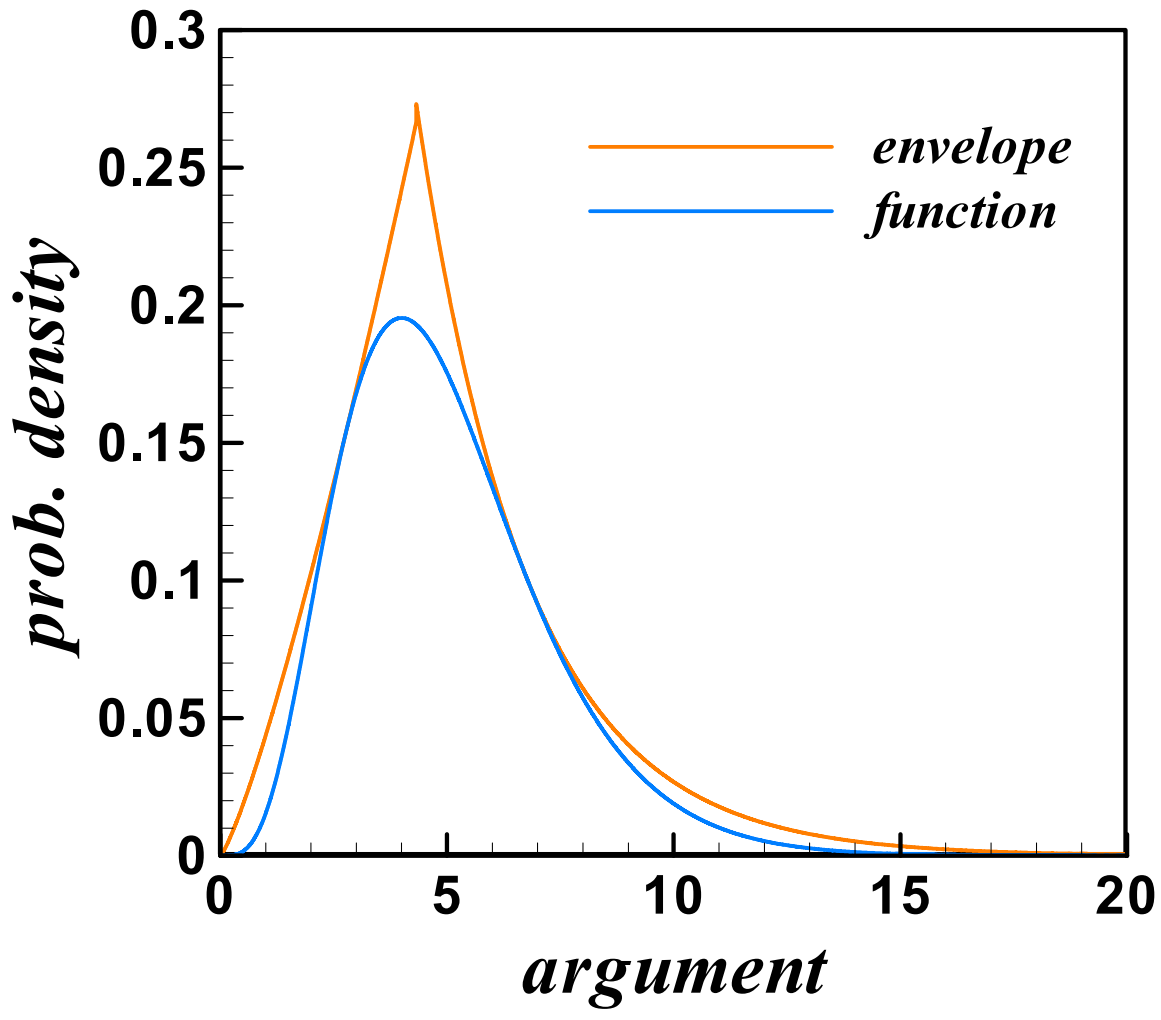

**Рис. 12.2-1**. Плотность вероятности гамма-распределения для 10 степеней свободы (голубая кривая), охватанная функцией-конвертом (коричневая кривая).

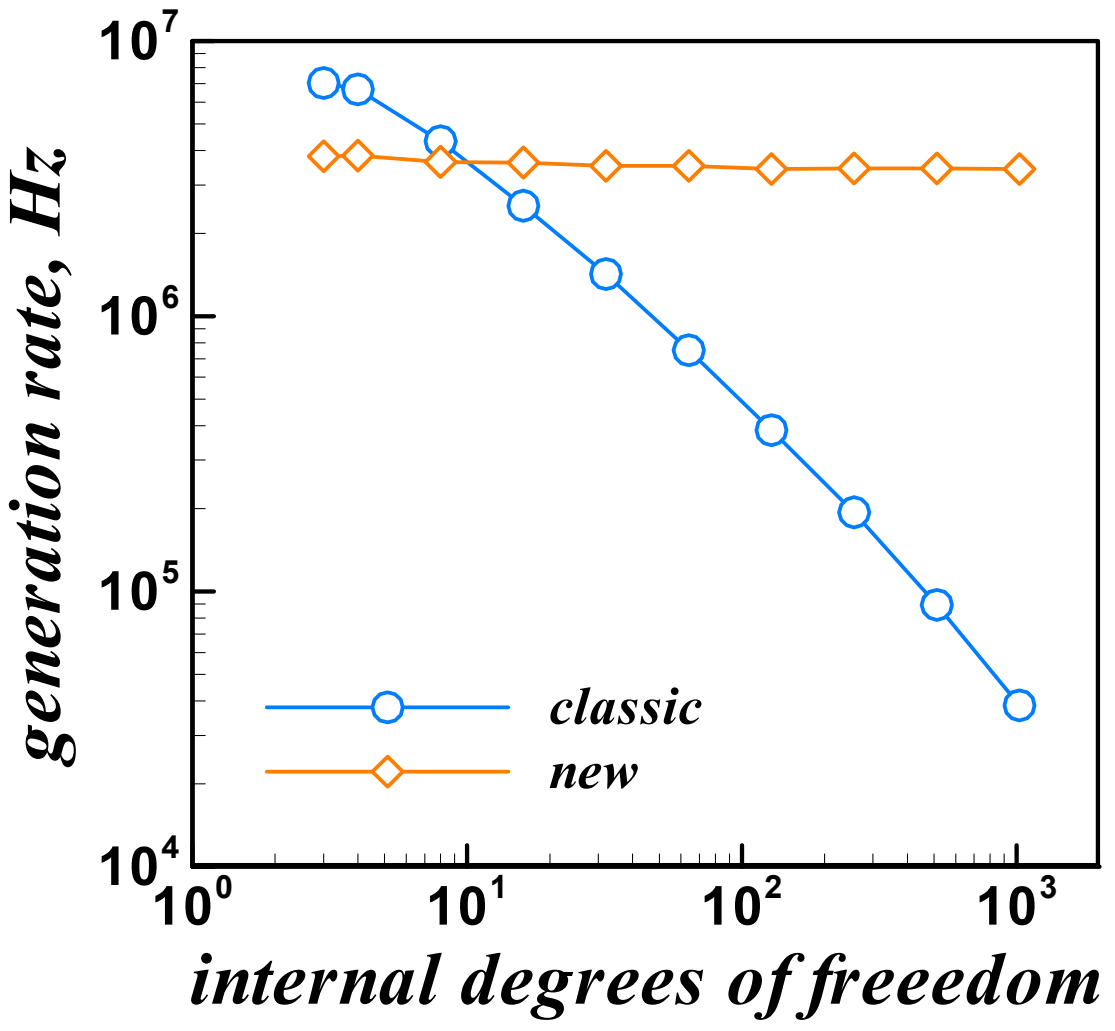

**Рис. 12.2-2**. Скорость генерации выборки из бета-распределения при разном числе внутренних степеней свободы. Число поступательных степеней свободы составляет 3.5 (ω = 0.75). Голубая кривая – классический подход, коричневая кривая – новый подход.

# Список литературы